%% file: main.tex
\documentclass[fleqn,usenatbib]{mnras}

\usepackage{newtxtext,newtxmath}

\usepackage[T1]{fontenc}

\DeclareRobustCommand{\VAN}[3]{#2}
\let\VANthebibliography\thebibliography
\def\thebibliography{\DeclareRobustCommand{\VAN}[3]{##3}\VANthebibliography}

\usepackage{graphicx}	
\usepackage{amsmath}	
\usepackage{booktabs}
\usepackage{multirow}

\newcommand{\bagpipes}{\mbox{\sc{Bagpipes}}}

\newcommand{\cloudy}{\mbox{\sc cloudy}}
\newcommand{\colibre}{\mbox{\sc Colibre}}
\newcommand{\consistenttrees}{\mbox{\sc Consistent Trees}}
\newcommand{\dhalo}{\mbox{\sc{Dhalo}}}
\newcommand{\dtrees}{\mbox{\sc{D-Trees}}}
\newcommand{\eagle}{\mbox{\sc{Eagle}}}

\newcommand{\flares}{\mbox{\sc Flares}}
\newcommand{\gadget}{\mbox{\sc Gadget}}
\newcommand{\galform}{\mbox{\sc Galform}}
\newcommand{\grasil}{\mbox{\sc Grasil}}

\newcommand{\HBTherons}{\mbox{\sc HBT-Herons}}
\newcommand{\lgalaxies}{\mbox{\sc L-Galaxies}}
\newcommand{\lhalotree}{\mbox{\sc LHaloTree}}
\newcommand{\pmill}{\mbox{\sc P-Millennium}}
\newcommand{\mill}{\mbox{\sc Millennium}}
\newcommand{\rockstar}{\mbox{\sc Rockstar}}

\newcommand{\scsam}{\mbox{\sc SC-SAM}}
\newcommand{\shark}{\mbox{\sc Shark}}
\newcommand{\subfind}{\mbox{\sc Subfind}}
\newcommand{\sublink}{\mbox{\sc Sublink}}

\title[FLARES XX: Comparing SAMs at high-z]{First Light And Reionisation Epoch Simulations (FLARES) XX: Comparing semi-analytic models at high-redshift}

\author[Seeyave et al.]{Louise T. C. Seeyave$^{1}$\thanks{E-mail: L.Seeyave@sussex.ac.uk. \newline Note that the order of the author list is explained in the acknowledgements.},
Carlton M. Baugh$^{2}$,
Ángel Chandro-Gómez$^{3}$, 
Claudia del P. Lagos$^{3}$,
\newauthor
Robert M. Yates$^{4}$, 
L. Y. Aaron Yung$^{5}$, 
Rachel S. Somerville$^{6}$, 
Stephen M. Wilkins$^{1,7}$, 
\newauthor
Christopher C. Lovell$^{8,9}$, 
William J. Roper$^{1}$, 
Aswin P. Vijayan$^{1}$, 
Cedric G. Lacey$^{2}$,
Chris Power$^{3}$, 
\newauthor
Shihong Liao$^{10}$, 
Maxwell G. A. Maltz$^{1}$,
Jack C. Turner$^{1}$
\\
$^{1}$Astronomy Centre, University of Sussex, Falmer, Brighton BN1 9QH, UK\\
$^{2}$Institute for Computational Cosmology, Department of Physics, Science Laboratories, South Road, Durham, DH1 3LE, UK\\
$^{3}$International Centre for Radio Astronomy Research, The University of Western Australia, M468, 35 Stirling Highway, Crawley, WA 6009, Australia\\
$^{4}$Centre for Astrophysics Research, University of Hertfordshire, Hatfield, AL10 9AB, UK\\
$^{5}$Space Telescope Science Institute, 3700 San Martin Dr., Baltimore, MD 21218, USA\\
$^{6}$Center for Computational Astrophysics, Flatiron Institute, 162 5th Avenue, New York, NY 10010, USA\\
$^{7}$Institute of Space Sciences and Astronomy, University of Malta, Msida MSD 2080, Malta\\
$^{8}$Kavli Institute for Cosmology, Madingley Road, Cambridge CB3 0HA, UK\\
$^{9}$Institute of Astronomy, Madingley Road, Cambridge CB3 0HA, UK\\
$^{10}$Key Laboratory for Computational Astrophysics, National Astronomical Observatories, Chinese Academy of Sciences, Beijing 100101,
China
}

\date{Accepted XXX. Received YYY; in original form ZZZ}

\pubyear{\the\year{}}

\begin{document}
\label{firstpage}
\pagerange{\pageref{firstpage}--\pageref{lastpage}}
\maketitle

\begin{abstract}
We explore how the choice of galaxy formation model affects the predicted properties of high-redshift galaxies. Using the \flares\ zoom resimulation strategy, we compare the \eagle\ hydrodynamics model and the \galform, \lgalaxies, \scsam\ and \shark\ semi-analytic models (SAMs) at $5\leq z \leq 12$. The first part of our analysis examines the stellar mass functions, stellar-to-halo mass relations, star formation rates, and supermassive black hole (SMBH) properties predicted by the different models. Comparisons are made with observations, where relevant. We find general agreement between the range of predicted and observed stellar mass functions. The model predictions differ considerably when it comes to SMBH properties, with \galform\ and \shark\ predicting between $1.5-3\ {\rm dex}$ more massive SMBHs ($M_{\rm BH}>10^6\ {\rm M_\odot}$) than \lgalaxies\ and \scsam, depending on redshift. The second half of our analysis focuses on passive galaxies. We show that in \lgalaxies\ and \scsam, environmental quenching of satellites is the prevalent quenching mechanism, with active galactic nuclei (AGN) feedback having little effect at the redshifts probed. On the other hand, $\sim40\%$ of passive galaxies predicted by \galform\ and \shark\ are quenched by AGN feedback at $z=5$. The SAMs are an interesting contrast to the \eagle\ model, in which AGN feedback is essential for the formation of passive galaxies, in both satellites and centrals, even at high redshift.
\end{abstract}




\input{sections/1_intro}
\input{sections/2_models}
\input{sections/3_method}
\input{sections/4_properties}
\input{sections/5_passive}
\input{sections/6_conclusion}

\section*{Acknowledgements}

We wish to thank the \eagle\ team for their efforts in developing the \eagle\ simulation code, as well as Scott Kay and Adrian Jenkins for their invaluable help getting up and running with the \eagle\ resimulation code. We also very grateful to John Helly for his advice with running \galform, and for allowing us to use the \dhalo\ code.

This work used the DiRAC@Durham facility managed by the Institute for Computational Cosmology on behalf of the STFC DiRAC HPC Facility (www.dirac.ac.uk). The equipment was funded by BEIS capital funding via STFC capital grants ST/K00042X/1, ST/P002293/1, ST/R002371/1 and ST/S002502/1, Durham University and STFC operations grant ST/R000832/1. DiRAC is part of the National e-Infrastructure. Additionally, we have had the assistance of resources from the National Computational Infrastructure (NCI Australia), an NCRIS enabled capability supported by the Australian Government. We wish to acknowledge the following open source software packages used in the analysis: \textsc{Matplotlib} \citep{Hunter:2007}, \textsc{Numpy} \citep{harris2020array}. 

LTCS, MGAM and JCT are supported by STFC studentships. CMB and CGL acknowledge support from the STFC through ST/X001075/1. The Flatiron Institute is supported by the Simons Foundation. WJR, APV and SMW acknowledge support from the Sussex Astronomy Centre STFC Consolidated Grant ST/X001040/1. CP acknowledges the support of the ARC Centre of Excellence for All Sky Astrophysics in 3 Dimensions (ASTRO 3D), through project number CE170100013.

The ordering of the author list was decided as follows: 1) main author, 2) authors who were highly involved in running the SAMs or helping the main author to do so (in alphabetical order), 3) remaining authors. We list here the roles and contributions of the authors according to the Contributor Roles Taxonomy (CRediT)\footnote{\url{https://credit.niso.org/}}.
\textbf{Louise T. C. Seeyave}: Conceptualization, Data curation, Methodology, Investigation, Formal Analysis, Visualization, Writing - original draft.
\textbf{Stephen M. Wilkins}: Conceptualization, Methodology, Writing - review \& editing.
\textbf{Christopher C. Lovell,  Aswin P. Vijayan}: Data curation, Methodology, Writing - review \& editing.
\textbf{Carlton M. Baugh, Ángel Chandro-Gómez, Claudia del P. Lagos, Chris Power, Rachel S. Somerville, Robert M. Yates, L. Y. Aaron Yung}: Data curation, Writing - review \& editing.
\textbf{Cedric G. Lacey, Shihong Liao, Maxwell A.G. Maltz, William J. Roper, Jack C. Turner}: Writing - review \& editing.

\section*{Data Availability}

Data is available on request from the corresponding author. Plots from the convergence tests may be found in the corresponding author's thesis, which will be publicly available shortly.



\bibliographystyle{mnras}
\bibliography{cite/algorithms, cite/flares, cite/general, cite/sams} 




\appendix

\input{sections/appendix}


\bsp	
\label{lastpage}
\end{document}

%% file: sections/1_intro.tex
\section{Introduction}\label{sec:intro}

Galaxy formation and evolution is a rich interplay of numerous physical processes. In hierarchical structure formation, a feature of the widely-used $\Lambda$CDM cosmological model, gravitationally-bound dark matter halos grow through repeated mergers. Baryonic matter traces the distribution of dark matter, with gas accreting onto dark matter halos and cooling to form stars. Over time, these clusters of gas and stars merge to form galaxies. Key baryonic processes in galaxies include: the aforementioned formation of stars from cold gas, stellar feedback heating up the interstellar medium (ISM) and enriching it with metals, the formation of black holes and their subsequent growth through accretion of matter, and feedback from active galactic nuclei (AGN) in the form of radiation and jets. Galaxies are also affected by their environment, for example, through tidal interactions or mergers with other galaxies \citep{Benson_2010, Somerville_2015}. 

The picture of how galaxies form and evolve is continually developing. In the last few years, observations by the \textit{James Webb Space Telescope} \cite[\textit{JWST},][]{Gardner_2006} have opened up new debate around our understanding of high-redshift ($z>5$) galaxies. Much emphasis has been placed on the large numbers of UV-luminous galaxies observed at $z>10$ \citep{Castellano_2023, Finkelstein_2023, Leung_2023, Adams_2024, Finkelstein_2024, Robertson_2024, Whitler_2025}, the abundance of passive galaxies measured at $z=4-5$ \citep{Alberts_2024, Russell_2024, Weibel_2024_RUBIES, Xiao_2024, Baker_2025, Zhang_2025}, as well as the populations of AGN discovered at $z=5-8$ \citep{Harikane_2023, Greene_2024, Matthee_2024}. Some findings, particularly those at higher redshifts, are in excess of a number of theoretical predictions \citep[][]{Lovell_2023_EVS, Adams_2024, Weibel_2024_RUBIES}. However, physical properties estimated from observations are often subject to large uncertainties \citep{Suess_2022, Desprez_2024, Trussler_2024, Narayanan_2024}. In addition, theoretical studies have suggested that the tension between observed and predicted UV luminosity functions (UVLFs) can be reduced with reasonable adaptations to the physics modelling, for example, by increasing the star formation efficiency \citep{Lu_2025_GALFORM, Somerville_2025} or invoking bursty star formation \citep{Sun_2023, Gelli_2024}.

Passive galaxies, also known as `quiescent' or `quenched' galaxies, are those in which star formation has ceased. This can occur through internal channels, such as feedback events that heat and expel cold gas, leaving little to form stars. In low-mass galaxies, stellar feedback is sufficient to drive such a process \citep{Merlin_2012}, while in high-mass galaxies, AGN feedback is thought to be required for gas to escape the deeper potential wells \citep{DiMatteo_2005, Croton_2006, Somerville_2008_SCSAM}. Environmental processes can also drive passivity. Satellites may be `starved' of cold gas \citep{Feldmann_2015, Trussler_2020}, or undergo ram pressure stripping as they travel through the intracluster medium  \citep[ICM;][]{Simpson_2018, Xu_2025}. Since dense galaxy clusters do not yet exist at $z>2$ \citep{Overzier_2016}, environmental processes are expected to play less of a role at high redshifts. The number density of passive galaxies in the early universe may thus offer useful information about the aforementioned internal feedback channels.

There are two main approaches to modelling the physics involved in galaxy formation and evolution -- semi-analytic models (SAMs) and numerical hydrodynamic simulations. In SAMs, galaxies are modelled analytically atop the merger histories of dark matter halos. These merger histories are encapsulated in merger trees, which can be obtained from dark matter-only (DMO) N-body simulations \citep[][]{Behroozi_2013_CTrees, Elahi_2019_TreeFrog, Roper_2020_MEGA} or Extended Press-Schechter (EPS) methods \citep[][]{Somerville_1999, Somerville_2008_SCSAM}. Physical processes in SAMs are described by equations that manipulate flows of baryonic material to and from different `reservoirs', typically representing stars, black holes, interstellar cold gas, circumgalactic hot gas and intracluster/intergalactic hot gas. Stellar and cold gas reservoirs may be split into bulge and disk components, or finer annuli \citep{Kauffmann_1996,Baugh_1996,Fu_2010_LGalaxies, Porter_2014,Stevens_2016_DARKSAGE, Henriques_2020_LGalaxies, Stevens_2024_DARKSAGE}. SAMs are simpler in design and hence computationally inexpensive compared to full hydrodynamical simulations, making them ideal for simulating large volumes of the universe \citep{Cowley_2018_GALFORM, Lu_2025_GALFORM} and exploring different physical models \citep[][]{Mutch_2016_MERAXES, Yung_2019_JWST_I, Hutter_2021_ASTRAEUS_III, Cueto_2024_ASTRAEUS_IX}.

In numerical hydrodynamic simulations, baryonic matter is modelled as a fluid. The equations of gravity, hydrodynamics, thermodynamics, and optionally magnetism and radiative transfer, are solved for individual particles or cells. Subgrid models are used to model physical processes that occur on scales smaller than can be resolved, for example the formation and feedback processes of stars and black holes. The higher spatial resolution of hydrodynamical simulations allows for more realistic representations of environmental processes, morphology, and galaxy kinematics than SAMs \citep{Mitchell_2018, Pandya_2020, Ayromlou_2021}. A consequence of this is increased computational expense, which leads to a strong trade-off between simulation resolution and volume.

Many comparison studies involving SAMs can be found in the literature, some solely focused on SAMs \citep{DeLucia_2010, Lu_2014, Saghiha_2017} and others including hydrodynamical simulations as a contrast, often with a view to understanding how the modelling in SAMs can be improved \citep{Hirschmann_2012, Monaco_2014, Guo_2016, Cui_2018, Mitchell_2018, Pandya_2020, Ayromlou_2021, Hadzhiyska_2021, Yates_2021b_LGalaxies, Gabrielpillai_2022, Lagos_2025}. To our knowledge, there has not been a comparison of different SAMs focused on the Epoch of Reionisation ($z>5$), though predictions by individual SAMs have certainly been made in this redshift regime \citep{Yung_2019_JWST_I, Yung_2019_JWST_II, Yung_2020_JWST_III, Yung_2021_JWST_V, Yung_2022_JWST_IV, Yung_2024_SCSAM, Lu_2025_GALFORM, Vani_2025, Somerville_2025}. Most models, be they hydrodynamical or semi-analytic, are calibrated on low-redshift ($z<3$) datasets. Going forward, it is likely that observational constraints from \textit{JWST} will play a part in informing the development of galaxy formation models, either as a direct constraint in the calibration process, or as a consideration when looking to improve the physics in these models. Thus, the more we can understand how and why our models differ at high redshift, the better.

In this paper, we use the re-simulation strategy of the First Light And Reionisation Epoch Simulations \citep[\flares,][]{Lovell_2021, Vijayan_2021} to perform a high-redshift ($5\leq z \leq 12$) comparison of the \eagle\ hydrodynamics model \citep{Crain_2015, Schaye_2015} and four SAMs -- \galform\ \citep{Baugh_2019_GALFORM}, \lgalaxies\ \citep{Yates_2024_LGalaxies}, the Santa Cruz SAM \citep[\scsam,][]{Yung_2024_SCSAM} and \shark\ \citep{Lagos_2024_SHARK}. The \flares\ zoom simulations are specifically designed for studying galaxy formation and evolution at high-redshift. The fiducial \flares\ suite, which features in previous \flares\ publications, adopts the \eagle\ model. The SAMs are run on merger trees extracted from a DMO simulation suite that uses the same zoom regions as the fiducial suite. These zoom regions are selected with a preference for highly overdense environments, so as to improve the statistics of massive galaxies. To obtain a globally representative sample, a weighting scheme is used to combine galaxy populations from the different regions. SAMs are not typically run on zoom simulations -- as mentioned earlier, they do not require much computational resource, and can thus very feasibly be run on large DMO simulations. The efficiency gained by taking the zoom simulation approach is more meaningful when simulating with hydrodynamics. However, running the SAMs on the DMO suite allows us to compare the SAMs with the fiducial \flares\ suite using the same region selection and volume, and ensures that the effective volume sampled is large enough for a robust comparison at high-redshift.

The structure of the paper is as follows: in Section \ref{sec:models}, we introduce the models used in this analysis and provide an overview of how relevant physical processes are implemented in each model. In Section \ref{sec:methods}, we introduce the fiducial \flares\ suite and the DMO simulation suite on which the SAMs are run. We also detail our procedure for running the SAMs. Section \ref{sec:global} contains our analysis of general galaxy properties predicted by the models, and in Section \ref{sec:passive} we focus on the properties of passive galaxies, before concluding in Section \ref{sec:conclusions}.

%% file: sections/2_models.tex
\section{Model descriptions}\label{sec:models}

The models that feature in our analysis are \eagle, \galform, \lgalaxies, \scsam\ and \shark. In this section, we specify their version, summarise how relevant physical processes are modelled, and list the observables used in their calibration. 

\subsection{EAGLE}\label{sec:eagle}

The \eagle\ hydrodynamics model is detailed in \cite{Crain_2015} and \cite{Schaye_2015}. We use the AGNdT9 model, which leads to less frequent, more energetic AGN feedback events than the fiducial model. This model better reproduces the observed properties of hot gas in groups and clusters, while maintaining similar mass functions to the fiducial \eagle\ model \citep{Barnes_2017_CEagle}. This is particularly important for the overdense environments sampled in \flares.

Radiative cooling and photoheating are modelled element-by-element using rates tabulated by \cite{Wiersma_2009_tables}. Star formation is modelled stochastically, with a dependence on pressure rather than density, following \cite{Schaye_2008}. Since gas is not divided into atomic and molecular phases, a metallicity-dependent density threshold \citep{Schaye_2004} is defined, above which star formation can occur. The stellar mass lost due to stellar winds and supernovae is distributed to neighbouring gas particles, with the exact amount depending on particle separation \citep{Wiersma_2009}. Stellar feedback is modelled as a stochastic injection of thermal energy into neighbouring gas particles when a stellar particle reaches an age of 30 Myr. Affected gas particles all experience a temperature increase of $10^{7.5}\ {\rm K}$. Higher densities and lower metallicities lead to more energy being injected into the ISM per supernova event. 

If a Friends-Of-Friends \citep[FOF,][]{Davis_1985} halo does not already contain a black hole, it is seeded with one when it reaches a mass of $M_{\rm FOF}=10^{10}\ h^{-1}{\rm M_\odot}$. The rate at which a black hole accretes mass is taken as the minimum of the Eddington accretion rate and the Bondi-Hoyle accretion rate \citep{Bondi_1944} times an efficiency factor, whichever of the two is lower (see \S 4.6.2 of \cite{Schaye_2015} for more detail). At each time step, the amount of energy stored in a `reservoir' for AGN feedback is proportional to the black hole accretion rate. Only when the reservoir contains sufficient energy can feedback occur, similarly to stellar feedback, via the stochastic injection of thermal energy into neighbouring gas particles. The resulting temperature increase of each gas particle is $\Delta T=10^{9}\ {\rm K}$ for the AGNdT9 model.

The model is calibrated at $z=0$, with stellar feedback and black hole accretion tuned using the stellar mass function (SMF) and galaxy disk sizes, and AGN feedback tuned using the stellar-to-black hole mass relation \citep{Crain_2015}.

\subsection{GALFORM}\label{sec:galform}

We run the same version of \galform\ \citep{Cole_2000_GALFORM} used in \cite{Cowley_2018_GALFORM, Baugh_2019_GALFORM, Lu_2025_GALFORM}. The model is similar to that described in \cite{Lacey_2016_GALFORM}, with just an updated merger prescription and parameters tuned to the \pmill\ simulation \citep{Baugh_2019_GALFORM}.

The cooling time is a characteristic timescale over which circumgalactic hot gas radiates away its thermal energy. It is calculated as in \cite{White_1991}, using the cooling function of \cite{Sutherland_1993}, but with a different halo gas density profile. The radius at which gas cools onto a galaxy (also known as the cooling radius) is the smaller of two quantities: 1) the radius at which the cooling time is equal to the time since the halo formation event (taken to occur every time a halo doubles in mass), and 2) the radius at which the time taken for a particle to fall to the halo centre under the effects of gravity (also known as free-fall time) is equal to the time since the halo formation event.

Star formation is modelled differently in the disk and spheroid. In the disk, cold gas is partitioned into atomic and molecular phases using the formulation by \cite{Blitz_2006}, which depends on the midplane hydrostatic pressure of the disk, obtained using an approximation by \cite{Elmegreen_1993} (see \citealt{Lagos_2011} for more detail). In the spheroid, all gas is assumed to be in the molecular phase. Episodes of star formation in the spheroid, also known as starbursts, are triggered by mergers and disk instabilities. The star formation rates (SFRs) in both the disk and spheroid are proportional to their respective molecular gas masses, by different tunable factors. Different stellar initial mass functions (IMFs) are assumed as well. In the disk, stars form with a \cite{Kennicutt_1983} IMF, such that the fraction of stellar mass instantaneously returned to the cold gas as a result of dying stars is $R=0.44$. In the spheroid, a single power law IMF is used with its slope as a tunable parameter, leading to a top-heavy IMF with a slope of $x=1$ (c.f. \cite{Salpeter_1955} $x=1.35$) and an instantaneous stellar mass return fraction of $R=0.54$. Stellar feedback ejects cold gas into an ejecta reservoir at a rate dependent on circular velocity.

A SMBH starts out with negligible mass and experiences its first significant growth during a galaxy's first merger-triggered starburst \citep{Malbon_2007}. When starbursts occur, a fraction of the cold gas in the spheroid is accreted by the SMBH. This `starburst mode' accretion is the dominant channel for SMBH growth. A separate channel, known as `radio mode' accretion \citep{Bower_2006, Croton_2006}, contributes little to the mass of a SMBH, but is associated with strong AGN feedback in the form of relativistic jets. Radio mode accretion occurs if: 1) hot gas cools slowly enough to be in quasi-hydrostatic equilibrium (determined by the ratio of the cooling time and free-fall time), and 2) the SMBH is sufficiently large, such that the power required to balance the radiative cooling luminosity of the hot gas can be produced at accretion rates much lower than the Eddington luminosity (assuming relativistic jets are only produced by low accretion rates). The power of these AGN jets is adjusted to balance the radiative cooling luminosity exactly.

When a galaxy becomes a satellite, it undergoes instantaneous ram pressure stripping -- all hot gas associated with its halo is transferred to the halo of the new central. As such, satellites are unable to accrete cold gas. Subhalo positions are used to track satellite galaxies until the subhalos have been stripped of enough mass that they are no longer resolved ($<20$ particles). At this point, the galaxy becomes an orphan, and its merger timescale is calculated analytically, accounting for the effects of dynamical friction and tidal disruption using the prescription developed by \cite{Simha_2017}.

\galform\ is calibrated using forward-modelled observables. Stellar SEDs are obtained using the \cite{Maraston_2005} stellar population synthesis (SPS) library, and dust is modelled self-consistently as molecular clouds embedded in a diffuse dust component, in a similar manner to the \grasil\ spectrophotometric model \citep{Silva_1998, Lacey_2016_GALFORM}.

In the calibration process, primary constraints take priority over secondary constraints. On the list of primary constraints are the present-day optical and near-IR luminosity functions, present-day HI mass function, present-day morphological fractions, present-day black hole-bulge mass relation, near-IR luminosity functions from $z=0-3$, sub-millimetre galaxy counts, far-IR galaxy counts, and far-UV luminosity functions of Lyman-break galaxies at $z=3$ and $z=6$. Secondary constraints consist of the Tully-Fisher relation at $z=0$, early- and late-type galaxy sizes at $z=0$, and early-type galaxy stellar metallicities at $z=0$ \citep{Lacey_2016_GALFORM}.

\subsection{L-GALAXIES}\label{sec:lgalaxies}

We use the publicly available\footnote{\url{https://github.com/LGalaxiesPublicRelease/LGalaxies2020_PublicRepository/tree/Yates2023}} \cite{Yates_2024_LGalaxies} version of \lgalaxies, which builds on the \cite{Yates_2021_LGalaxies} version by implementing binary stellar evolution and a detailed dust model. \cite{Yates_2021_LGalaxies} is a version of \lgalaxies\ 2020 \citep{Henriques_2020_LGalaxies} with a modified stellar feedback scheme that increases the fraction of metals directly ejected into the circumgalactic medium (CGM), as opposed to mixing with the ISM. This leads to an improved agreement with low-redshift observations of mass-metallicity relations and radial metallicity profiles for both stars and gas, while retaining previous agreement with other key properties.

The cooling time is calculated following \cite{White_1991}, using the cooling function of \cite{Sutherland_1993}. The cooling radius is given by the radius at which the cooling time is equal to the halo dynamical time. Two accretion modes are defined -- hot (cold) mode accretion occurs when the cooling radius is smaller (larger) than the halo radius, which is defined as $R_{\rm 200c}$.

\lgalaxies\ 2020 (and its offshoots) adopts a spatially-resolved treatment for gas and stellar disks, using layers of concentric rings \citep{Fu_2013_LGalaxies}. In each ring, cold gas is divided into atomic and molecular phases using a fitting formula by \cite{Mckee_2010}, based on work by \cite{Krumholz_2009}, who model the fraction of molecular hydrogen as a function of metallicity and the local surface density of cold gas. The star formation surface density is then proportional to the surface density of molecular hydrogen. Mergers can lead to starbursts, with the stellar mass formed in each ring dependent on the baryonic mass ratio of the merging galaxies and the cold gas mass in the ring \citep{Somerville_2001}. Stellar feedback, dependent on the maximum circular velocity, can reheat cold gas into the hot halo, and also eject hot gas into the ejecta reservoir, if there is energy left over from heating the cold gas. The release of energy and metals by stellar populations is tracked separately for each ring, assuming a \cite{Chabrier_2003} IMF.

SMBHs are seeded with negligible mass. SMBH growth in \lgalaxies\ occurs primarily through `starburst mode' accretion, triggered by galaxy mergers, when cold gas is driven towards the central region of the new system. SMBHs also grow via inflows of gas from the hot halo in `radio mode' accretion. The radio mode accretion rate is proportional to the hot halo mass and SMBH mass by a tunable factor. This is almost identical to the model presented in \cite{Croton_2006}, but with no dependence on $H(z)$, leading to enhanced accretion at low redshift. Radio mode feedback is the only mode of AGN feedback in \lgalaxies, and is modelled by using a fraction $\epsilon_{\rm heat}=0.1$ of the energy obtained from radio mode accretion to suppress the cooling of hot gas. One could consider that other forms of AGN feedback may be accounted for in the modelling of stellar feedback (this same perspective could be applied to the highly similar modelling of AGN feedback in \galform\ as well).


Subhalo positions are used to track satellite galaxies until the subhalos become sub-resolution ($<20$ particles). Satellites are only subject to tidal and ram pressure stripping if situated within the virial radius of their host halo. Tidal forces are modelled by the removal of hot gas beyond a minimum radius in proportion to the amount of dark matter being lost from the subhalo. Satellites situated within larger parent halos with masses greater than $\sim10^{14}\ \rm{M_\odot}$ are additionally affected by ram pressure stripping (note, though, that halos do not reach this size at the redshifts probed in this paper). Hot gas removed from a satellite via the aforementioned mechanisms is transferred to the hot gas halo of the central galaxy, and any hot gas that remains is allowed to cool onto the satellite. By construction, orphan galaxies do not have any associated dark matter or hot gas. Their merger timescale is calculated following \cite{Binney_1987}, but they may be tidally disrupted beforehand if their baryonic density is less than the dark matter density of the host halo.

\lgalaxies\ 2020 is calibrated using observations of the stellar mass function at $z=0$ and $z=2$, passive fractions at $z=0$ and $z=2$, and the HI mass function at $z=0$. \cite{Yates_2021_LGalaxies} find that differences in the general galaxy properties between their new model and \lgalaxies\ 2020 are small at $z<2$. Thus, the \cite{Yates_2021_LGalaxies} and \cite{Yates_2024_LGalaxies} models are not recalibrated -- they use the same default parameter values as \lgalaxies\ 2020.

\subsection{Santa Cruz SAM}\label{sec:scsam}

We use the same version of \scsam\ as in \cite{Yung_2024_SCSAM}. Details of the model can be found in \cite{Somerville_2008_SCSAM, Somerville_2015_SCSAM, Yung_2024_SCSAM}.

As in \galform\ and \lgalaxies, the cooling time is calculated using the formula by \cite{White_1991} and the temperature- and metallicity-dependent cooling function of \cite{Sutherland_1993}. Similarly to \lgalaxies, the cooling radius is defined as the radius within which the cooling time is equal to the halo dynamical time. When the cooling radius is less (greater) than the virial radius of the halo, gas accretion occurs in hot (cold) mode.

Cold gas is partitioned into atomic, molecular and ionised components, using fitting formulae based on the results of simulations by \cite{Gnedin_2011}, which depend on gas phase metallicity, gas phase surface density and UV-background. Star formation efficiency is assumed to increase with molecular hydrogen surface density, using a double power law motivated by observations \cite[for details, see][]{Somerville_2015_SCSAM}. Galaxy mergers lead to episodes of star formation known as starbursts, which have a star formation efficiency and timescale determined by the mass ratio and gas fraction of the merging galaxies, based on predictions from hydrodynamic simulations by \cite{Robertson_2006} and \cite{Hopkins_2009_mergers} \citep{Somerville_2008_SCSAM}. A universal \cite{Chabrier_2003} IMF is assumed, with a recycled fraction of $R=0.43$. The outflow rate of cold gas due to supernova feedback is assumed to depend on the circular velocity of the galaxy. Some of the ejected gas is transferred to the hot halo, and the remainder is deposited in the ejecta reservoir. 

In \scsam, host halos are seeded with a black hole of mass $M_{\rm BH}=10^4\ \rm{M}_\odot$. 
The treatment of black hole growth resulting from galaxy mergers is informed by the results of hydrodynamic simulations of binary mergers of idealized galaxies. Just after a merger, the SMBH mass of a galaxy is equal to the summed SMBH masses of its progenitors. The SMBH then grows at the Eddington rate until reaching a critical mass, at which point it produces feedback in the form of a pressure-driven outflow -- we will call this `wind mode' feedback. A radiatively-efficient `blow-out' phase occurs, where the SMBH accretion rate declines as a power law, following light curves from simulations by \cite{Hopkins_2006}. Disk instabilities are another channel for rapid SMBH growth. In such cases, a fraction $f_{\rm BH,disk}=0.001$ of the disk mass is allowed to be accreted onto the black hole, over a timescale guided by the light curves of \cite{Hopkins_2006}. When a halo is cooling in hot mode, radio mode accretion occurs via a cooling flow from the hot halo, assuming Bondi-Hoyle accretion (the exact strength of which is tuned by a free parameter) and the isothermal cooling flow model of \cite{Nulsen_2000}. The associated radio mode feedback is modelled by using a fraction of the energy gained through radio mode accretion to offset cooling from the hot halo to the ISM.

Satellites are modelled analytically rather than relying on dark matter substructure, i.e. all satellites are treated as orphans. Like \galform, satellite galaxies do not accrete new cold gas. The merger timescale of a satellite is calculated using an updated version of the Chandrasekhar formula for dynamical friction that considers the energy and angular momentum of the satellite's orbit, as well as tidal stripping effects on subhalos \citep{BK_2008,Somerville_2008_SCSAM}. Satellites lose between $30-40\%$ of their mass every orbital period, and are considered to be tidally disrupted if their mass is less than that contained within the \cite{Navarro_1996} scale radius at the time of infall. The cold gas in tidally-disrupted satellites is transferred to the hot gas reservoir of the central galaxy.

The model parameters used here are those obtained in the study by \cite{Gabrielpillai_2022}, with the following $z=0$ quantities used as constraints: the stellar mass function, stellar-to-halo mass relation, stellar mass-metallicity relation, cold gas fraction-to-stellar mass relation for disk galaxies, and black hole mass-bulge relation.

\subsection{SHARK}\label{sec:shark}

We use the publicly available\footnote{\url{https://github.com/ICRAR/shark}} \shark\ 2.0, which is an updated version of the original \shark\ 2.0 model described in \cite{Lagos_2024_SHARK}. The full list of changes can be found in Oxland et al. (2025; submitted). We include a description of the updates, where relevant.

The cooling time is computed as in \cite{White_1991}, using tabulated cooling functions obtained with \cloudy\ \citep{Ferland_1998}. Cooling is modelled using the \cite{Croton_2006} model, with the same definitions of accretion modes as \lgalaxies\ and \scsam\ -- hot (cold) accretion takes place when the cooling radius is smaller (greater) than the virial radius of the halo.

Star formation in the disk is calculated similarly to \galform, using the \cite{Blitz_2006} model to calculate the ratio of molecular to atomic hydrogen, and with the hydrostatic mid-plane pressure calculated following \cite{Elmegreen_1993}. The SFR surface density is proportional to the molecular hydrogen surface density. 
Star formation in the spheroid is modelled as starbursts, triggered by galaxy mergers or disk instabilities. The same gas partitioning recipe is used in the disk and spheroid. However, the star formation efficiency in the spheroid is boosted by a multiplicative factor, tuned to observations. A \cite{Chabrier_2003} IMF is used throughout, with a recycled fraction of $R=0.46$. Stellar feedback causes gas to be expelled from a galaxy, some of which is also expelled from the halo. Feedback is modelled following the implementation of \cite{Lagos_2013}, whereby the rate at which gas is expelled from a galaxy is proportional to circular velocity. Unlike in \cite{Lagos_2024_SHARK}, stellar feedback is dependent on the age of the universe rather than $(1+z)$ (see Equation 29 in \cite{Lagos_2018_SHARK}). This change was made to improve supernova feedback at high redshifts.

SMBHs are seeded with a mass of $M_{\rm BH}=10^{4}\ h^{-1}{\rm M_\odot}$ in halos with a minimum mass of $M_{\rm halo}>10^{10}\ h^{-1}{\rm M_\odot}$.
The accretion rate of a SMBH when its host galaxy is undergoing hot mode accretion is computed following \cite{Croton_2016_SAGE}, assuming a Bondi-Hoyle-like accretion mode that can be tuned by a free parameter. Black holes also grow when starbursts occur, according to the prescription by \cite{Kauffmann_2000}, with the accreted mass proportional to that of the cold gas reservoir. A normalised accretion rate is obtained by dividing the total accretion rate by the Eddington accretion rate. This is used to differentiate between accretion regimes, in order of decreasing normalised accretion rate: Super-Eddington (SE) accretion; a thin disk (TD) regime \citep{Shakura_1973}; a high accretion rate advection-dominated accretion flow (ADAF) regime, in which electrons are predominantly heated by viscous heating; a low accretion rate ADAF regime, in which electrons are predominantly heated by ion-heating. AGN bolometric luminosities are calculated following \cite{Griffin_2019} and \cite{Griffin_2020}, with the calculation depending on which of four accretion regimes the black hole is in. AGN feedback is modelled in two modes, known as `jet mode' and `wind mode'. Unlike in other SAMs, these modes are not directly related to the gas accretion mode. Jet mode feedback occurs when a hot gas halo has formed, according to conditions defined in \cite{Correa_2018}. The jet power is calculated following \cite{Meier_2002}, who differentiate between AGN in an ADAF-dominated regime and AGN with higher accretion rates, using the aforementioned classification. The total accretion rate of the SMBH is used to compute the jet power, and a fraction of the jet power is used to counteract the cooling of hot gas. Wind mode feedback is triggered when the bolometric luminosity of the AGN is above a critical value such that radiation pressure is able to deposit momentum onto dust grains, driving galactic winds.

Satellites in \shark\ are tracked using the information of their subhalo. Stars and cold gas in satellites are tidally stripped following a formula presented in \cite{Errani_2015} that depends on the ratio between the subhalo mass at present and its value at infall. Ram pressure stripping is modelled following the results of hydrodynamical simulations \citep{Font_2008, McCarthy_2008}, and affects both the hot and cold gas components in satellites, with different stripping radii assumed for each component. Satellite galaxies are still able to accrete cold gas from their hot halo. If the satellite contains an AGN in jet mode, any AGN feedback power leftover from counteracting the cooling of hot gas is used to heat the hot gas atmosphere associated with the central galaxy. When the satellite becomes an orphan, its cold gas component is subject to ram pressure stripping as per usual, while all of its hot gas is immediately moved to the hot gas reservoir of the central galaxy. The merger timescale of the orphan is obtained using a formula by \cite{Poulton_2021}, which compared to other models has a reduced dependence on the subhalo-to-host halo mass ratio and differentiates between galaxies within and beyond the virial radius of the host halo. Note that since the initial release of \shark\ 2.0 in \cite{Lagos_2024_SHARK}, a new parameter (\verb|alpha_cold|) has been introduced to control the stripping of cold gas. Additionally, parameters controlling the stripping of hot gas have been updated so that stripping occurs more gradually. 

An additional update to this version of \shark\ is the inclusion of fixes at the SAM level to deal with numerical artefacts in merger tree data \citep{Chandro_2025}.


The version of \shark\ we use is calibrated manually and fine-tuned using an automated parameter search, via the \texttt{optim} Particle Swarm Optimisation (PSO) package in \shark, with the stellar mass functions at $z=0,1,2,4$ and the HI mass function at $z=0$ used as direct constraints. 


%% file: sections/3_method.tex
\begin{figure*}
    \includegraphics[width=2\columnwidth]{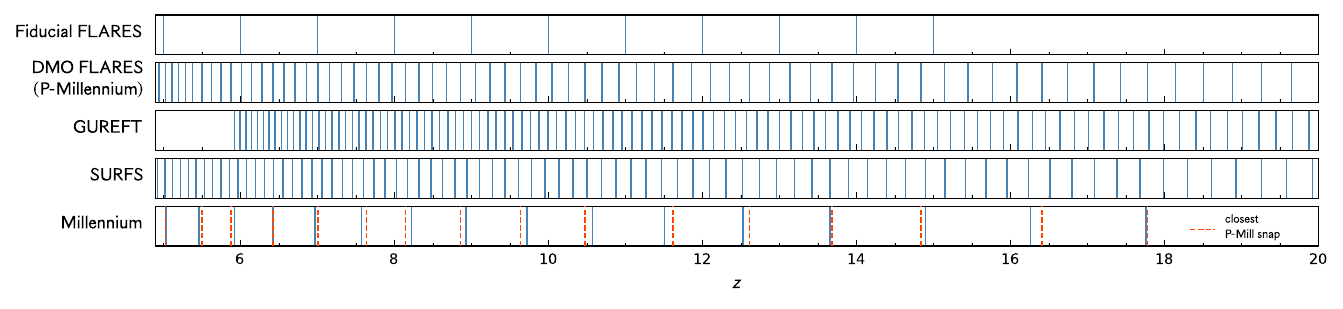}
	\caption{Snapshot cadence used in various simulations, from top to bottom: the fiducial \flares\ suite, the \flares\ DMO suite (resimulated from the parent simulation with the same cadence as the \pmill\ simulation \citep{Baugh_2019_GALFORM}), the GUREFT simulation suite \citep{Yung_2024_GUREFT}, the SURFS simulation suite \citep{Elahi_2018_SURFS}, and the \mill-I and II simulations \citep{Springel_2005_LHaloTree, BK_2009_MRII}. The dashed red lines in the bottom-most row represent the snapshots used when running \lgalaxies\ on the \flares\ DMO suite -- these are subsampled from the SAM DMO snapshot cadence (same as \pmill).} \label{fig:snaps}
\end{figure*}

\section{Methods}\label{sec:methods}

In this section, we introduce the simulation suites that feature in our analysis and detail our procedure for running the SAMs.

\subsection{First Light And Reionisation Epoch Simulations}\label{subsec:flares_intro}

We distinguish between the \flares\ simulation suite run with the \eagle\ hydrodynamics model and the \flares\ simulation suite run with dark matter only by labelling the former as the fiducial suite and the latter as the DMO suite. Note that all mentions of the \eagle\ model in this paper reference results from the fiducial \flares\ suite, rather than the flagship \eagle\ simulations. However, to facilitate comparison between the different models in this paper, we will refer to galaxies in the fiducial \flares\ suite as \eagle\ galaxies.

The fiducial \flares\ suite is introduced in \cite{Lovell_2021} and \cite{Vijayan_2021}. The suite comprises 40 spherical zoom-in simulations, each of radius $14\ h^{-1}\rm{Mpc}$ at $z=4.67$. The zoom regions are selected from a large $(3.2\ \rm{cGpc})^3$ DMO parent simulation \citep{Barnes_2017_MACSIS}, allowing a wide range of overdensities to be sampled ($\delta=-0.497\rightarrow0.970$). Highly overdense regions are preferentially selected in order to simulate a greater number of massive galaxies \citep{Chiang_2013, Lovell_2018}. The regions are selected at $z=4.67$, when density perturbations are only mildly non-linear. This helps to ensure that the rank ordering of overdensities is still preserved at high-redshift, ensuring the weighting scheme remains robust. The hydrodynamics model used is the AGNdT9 variant of the \eagle\ subgrid model \citep{Crain_2015, Schaye_2015}, as mentioned in \S \ref{sec:eagle}. The fiducial \flares\ suite has the same resolution as the flagship \eagle\ run, with a dark matter particle mass $M_{\mathrm{DM}}=9.7\times10^6$ M$_{\odot}$, an initial gas particle mass $M_{\mathrm{gas}}=1.8\times10^6$ M$_{\odot}$ and a softening length of $2.66\ \rm{ckpc}$. Outputs are stored at integer redshift, from $z=15\rightarrow5$.

The SAM DMO suite used in this study is a resimulation of the same 40 zoom regions, as in the fiducial \flares\ suite, 
using a dark matter particle mass of $M_{\rm DM}=8.36\times10^7\ \rm{M_\odot}$ and a softening length of $4.32\ \rm{ckpc}$. This particle mass was chosen as a resolution at which the four SAMs in this study are sufficiently converged. More information on the convergence tests is provided in \S \ref{sec:part_res} and \S \ref{sec:conv}. Outputs are stored at the same redshifts as in the \pmill\ simulation \citep{Baugh_2019_GALFORM}. This output frequency, much higher than the fiducial \flares\ suite, is necessary to accurately capture the halo merger histories required for SAMs. The first two rows of Figure \ref{fig:snaps} show the output snapshots of the fiducial and DMO suites. More discussion on the choice of snapshot cadence can be found in \S \ref{sec:snap_cadence}.

As our region selection contains many more highly overdense regions than one would find in a periodic box of the same simulated volume, the zoom regions on their own do not provide a statistical representation of galaxies in the universe. To study the global population of galaxies, we combine the statistics of individual regions using a weighting scheme that reduces the contribution from galaxies in rarer regions (i.e. highly overdense and underdense ones). Since the same region selection is used in the fiducial and DMO suites, the same weightings apply. All composite distribution functions and scaling relations shown in this paper are obtained using this weighting scheme. More detail on how the weights are obtained can be found in \cite{Lovell_2021}.

The fiducial and DMO suites assume a \cite{Planck_2014} cosmology, with $\Omega_\Lambda=0.693,\ \Omega_m=0.307,\ h=0.6777$.

\subsection{Running the SAMs}\label{subsec:conv}


\input{tables/sims}

We note that the SAMs are not recalibrated for this analysis, and that the predictions of a SAM can be sensitive to several factors: cosmology, the resolution of the DMO simulation from which merger histories are obtained, the halo-finding and merger tree codes used, and the time resolution of the merger histories. Here, we detail our methodology, and explain where relevant how we have checked for convergence.

\subsubsection{Halo-finding and merger tree codes}\label{subsec:codes}


\galform\ and \shark\ are run on the same set of merger trees, obtained using the \subfind\ \citep{Springel_2001_Subfind} halo-finder, \dtrees\ \citep{Jiang_2014_Dhalo} merger tree builder and \dhalo\ \citep{Jiang_2014_Dhalo} post-processing tool. \lgalaxies\ is run on merger trees obtained using the \subfind\ and \lhalotree\ \citep{Springel_2005_LHaloTree} algorithms present in the \gadget-4 \citep{Springel_2021_Gadget4} code. \scsam\ is run on merger trees obtained using the \rockstar\ \citep{Behroozi_2013_Rockstar} and \consistenttrees\ \citep{Behroozi_2013_CTrees} algorithms. 


Using the same pipeline for each SAM would lead to a cleaner comparison of the modelling. However, the performance of each SAM likely depends, to varying extent, on the algorithms used. \cite{Gabrielpillai_2022} mention that \scsam\ performs less well on \sublink\ \citep{RG_2015_SubLink} merger trees, due to differences in the number of identified mergers. \cite{Lee_2014} find that using different tree codes, \dtrees\ and \consistenttrees\ included, can lead to notable differences in the mean star formation histories (SFHs), satellite numbers and galaxy merger rates predicted by a SAM. On the other hand, \cite{Gomez_2022} show that predictions by \galform\ are robust to the use of different algorithms, for the same underlying DMO simulation. \cite{Chandro_2025} find that SAMs relying on subhalo properties, such as previous versions\footnote{The updated version of \shark\ includes fixes at the SAM level to reduce numerical artefacts in merger tree data.} of \shark, are susceptible to numerical issues in merger tree data. In contrast, halo-based SAMs, such as \galform, are less affected.

Since the focus of this paper is on the physical predictions of different SAMs at high-redshift, rather than the details of their modelling, we endeavour to run halo-finding and tree-building algorithms that have been used to run the SAMs in prior studies. In the case of \galform, \lgalaxies\ and \scsam, we use the same codes employed in their calibration runs (\cite{Baugh_2019_GALFORM}, \cite{Henriques_2020_LGalaxies} and \cite{Gabrielpillai_2022}, respectively). \shark\ is an exception, as we use \subfind+\dtrees+\dhalo\ in this analysis, as was done in \cite{DSilva_2025} and \cite{Lagos_2025}. This is different from the \HBTherons+\dhalo\ pipeline employed in calibrating the \shark\ model (Oxland et al., submitted).

\subsubsection{Particle resolution}\label{sec:part_res}

The fourth column in Table \ref{table:sims} shows the particle mass of the DMO simulation(s) on which the calibrated model of each SAM is run. Both \lgalaxies\ and \scsam\ are calibrated to work at more than one resolution. We list these simulations to indicate the range of resolutions at which the models can be applied, guided by previous work. That said, it is not strictly true that the SAMs are fully converged for all properties in the stated resolution range. For example, gas-phase metallicities in \lgalaxies\ are not fully converged between the \mill-I and II resolutions \citep{Yates_2024_LGalaxies}. 

We run detailed tests to determine the resolutions at which the SAMs are sufficiently converged, considering the galaxy properties studied in Section \ref{sec:global}. A description of the test results can be found in Section \ref{sec:conv} of the appendix. In short, to identify a common particle mass that works for the SAMs, we rerun the \flares\ DMO suite at different resolutions, using a particle mass of $1.15\times10^7$, $8.36\times10^7$, $1.58\times10^8$ and $3.22\times10^8\ \rm{M_\odot}$ -- these approximately correspond to the \mill-II, GUREFT-90, \pmill\ and SURFS resolutions, and serve as benchmarks for \lgalaxies, \scsam, \galform\ and \shark\ respectively. We run all four SAMs at these resolutions and assess convergence at a given resolution by comparing the predictions of each SAM with those of its benchmark run. Where possible, we also compare with previously published results. We find that a particle mass of $8.36\times10^7\ \rm{M_\odot}$, similar to that of GUREFT-90, works well across all SAMs.

\subsubsection{Snapshot cadence}\label{sec:snap_cadence}

\galform, \scsam\ and \shark\ are run on merger trees constructed using the full set of snapshots stored from the \flares\ DMO suite. As mentioned in \S \ref{subsec:flares_intro}, outputs of the DMO suite are saved at the same redshifts as in the \pmill\ simulation, ensuring a high enough time resolution to resolve the merger histories of halos \citep{Baugh_2019_GALFORM}. The GUREFT and SURFS simulations, used in the calibration runs of \scsam\ and \shark\ respectively, have a similarly high time resolution (see second, third and fourth rows of Figure \ref{fig:snaps}). \lgalaxies\ is calibrated on the \mill-I and II simulations, which have a coarser time resolution at high-redshift. To maintain consistency with previous works, we run \lgalaxies\ on a sub-sample of the DMO outputs by constructing merger trees from snapshots closest in redshift to the output snapshots in the \mill-I and II simulations (dashed red lines in Figure \ref{fig:snaps}).

\subsubsection{Cosmology}

The SAMs are run assuming the same cosmology as \flares\ (see \S \ref{subsec:flares_intro}). All SAMs adopt a similar $\Lambda$CDM Planck cosmology to the fiducial \flares\ suite in their calibration runs (see last three columns of Table \ref{table:sims}). Hence, issues of convergence arising from the use of different cosmologies are likely to be minimal.

\begin{figure*}
	\includegraphics[width=2\columnwidth]{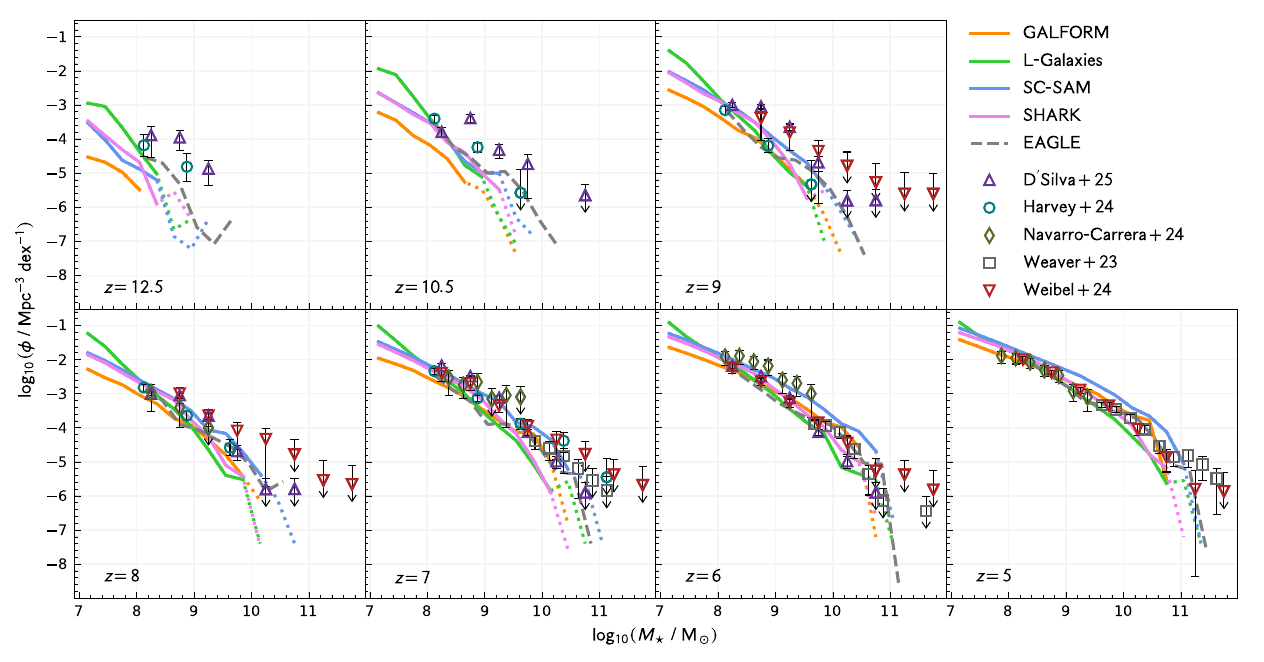}
	\caption{Stellar mass function predicted by the four SAMs (solid lines, or dotted to denote bins containing fewer than 10 galaxies) and \eagle\ (dashed grey line). A stellar mass cut of $M_\star>10^8\ \rm{M_\odot}$ is applied to the \eagle\ results. As the outputs of the fiducial \flares\ suite are saved at integer redshift, the \eagle\ predictions in the $z=10.5$ and $z=12.5$ panels are in fact at $z=10$ and $z=12$ respectively. Observations are plotted as symbols with error bars.} \label{fig:smf}
\end{figure*}

\subsection{Measuring galaxy properties}

In the fiducial \flares\ suite, a FOF algorithm \citep{Davis_1985} is used to group particles into halos. Substructure, which includes galaxies, is then identified using the \subfind\ \citep{Springel_2001_Subfind} algorithm. Galaxy properties are computed using particles situated within a 30 pkpc (physical kpc) radius of the most bound particle in each substructure. The star formation rate (SFR) of a galaxy is averaged over the most recent 20 Myr. Throughout this work, we impose a stellar mass criteria of $\log_{10}(M_\star/\rm{M_\odot})>8$ on \eagle\ galaxies, which roughly translates to the requirement that galaxies contain at least 50 star particles. 

In the SAMs, galaxy properties are summed over spheroid and disk components (where split). The timescale over which the SFR is measured, $t_{\rm SFR}$, differs across SAMs. For \scsam, we choose to use $t_{\rm SFR}=20$ Myr. In \lgalaxies, we calculate the SFR from the output SFHs, and obtain a timescale of $t_{\rm SFR}=23.7,24.1,13.5,16.4,20.6,25.1\ {\rm Myr}$ at $z=5,6,7,8,9,10.5$, respectively. In \galform\ and \shark, the SFR is averaged over the time between the current and previous snapshots, such that $t_{\rm SFR}=24.6,26.8,21.4,17.5,15.6,12.5\ {\rm Myr}$ at $z=5,6,7,8,9,10.5$, respectively. SAMs can probe relatively lower baryon masses at fixed DM resolution -- we include all galaxies in halos with at least 20 DM particles, which corresponds to halos with mass $M_{\rm halo}\geq1.67\times10^9\ {\rm M_\odot}$.

The halo masses shown in our analysis are defined differently in the models. \galform\ and \shark\ use the \dhalo\ \citep{Jiang_2014_Dhalo} definition of halo mass, \scsam\ use the \cite{Bryan_Norman_1998} definition based on virial mass, and both \lgalaxies\ and the fiducial \flares\ suite (i.e. \eagle) use $M_{200c}$. We find that the HMFs obtained with the different halo definitions are very similar, and thus unlikely to affect the conclusions of our analysis.

%% file: tables/sims.tex
\begin{table*}
\centering
\footnotesize
\begin{tabular}{@{}lllccccc@{}}
\toprule
SAM & Version & DMO & $M_{\rm DM}$ ($ \rm{M_\odot}$) & Volume ($\rm{cMpc^3}$) & $\sigma_8$ & $h$ & $\Omega_m$ \\ \toprule
(All) & (See below) & \flares\ DMO & $8.36\times10^7$ & $40\times33.3^3$ & 0.829 & 0.678 & 0.307 \\ \midrule
GALFORM & \cite{Baugh_2019_GALFORM} & P-Millennium & $1.56\times10^8$ & $800^3$ & 0.829 & 0.678 & 0.307 \\ \midrule
\multirow{2}{*}{L-Galaxies} & \multirow{2}{*}{\cite{Yates_2024_LGalaxies}} & Millennium I & $1.43\times10^9$ & $714^3$ & \multirow{2}{*}{0.829} & \multirow{2}{*}{0.673} & \multirow{2}{*}{0.315} \\
 &  & Millennium II & $1.14\times10^7$ & $143^3$ &  &  &  \\ \midrule
\multirow{4}{*}{SC-SAM} & \multirow{4}{*}{\cite{Yung_2024_SCSAM}} & GUREFT-05 & $1.5\times10^4$ & $7.4^3$ & \multirow{4}{*}{0.823} & \multirow{4}{*}{0.678} & \multirow{4}{*}{0.308} \\
 &  & GUREFT-15 & $4.0\times10^5$ & $22.1^3$ &  &  &  \\
 &  & GUREFT-35 & $5.0\times10^6$ & $53.6^3$ &  &  &  \\
 &  & GUREFT-90 & $8.5\times10^7$ & $132^3$ &  &  &  \\ \midrule
SHARK & \cite{Lagos_2024_SHARK} & SURFS & $3.27\times10^8$ & $311^3$ & 0.815 & 0.675 & 0.312 \\ \bottomrule
\end{tabular}
\caption{Columns contain, from left to right: (1) the name of the SAM, (2) the publication associated with the model of the SAM, (3) the name of the DMO simulation on which the SAM was run in the publication stated in (2), (4) the particle mass in the DMO simulation, (5) the volume of the DMO simulation, (6) the amplitude of the linear power spectrum on scales of 8 $h^{-1}$Mpc, (7) the reduced Hubble parameter, (8) the present-day energy density of matter in units of the critical density. The 40 regions in the \flares\ DMO suite are spherical, but their individual volume is expressed as a cubed length to facilitate comparison with the other simulations.}
\label{table:sims}
\end{table*}

%% file: sections/4_properties.tex
\section{Physical properties}\label{sec:global}

We now turn to the model predictions, focusing in this section on general trends in physical properties. 

\begin{figure*}
    \includegraphics[width=2\columnwidth]{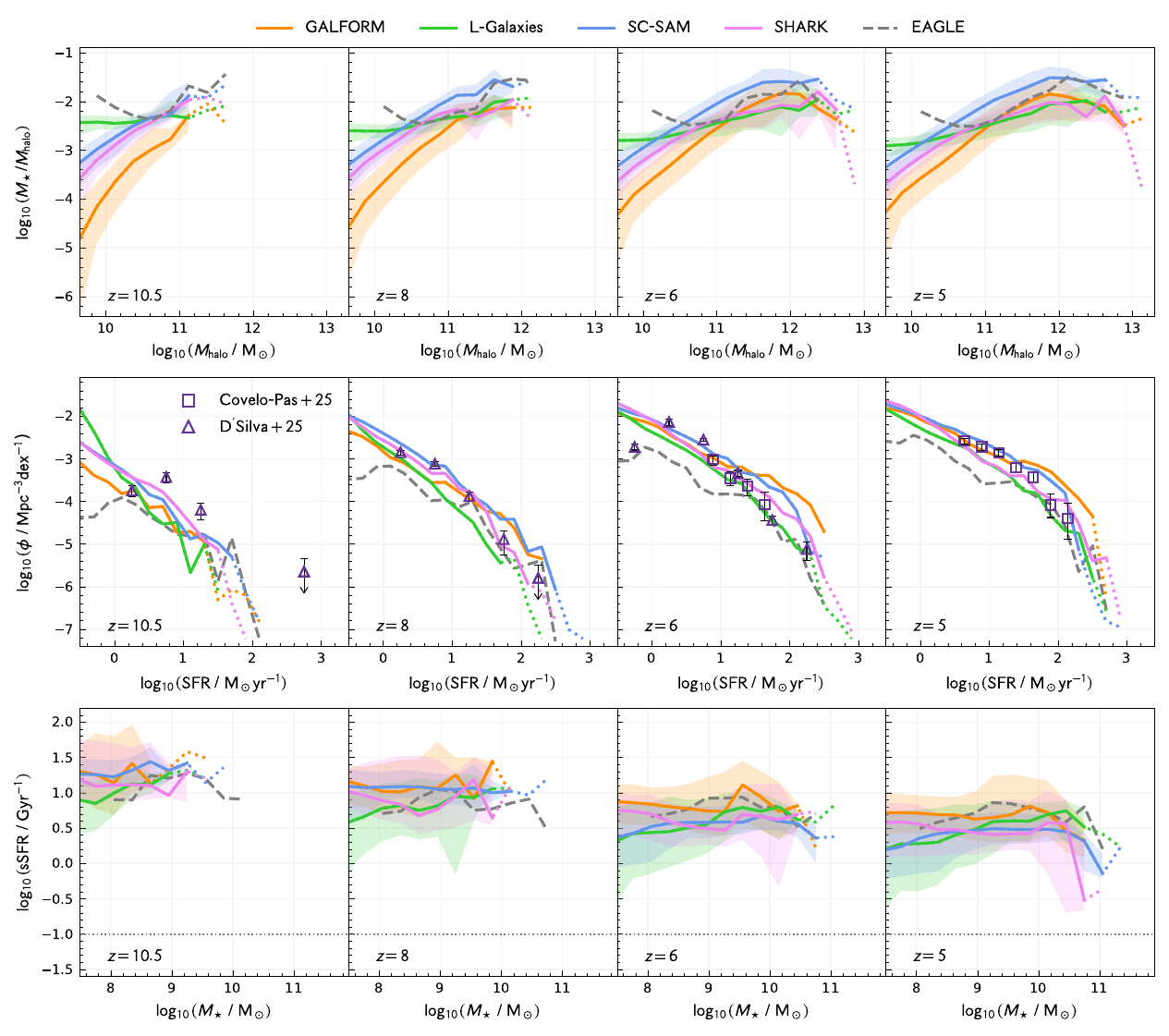}
	\caption{Each row shows a different quantity predicted by the four SAMs and \eagle\ across a range of redshifts. From top to bottom: stellar-to-halo mass relation, SFR distribution function, star-forming main sequence. Note that a stellar mass cut of $M_\star>10^8\ \rm{M_\odot}$ is applied to the \eagle\ results, and that the \eagle\ predictions at $z=10.5$ are at $z=10$, as outputs of the fiducial \flares\ suite are saved at integer redshift. For the SAMs, dotted lines denote bins containing fewer than 10 galaxies. Shaded regions denote the 1$\sigma$ range of the SAMs. The dotted horizontal line in the final row marks the sSFR threshold below which galaxies are considered passive (relevant to \S \ref{sec:passive}).}
    \label{fig:properties}
\end{figure*}

\subsection{Stellar mass function}\label{subsec:smf}


The redshift evolution of the stellar mass function (SMF) is shown in Figure \ref{fig:smf}, alongside recent measurements utilising \textit{JWST} NIRCam data \citep{Harvey_2024, Navarro_2024, Weibel_2024, DSilva_2025} and the COSMOS2020 catalogue \citep{Weaver_2023}. 

Comparing the models, we find that \lgalaxies\ has a relatively steep SMF, predicting a large number of low-mass galaxies ($\log_{10}(M_\star/\rm{M_\odot})<8$), especially at the highest redshifts, while also predicting fewer high-mass galaxies ($\log_{10}(M_\star/\rm{M_\odot})>10$). At $z\leq9$, \scsam\ tends to have the highest SMF of the models, with \galform\ sitting $\sim0.3$ dex lower across the range of stellar masses probed. \shark\ predicts numbers of high-mass galaxies in between that of \galform\ and \lgalaxies, and has a similar slope to \scsam\ at lower stellar masses. As for the \eagle\ SMF, it has a steep slope initially, resembling that of \lgalaxies\ at $\log_{10}(M_\star/\rm{M_\odot})<9$. As stellar mass increases, it follows the higher \galform\ SMF more closely.

All in all, the spread of predictions from the models is generally consistent with observational constraints. \scsam\ appears to slightly overestimate the SMF at $z=5$. At first glance, \lgalaxies\ tends to slightly under-predict the number of high-mass galaxies ($\log_{10}(M_\star/\rm{M_\odot})>10$) relative to the observations. However, we find that the discrepancy is no longer significant after adding a small margin of error to the stellar masses, to account for Eddington bias (see appendix). At the highest redshifts ($z>10$), measurements by \cite{Harvey_2024} and \cite{DSilva_2025} suggest a higher SMF than predicted by most models, in particular \galform, which has the lowest SMF in this redshift regime. 

We note that the assumptions made in converting between observed data and galaxy properties may strongly affect our conclusions. Although \galform\ predicts the fewest galaxies at $z>10$, \cite{Lu_2025_GALFORM} show that \galform\ is within the range of observations of the UVLF at $z=10$. This may be due to a combination of the extremely top heavy IMF used by \galform\ for starbursts, as well as the forward-modelling choices made in \cite{Lu_2025_GALFORM}. It is also worth mentioning that observational constraints are less robust as redshift increases. \cite{Harvey_2024} find that for their $z\sim10.5$ galaxy sample, stellar masses can increase by $>1$ dex when switching from their fiducial parametric SFH model to a non-parametric model. They also find that using the redshift-dependent, top heavy IMF by \cite{Steinhardt_2023} reduces their estimated stellar masses at $z>8$ by $0.3-0.5\ {\rm dex}$, with a greater reduction at higher redshift. 

\subsection{Stellar-to-halo mass relation}\label{subsec:shmr}

The first row of Figure \ref{fig:properties} shows the stellar-to-halo mass relation (SHMR) of central galaxies, as a function of halo mass. We find little redshift evolution in the model predictions, except that they extend to higher halo masses as redshift decreases, as a result of hierarchical structure formation. \lgalaxies\ tends to populate low-mass halos ($\log_{10}(M_{\rm halo}/\rm{M_\odot})<10$) with the most massive galaxies. However, compared to other models, \lgalaxies\ predicts a slower growth of stellar mass as halo mass increases. This explains the steep SMF that we saw previously. At $\log_{10}(M_{\rm halo}/\rm{M_\odot})>10.5$, it is generally \scsam\ that predicts the highest stellar masses for a given halo mass, likely due to the assumption of a steeper slope for the Kennicutt–Schmidt relation \citep{Kennicutt_1989} above a critical gas mass density \citep{Somerville_2015_SCSAM, Yung_2019_JWST_II}. \galform\ has the steepest SHMR, predicting the lowest stellar masses in the smallest halos, 
and catching up to the other models at $\log_{10}(M_{\rm halo}/\rm{M_\odot})=11$.
At $z\leq8$, the \eagle\ SHMR tends to sit between the \galform\ and \scsam\ relations, with an initial gradient close to that of \galform, but eventually reaching slightly higher stellar masses as halo mass increases. Note that the increased stellar-to-halo mass ratio at low halo masses is due to the $M_\star>10^8\ \rm{M_\odot}$ cut applied to the \eagle\ predictions. At the highest halo masses ($\log_{10}(M_{\rm halo}/\rm{M_\odot})>11.5$), the stellar-to-halo mass ratios predicted by \eagle, \galform, \scsam\ and \shark\ increase more gradually. In \galform, there is a clear downturn in the relation, as AGN feedback starts to suppress the growth of massive galaxies. On the other hand in \lgalaxies, stellar mass increases at high halo masses without turning over.

\subsection{Star formation rates}\label{subsec:sfrs}

\begin{figure*}
	\includegraphics[width=2\columnwidth]{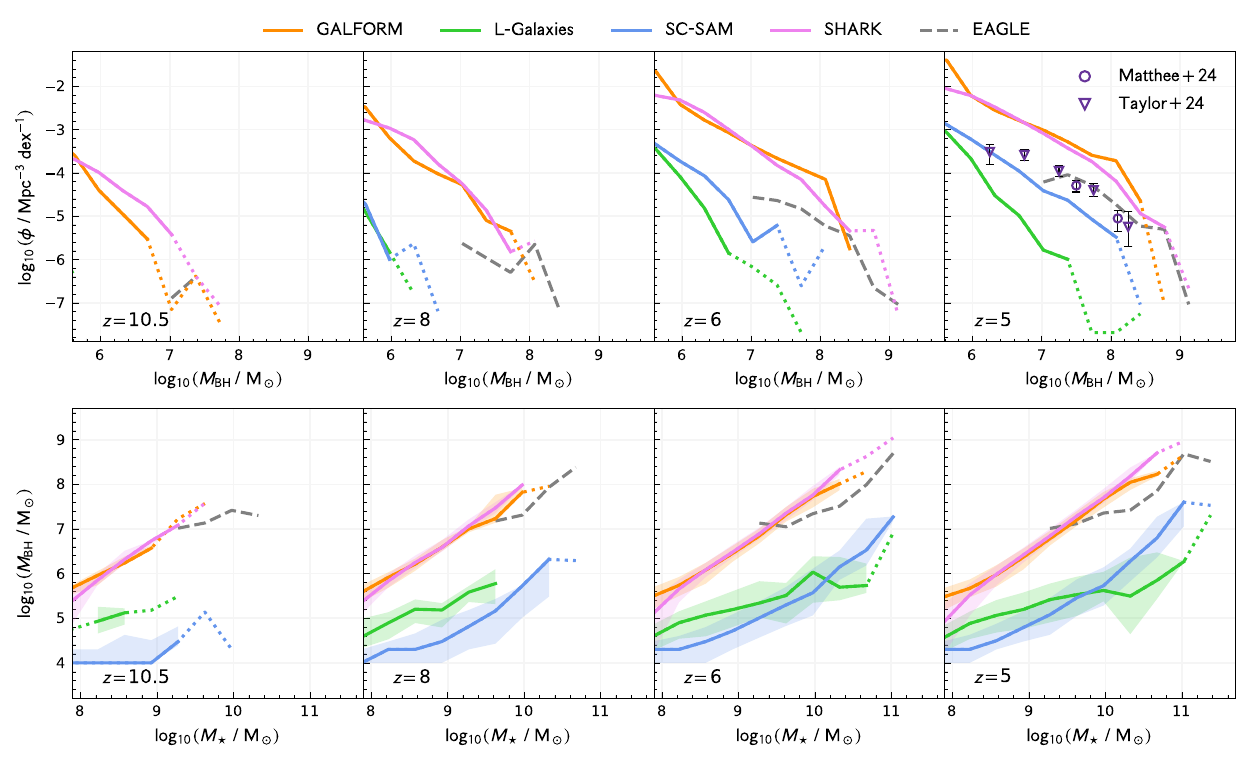}
	\caption{Each row shows a different quantity predicted by the four SAMs and \eagle\ across a range of redshifts. The top row shows the black hole mass function, and the bottom row shows the stellar-to-black hole mass relation. Note that a black hole mass selection of $M_{\rm BH}>10^7\ \rm{M_\odot}$ is applied to the \eagle\ results, and that the \eagle\ predictions at $z=10.5$ are at $z=10$, as outputs of the fiducial \flares\ suite are saved at integer redshift. For the SAMs, dotted lines denote bins containing fewer than 10 galaxies, and shaded regions denote the 1$\sigma$ range.}
    \label{fig:bh}
\end{figure*}

The second row of Figure \ref{fig:properties} shows the star formation rate function (SFRF) predicted by the models. At $z\leq8$, the SFRFs are fairly similar at lower SFRs, differing by not more than $\sim0.5\ {\rm dex}$ at ${\rm SFR}<10\ {\rm M_\odot yr^{-1}}$. The predictions diverge more at higher SFRs, with the highest and lowest SFRF predictions differing by more than $1\ {\rm dex}$ at ${\rm SFR}>10^{2}\ {\rm M_\odot yr^{-1}}$. 

\eagle\ and \lgalaxies\ tend to predict the fewest galaxies at high SFRs (${\rm SFR}>10\ {\rm M_\odot Gyr^{-1}}$). This aligns with our previous observation of \lgalaxies\ having a more gently-sloped SHMR, which is indicative of lower SFRs. However, it is interesting that \eagle\ should predict few galaxies with high SFRs, given that the \eagle\ SMF tends to be higher than that of \lgalaxies\ and \shark, and comparable to \galform\ at the high-mass end. 
Both \galform\ and \scsam\ predict larger populations of star-forming galaxies than the rest of the models at ${\rm SFR}>10\ {\rm M_\odot Gyr^{-1}}$. These are also the two models that have higher SMFs at $M_\star>10^9\ {\rm M_\odot}$. At lower SFRs (${\rm SFR}<10^{1.5}\ {\rm M_\odot Gyr^{-1}}$), \scsam\ tends to predict higher abundances, while \eagle\ predicts the lowest abundances (note though that a stellar mass cut of $M_\star>10^8\ {\rm M_\odot}$ is applied to the \eagle\ galaxies, thus the sample is incomplete at ${\rm SFR}<1\ {\rm M_\odot Gyr^{-1}}$). Also shown are SFRFs estimated from \textit{JWST} observations. \cite{Covelo_2025} obtain SFR values by converting from H$\alpha$ luminosities, and \cite{DSilva_2025} do so by by fitting SEDs to photometry. The estimated SFRFs favour models with high number counts at ${\rm SFR}<10\ {\rm M_\odot Gyr^{-1}}$, and a subsequent steep decline, such as \shark. We note that the estimated SFRs likely have large uncertainties than portrayed due to uncertainties in dust modelling and SED fitting. 

The final row of Figure \ref{fig:properties} shows the sSFR as a function of stellar mass. Across the SAMs, there is a gradual decrease in the sSFR with redshift, for fixed stellar mass. The model predictions are generally comparable. Interestingly, we find that \galform\ galaxies are slightly more star-forming than the other SAMs at $z=5-6$, despite the \galform\ SMF being consistently lower than that of \scsam. This may be due to \galform\ adopting a top heavy initial mass function (IMF), which leads to a larger fraction of stellar mass being instantaneously returned to the ISM when modelling starbursts \citep{Lacey_2016_GALFORM}. This would also explain why, when looking at the SFRF, \galform\ produces more highly star-forming galaxies than \scsam. In \lgalaxies, the median sSFR increases by $\sim0.5\ {\rm dex}$ between $\log_{10}(M_\star/{\rm M_\odot})=7.5-10$, while the other SAMs predict a flatter trend. In the appendix, we show the star-forming main sequence (SFS) predicted by the models, alongside several observations. The predicted relations are generally in good agreement with the observations, despite the differences seen here in Figure \ref{fig:properties}.

\subsection{Black hole properties}\label{subsec:bhmf}


\begin{figure*}
	\includegraphics[width=2\columnwidth]{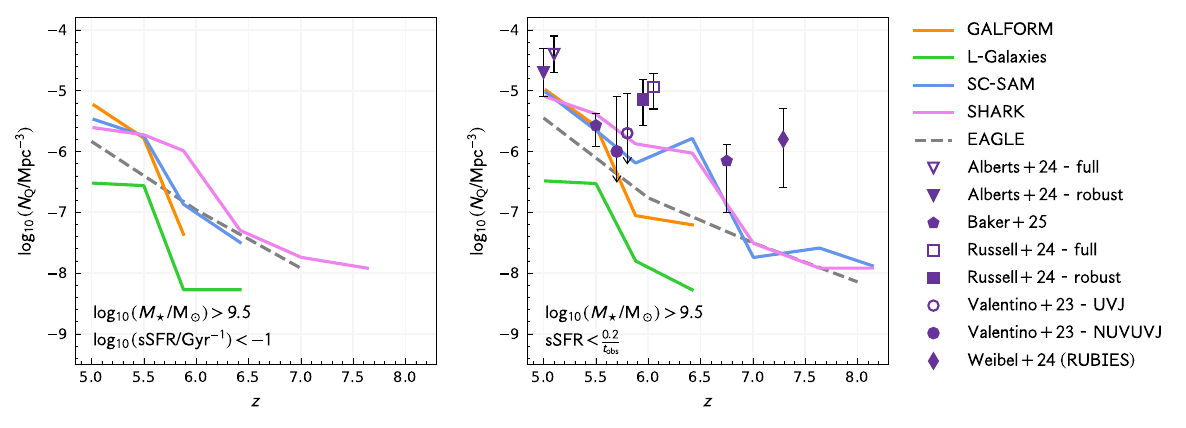}
	\caption{Number density of passive galaxies predicted by the four SAMs and \eagle\ when run on the \flares\ suite. In the left panel, the passive galaxy population is obtained using an sSFR cut of $\log_{10}(\rm{sSFR}/\rm{Gyr^{-1}})<-1$. For a more appropriate comparison with observations, the right panel uses an sSFR cut of $\log_{10}({\rm sSFR}/{\rm Gyr^{-1}})<0.2/t_{\rm obs}$. A mass cut of $\log_{10}(M_\star/\rm{M_\odot})>9.5$ is applied to facilitate comparison with observations.}
    \label{fig:passive_df}
\end{figure*}

The SAMs predict vastly different black hole mass functions (BHMFs), with a difference of $2-3\ {\rm dex}$ between the highest and lowest predictions, as shown in the first row of Figure \ref{fig:bh}. \galform\ and \shark\ predict the highest number of SMBHs across the mass range shown, and \eagle\ predicts $\sim1\ {\rm dex}$ fewer SMBHs with masses $M_{\rm BH}>10^7\ {\rm M_\odot}$. At lower masses, $M_{\rm BH}\sim10^6\ {\rm M_\odot}$, \lgalaxies\ and \scsam\ predict number densities that are $\sim2\ {\rm dex}$ lower than \galform\ and \shark. The \lgalaxies\ BHMF has a steeper negative slope than that of \scsam, due to a combination of \lgalaxies\ predicting fewer galaxies at $M_\star>10^8\ {\rm M_\odot}$, and also predicting lower mass black holes at a given stellar mass. 

The second row of Figure \ref{fig:bh} shows the black hole mass as a function of stellar mass. \galform\ and \shark\ exhibit similar stellar-to-black hole mass ratios, despite having rather different black hole models. In \lgalaxies, galaxies with a SMBH mass of $M_{\rm BH}>10^6\ {\rm M_\odot}$ tend to have higher stellar masses than the other models. This could be in part due to the fact that black holes in the \cite{Yates_2024_LGalaxies} and \cite{Henriques_2020_LGalaxies} \lgalaxies\ models do not grow through disk instabilities, which are known to contribute significantly to black hole growth in other versions of \lgalaxies\ \citep{Izquierdo_2020}. \cite{Yung_2021_JWST_V} mention that disk instabilities are an important mode of black hole growth in \scsam\ at early times \cite[see also][]{Hirschmann_2012}, and the same point is made by \cite{Lacey_2016_GALFORM} regarding the \galform\ model. 

We compare the predicted BHMFs at $z=5$ with observations by \cite{Matthee_2024} and \cite{Taylor_2024}. Both studies identify broad H$\alpha$ emitters (HAEs), and assume them to be broad-line AGN (BLAGN). The BHMFs of \scsam\ and \eagle\ roughly agree with these observations. However, as BLAGN are a subset of the total black hole population, the observations serve as lower limits rather than an estimate of the actual BHMF. The predictions by \galform\ and \shark\ may thus be closer to the truth.


%% file: sections/5_passive.tex
\section{Quiescent galaxies}\label{sec:passive}

In this section, we study the characteristics of quiescent galaxies predicted by the models. For the most part, we adopt the following definition of passivity:

\begin{equation}\label{eq:passive_theory}
    \log_{10}(\rm{sSFR}/\rm{Gyr^{-1}})<-1.
\end{equation}

We explored using a selection based on distance from the SFS, and found that for a difference of 1.5 dex, the results obtained are similar. In \S \ref{sec:passive_df}, we also use an alternative, time-dependent definition:

\begin{equation}\label{eq:passive_obs}
    {\rm sSFR}<\frac{0.2}{t_{\rm obs}},
\end{equation}
where $t_{\rm obs}$ is the age of the universe at the observed redshift of the galaxy \citep{Pacifici_2016}. This definition is more commonly used amongst observers.

\subsection{Number density}\label{sec:passive_df}


Figure \ref{fig:passive_df} shows the number density of massive quiescent galaxies as a function of redshift for two definitions of passivity, given by Equations \ref{eq:passive_theory} and \ref{eq:passive_obs} (left and right panels, respectively). Observations are also plotted, although the comparison is not quite apples-to-apples, given systematics such as the SFR timescale, and inherent differences in how simulated and observed galaxies are identified. Still, Equation \ref{eq:passive_obs} allows for a more appropriate comparison with recent observational estimates from JWST: 
\begin{itemize}
\item \cite{Alberts_2024} use NIRCam and MIRI data from the JADES field in their analysis. They select their full sample by UVJ classification \citep{Antwidanso_2023}, and identify robust passive candidates by requiring that 97.5\% of a galaxy's sSFR posterior, obtained from SED-fitting with \bagpipes\ \citep{Carnall_2018}, satisfies Equation \ref{eq:passive_obs}.
Their number densities are reported for stellar masses $\log_{10}(M_\star/\rm{M_\odot})>9.5$.
\item \cite{Baker_2025} identify passive galaxies by analysing NIRCam imaging from multiple JWST fields. A preliminary sample is identified using a relaxed UVJ selection, and processed using the \bagpipes\ SED-fitting code. Galaxies with sSFRs that satisfy Equation \ref{eq:passive_obs}
are considered to be passive. We show their predictions for galaxies with stellar masses $\log_{10}(M_\star/\rm{M_\odot})>9.5$.
\item \cite{Russell_2024} identify passive galaxies in the NEP, CEERS and JADES fields using primarily NIRCam imaging. They make their selection following Equation \ref{eq:passive_obs} -- at least 50\% (97.5\%) of a galaxy's sSFR posterior must meet the sSFR criteria to be included in the full (robust) sample. We show their estimated number densities for galaxies in the stellar mass range $\log_{10}(M_\star/\rm{M_\odot})>9.5$ by summing the number densities obtained for their two highest stellar mass bins.
\item \cite{Valentino_2023} analyse NIRCam, NIRISS and MIRI data from 11 JWST fields in their search for passive galaxies. Two selections are made, the first in relaxed UVJ space, and the second using Gaussian mixture modelling to identify candidates in NUV--U, U--V, V--J space. Their predictions are shown for the stellar mass range $\log_{10}(M_\star/\rm{M_\odot})>9.5$.
\item \cite{Weibel_2024_RUBIES} provide an estimate of the passive galaxy number density based on the discovery of RUBIES-UDS-QG-z7, a passive galaxy in the UltraDeep Survey (UDS) field, assuming a survey volume of $\sim300\ \rm{arcmin}^{-1}$ (PRIMER-UDS and CEERS combined).
\end{itemize}

\begin{figure*}
	\includegraphics[width=2\columnwidth]{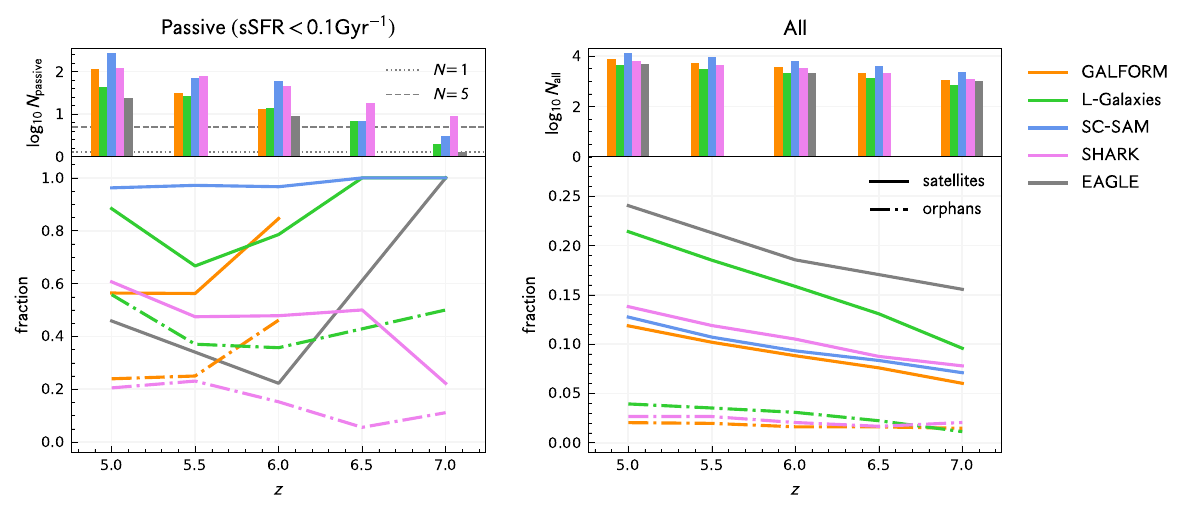}
	\caption{The left and right plots show the same statistics for different galaxy populations -- the passive galaxy population ($\log_{10}(M_\star/\rm{M_\odot})>9$ and $\log_{10}(\rm{sSFR}/\rm{Gyr^{-1}})<-1$) on the left, and the general galaxy population ($\log_{10}(M_\star/\rm{M_\odot})>9$)  on the right. The top panels show, for each population, the raw counts of galaxies predicted by the four SAMs and \eagle, as a function of redshift. The dotted and dashed grey lines in the top left panel indicate the height of a bin containing 1 and 5 galaxies respectively. Note that bins containing one galaxy are displayed as slightly larger than one, so that they appear on the plot. The bottom panels show the fraction of satellites and orphans (a subset of satellites), represented by solid and dot-dashed lines respectively. The orphan class of galaxies is applicable to neither \scsam\ nor \eagle\ -- in \scsam, satellites are modelled analytically, and \eagle\ is a hydrodynamics model. No information is shown for the \eagle\ run at $z=5.5$ and $6.5$, as fiducial \flares\ snapshots are saved at integer redshifts only.}
    \label{fig:passive_type}
\end{figure*}

Comparing both panels of Figure \ref{fig:passive_df}, we find that the fixed cut-off described by Equation \ref{eq:passive_theory} results in slightly lower number densities predicted by the models compared to the time-dependent cut-off described by Equation \ref{eq:passive_obs}, which is more relaxed at higher redshifts. In general, the passive abundances predicted by \galform, \scsam\ and \shark\ are comparable at $z=5$, but the \galform\ distribution function drops off more quickly as redshift increases. \lgalaxies\ predicts the fewest passive galaxies, $\sim1-2$ dex below \scsam\ and \shark\ at $z=5-6$. Note that the \eagle\ passive galaxies studied in this paper are not necessarily the same as those in \cite{Lovell_2023}, due to the differing physical apertures (and timescales, in the case of SFRs) over which galaxy properties are averaged.

Observations shown in the right panel of Figure \ref{fig:passive_df} favour the higher passive abundances predicted by \scsam\ and \shark. None of the models in our analysis predict the abundance estimated by \cite{Weibel_2024_RUBIES} -- reproducing this has been a struggle for most models in general, although the \colibre\ simulation is one exception \citep[][Chandro-Gomez et al. \textit{in prep}]{Chaikin_2025, Schaye_2025}. \lgalaxies\ appears to consistently underestimate the number density of passive galaxies at $z\geq5$. \cite{Araya-Araya_2025_LGalaxies}, who explore different calibrations of \lgalaxies, also find a similar result at high redshift. More observations will help to clarify the effect of cosmic variance on these observational estimates -- studies have shown that the size and depth of a survey may play a large role in the scatter of observed passive abundances \citep{Thomas_2023, Valentino_2023, Jespersen_2025}. This seems to be the case at lower redshifts ($3<z<5$), where passive abundances can vary by more than 1 dex from study to study \citep{Carnall_2023, Gould_2023, Valentino_2023, Alberts_2024, Russell_2024}.

\subsection{Satellite fractions}\label{sec:sat_frac_passive}


\begin{figure*}
	\includegraphics[width=2\columnwidth]{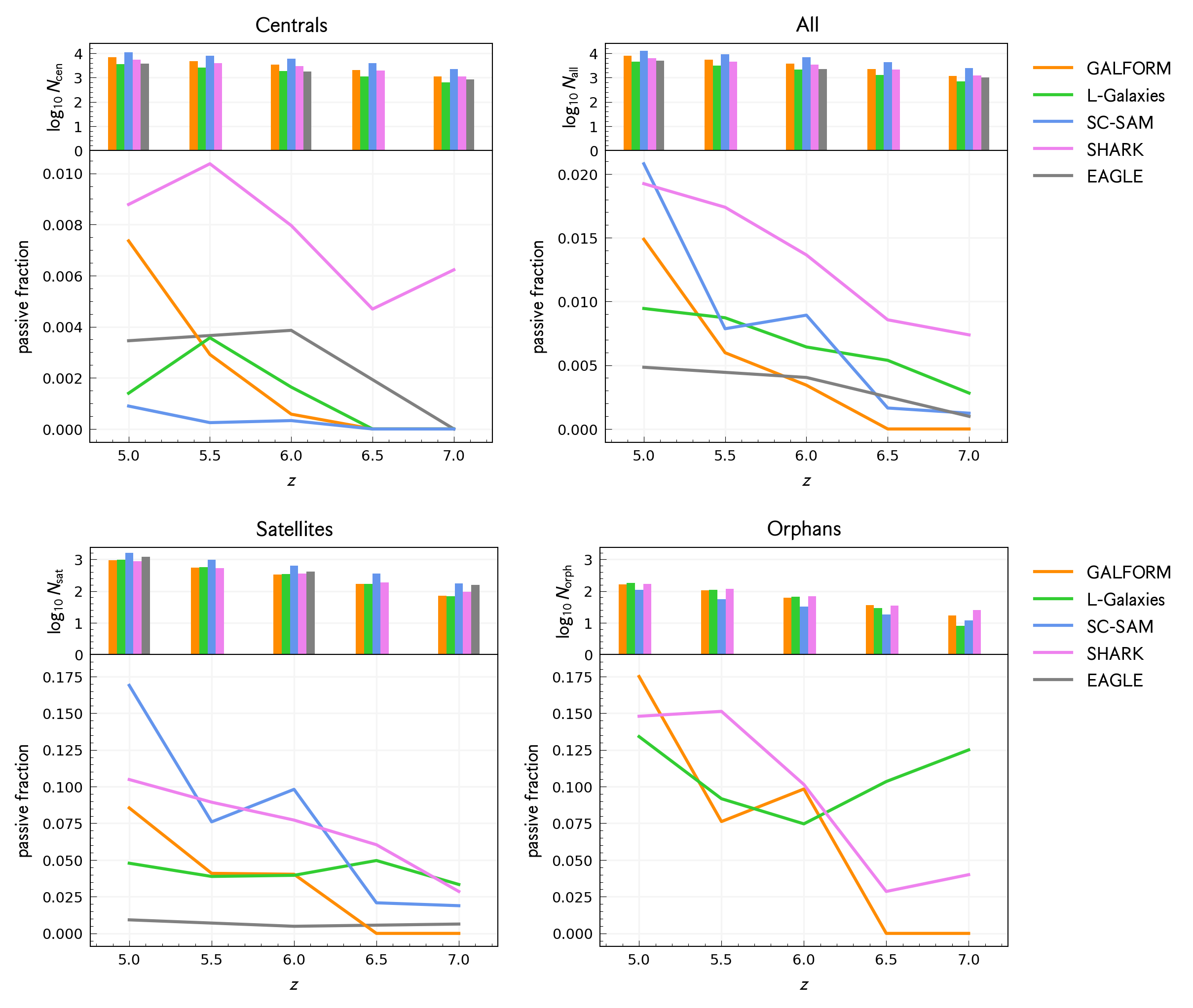}
	\caption{Each of the four subplots shows the same statistics for different galaxy populations (clockwise, from the top left): centrals, all galaxies, satellites, and orphans (a subset of satellites). A stellar mass cut of $\log_{10}(M_\star/\rm{M_\odot})>9$ is applied in all cases. The top panel of each subplot shows, for the galaxy population, the raw counts of galaxies predicted by the four SAMs and \eagle, as a function of redshift. The bottom panel shows the fraction of the population that are passive, i.e. with $\log_{10}(\rm{sSFR}/\rm{Gyr^{-1}})<-1$ at a given redshift. 
    No information is shown for the \eagle\ run at $z=5.5$ and $6.5$, as fiducial \flares\ snapshots are saved at integer redshift only.}
    \label{fig:passive_frac}
\end{figure*}

To investigate the nature of these passive galaxies, we first look at the proportion of centrals and satellite galaxies in the passive galaxy population. Generally, passive satellites are thought to be quenched more by environmental processes, and passive centrals more by AGN feedback. As we will see in the next section (\S \ref{sec:quench}), this distinction is valid for the SAMs, with the exception that AGN feedback in \lgalaxies\ and \scsam\ is not effective at the redshifts probed in this paper, perhaps due to SMBHs not growing quickly enough at high redshift (see Figure \ref{fig:bh}). In \eagle, it turns out that AGN feedback is the primary driver of passivity (also discussed in \S \ref{sec:quench}). From here on, a slightly lower stellar mass threshold of $\log_{10}(M_\star/\rm{M_\odot})>9$ is used, as we are no longer comparing with observations, but still focused on the massive galaxy population.

Figure \ref{fig:passive_type} shows that for all models in this analysis, across the redshifts probed, we find higher fractions of satellites in the passive galaxy population compared to the general galaxy population. This is particularly the case for the SAMs. At $z=5$, $>55\%$ of the passive galaxy population predicted by the SAMs are satellites, in contrast to $<25\%$ of the general galaxy population. \lgalaxies\ and \scsam\ predict notably high satellite fractions ($>90\%$ at $z=5$). Similar trends are observed at higher redshifts, though we caution that the statistics are less robust as redshift increases. Compared to the SAMs, \eagle\ predicts a smaller contrast, with satellites making up $45\%$ of the passive population and $25\%$ of the general population at $z=5$.

Satellites in SAMs are modelled in a highly idealised manner, which may lead to more rapid quenching, and thus higher passive fractions. For example, all satellites in \galform\ and \scsam\ are instantaneously stripped of hot gas, such that none can cool and form stars, whereas gradual hot gas stripping is applied in \lgalaxies\ and \shark. In reality, galaxies should be able to retain their hot halo for a period of time after becoming a satellite. \cite{Guo_2016} find that adopting a gradual stripping of hot gas in \galform\ leads to passive satellite numbers in closer agreement with \eagle. It is also possible for gas stripping to be overly strong in the SAMs that do not implement it instantaneously -- \cite{Harrold_2024} suggest that the discrepancy between the predicted and observed number of low-mass ($9<\log_{10}(M_\star/\rm{M_\odot})<10$) passive galaxies in \lgalaxies\ at $z<3$ could be solved if orphan galaxies were allowed to retain a larger cold gas supply.

Having noted a tendency for passive galaxies to be satellites rather than centrals, we can ask the question of how likely it is that a galaxy is passive, if it is a central, satellite or orphan. Figure \ref{fig:passive_frac} shows the passive fractions of the various galaxy populations. At $z=5$, \eagle\ predicts the lowest passive fraction overall ($0.5\%$), and the SAMs predict passive fractions between $1-2\%$. For all models, we find higher passive fractions in the satellite population compared to the central galaxy population, at a given redshift. In the SAMs, this suggests that the satellite-related processes are more effective at quenching galaxies than AGN. As mentioned in the opening paragraph of this section, this is less obviously the case in the \eagle\ model.

\begin{figure*}
    \includegraphics[width=2\columnwidth]{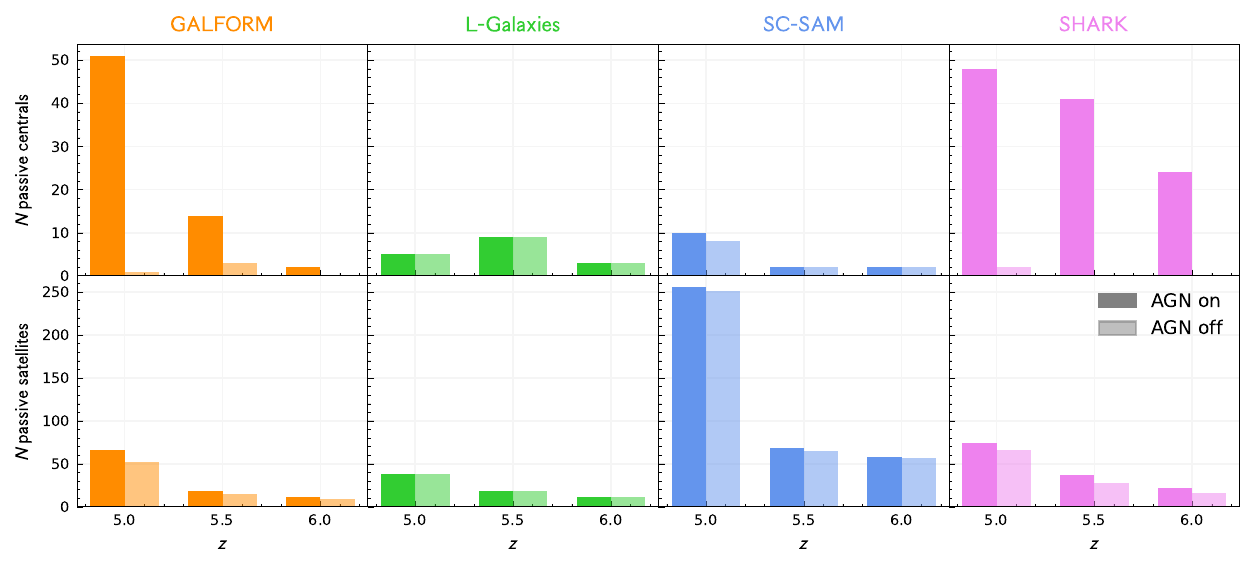}
    \caption{The top (bottom) row shows, the raw number count of central (satellite) galaxies that are passive when AGN feedback is left on (solid bars) or turned off (translucent bars), as a function of redshift. Each column shows the data for a different SAM, from left to right, \galform, \lgalaxies, \scsam\ and \shark. Passivity is defined using the sSFR cut $\log_{10}(\rm{sSFR}/\rm{Gyr^{-1}})<-1$. A stellar mass cut of $\log_{10}(M_\star/\rm{M_\odot})>9$ is applied as well.}\label{fig:count_passive}
\end{figure*}

Passive fractions do not exceed $20\%$ for the satellite populations, indicating that it is still a relatively small subset of satellites that are quenched. Focusing on satellites, we see that \eagle\ predicts the lowest passive fraction at $z=5$ ($1\%$), while \scsam\ predicts the highest ($17\%$), suggesting that the adopted satellite prescription is highly conducive to quenching. \shark, \galform\ and \lgalaxies\ fall in between, with passive fractions of $\sim10\%$, $8\%$ and $5\%$ in their satellite populations respectively. When it comes to the population of centrals at $z=5$, it is \lgalaxies\ and \scsam\ with the lowest passive fractions ($\sim0.1\%$), \galform\ and \shark\ with highest passive fractions ($\sim0.8\%$), and \eagle\ in between ($0.35\%$). \shark\ has a central galaxy passive fraction consistently higher than the rest ($>0.5\%$) at $z=5-7$, which suggests that AGN feedback is a particularly strong driver of passivity and effective out to higher redshifts, compared to the other models.

\subsection{Quenching mechanisms}\label{sec:quench}

In the previous section (\S \ref{sec:sat_frac_passive}), we looked at the occurrence of passivity in central and satellite galaxies, with the idea that passivity in these two populations may be caused by different physical processes. In this section, we explore quenching pathways in more detail. Due to the fundamental differences between hydrodynamic and semi-analytic models, we consider the two types of models separately.

\subsubsection{Quenching in the \eagle\ model}\label{sec:quench_eagle}

Passivity in the fiducial \flares\ suite (which uses the \eagle\ hydrodynamics model) has been studied in detail by \cite{Lovell_2023}. Their main finding is that passivity is strongly driven by SMBH feedback, regardless of whether the galaxy is a satellite or central. When black holes in \eagle\ accrete mass, an amount of energy proportional to the accretion rate is stored in a reservoir, to be used for AGN feedback. Once sufficient energy is available, feedback occurs via the stochastic heating of neighbouring gas particles. \cite{Lovell_2023} investigate the evolutionary histories of quiescent galaxies in the fiducial \flares\ suite and find that all exhibit a rapid decrease in SFR alongside a sharp spike in SMBH accretion rate. Thus, \cite{Lovell_2023} suggest that a galaxy is quenched when strong SMBH accretion triggers a feedback event that heats up the ISM, preventing further star formation.

With this in mind, it is interesting that at $z=5$, we observe quiescent galaxies in \eagle\ to be slightly biased towards being satellite galaxies (see discussion in \S \ref{sec:sat_frac_passive}, and Figures \ref{fig:passive_type} and \ref{fig:passive_frac}). As \eagle\ is a hydrodynamics model, there are no direct restrictions on the gas available for AGN in satellites to accrete. Thus, satellites are able to retain some hot gas that is then accreted onto the SMBH, allowing the transition from central to satellite occur smoothly, with no changes to the physics modelling. It is important to mention that we are limited by low number statistics -- at $z=5$, the \eagle\ model predicts a total of 24 passive galaxies in \flares\ (defined by $\log_{10}(M_\star/\rm{M_\odot})>9$ and $\log_{10}(\rm{sSFR}/\rm{Gyr^{-1}})<-1$). If the bias is indeed a feature of the model, a study by \cite{Visser_2025} on the quenching timescale of $z=0$ satellite galaxies in \eagle\ could provide some insight. By tracking gas particles, \cite{Visser_2025} find that satellites and centrals are similarly subject to outflows, likely from stellar or AGN feedback \citep{Mitchell_2020}. Crucially, while central galaxies are continuously replenished by inflows of fresh gas, satellites are not, and eventually run out of star-forming gas. It may be that a similar scenario is occurring in the fiducial \flares\ suite at high-redshift, contributing to the higher fraction of satellites in the passive galaxy population. This is speculative -- at higher redshifts, we would expect increased gas accretion rates onto galaxies due to increased accretion rates onto halos \citep{Vandevoort_2017}, and so cannot confidently assume that the `starvation' of satellites is occurring. Regardless of the role played by environmental processes, \cite{Lovell_2023} show that passive galaxies do not form when AGN feedback is turned off, which strongly suggests that AGN feedback is the dominant quenching mechanism in \eagle\ at $z>5$.

\subsubsection{Quenching in SAMs}

SAMs are, by definition, more prescriptive than hydrodynamic simulations. AGN feedback and environmental effects are generally understood to cause passivity in centrals and satellites respectively, at least at low redshift. Here, we test this hypothesis at high redshift.

To evaluate the impact of AGN feedback on passivity, we rerun the SAMs with AGN feedback turned off\footnote{Where models implement more than one mode of AGN feedback, all modes are turned off.} and examine how the number of passive satellites and centrals changes. Our results are summarised in Figure \ref{fig:count_passive}. When AGN feedback is turned off, the number of passive centrals predicted by \galform\ and \shark\ is reduced to zero, or close to zero. This suggests that in the two models, AGN feedback is the primary driver of passivity in central galaxies. Turning off AGN feedback makes little or no difference to the number of passive centrals predicted by \lgalaxies\ and \scsam, suggesting that central galaxies are not quenched by AGN feedback at these redshifts. We note that both models predict relatively few passive centrals ($<10$) compared to \galform\ and \shark\ ($\sim 50$ at $z=5$). Our finding for \scsam\ agrees with that of \cite{Yung_2019_JWST_I}, who observe no difference in their $z>4$ analysis when turning off AGN feedback. \cite{Henriques_2017} mention AGN feedback as a cause of passivity in \lgalaxies\ at low-redshift, but it seems that it is not yet effective at high-redshift, perhaps due to the slow growth rate of SMBHs, as observed in Figure \ref{fig:bh} (and discussed in \S \ref{subsec:bhmf}).

In the SAMs, the impact of AGN feedback on satellite galaxies is minimal. Having seen that AGN feedback has little effect on the number of passive centrals predicted by \lgalaxies\ and \scsam, it is not surprising that we should observe the same trend for passive satellites in these two models. In \galform, satellite galaxies do not undergo AGN feedback, as they do not retain their hot halo (a prerequisite for AGN feedback). The small reduction in the number of passive satellites when AGN feedback is turned off may be attributed to the effects of AGN feedback prior to the transition from central to satellite. In \shark, AGN feedback continues as per normal in satellite galaxies (provided the necessary conditions are met), however, the similar numbers of passive satellites observed in runs with and without AGN feedback show that AGN feedback is not a primary quenching mechanism of satellites in \shark.

The link between passivity and satellite-related processes in SAMs is well-established -- it has been mentioned and explored in previous works, e.g. \cite{Guo_2016} for \galform, \cite{Guo_2011_LGalaxies, Harrold_2024, Vani_2025} for \lgalaxies, \cite{Somerville_2008_SCSAM, Porter_2014} for \scsam, \cite{Lagos_2025} for \shark. While the exact treatment of satellites differs between the models, all prescriptions essentially reduce the amount of cold gas available for forming stars by stripping away hot gas (which would cool into cold gas), and in some cases by stripping away cold gas as well. All the SAMs in this analysis have $\rm H_2$-dependent SFRs. Thus, the SFR of a galaxy tends to trace its cold gas mass. In Section \ref{sec:app_passive_sats} of the Appendix, we include a plot of the evolutionary histories of several passive galaxies predicted by the SAMs, which highlights how the cold gas mass and sSFR decrease in tandem when a galaxy becomes a satellite. 

Our findings suggest that the vast majority of passive satellites in the SAMs are indeed quenched by satellite-related processes. AGN feedback appears to play a significant role in the quenching of centrals in \galform\ and \shark, but not in \lgalaxies\ and \scsam.

\subsection{Discussion}

We have built up a general picture of AGN feedback being the main driver of passivity in \eagle, and environmental quenching being more prevalent in the SAMs. In \eagle, AGN feedback leads to outflows of gas in both centrals and satellites alike. It is possible that satellites in \eagle\ are additionally quenched by a lack of fresh gas, as inflows tend to draw towards central galaxies \citep{Visser_2025}. This is quite different from what we see in the SAMs, where AGN feedback can drive passivity in centrals (in the case of \galform\ and \shark), but contributes minimally to the quenching of satellites. In the SAMs, passivity in satellites is mainly attributed to environmental processes such as ram pressure stripping and tidal stripping. While passive galaxies have been observed at high-redshift, not much information is available as to how they are quenched, or whether they are centrals or satellites. At lower redshift ($1.5<z<3$), there is some evidence that \lgalaxies\ tends to over-predict the number of low-mass ($9<\log_{10}(M_\star/\rm{M_\odot})<10$) passive galaxies \citep{Asquith_2018, Harrold_2024} and under-predict the number of high-mass ($\log_{10}(M_\star/\rm{M_\odot})>10.5$) passive galaxies \citep{Henriques_2020_LGalaxies, Vani_2025}, which suggests that environmental quenching may be overly strong and AGN feedback not strong enough. Figure \ref{fig:passive_df} showed that the SAMs predict passive galaxy abundances that are either in line with or lower than observational estimates. Thus, if any of the SAMs, particularly those with high satellite fractions in their passive populations, do predict too many environmentally-quenched satellites at $z>5$, then quenching by AGN feedback would have to occur more frequently to match the observed passive abundances. In \eagle, satellite evolution is already self-consistently modelled. Hence, the modelling of AGN would have to be adjusted to better match high-redshift observations \citep{Turner_2025}.

%% file: sections/6_conclusion.tex
\section{Conclusions}\label{sec:conclusions}

We have run four SAMs (\galform, \lgalaxies, \scsam\ and \shark) at $5\leq z \leq12$ using the \flares\ simulation strategy. The predictions of the SAMs are compared with recent observations from \textit{JWST}, as well as results from the fiducial \flares\ suite, which is run with the \eagle\ hydrodynamics model. In this paper, we have explored the general properties of the simulated galaxies and also taken a closer look at how passive galaxies form in the different models. Our results are summarised below:

\begin{itemize}

\item At $z<10$, there is generally good overlap between the predicted SMFs and recent observational estimates. \scsam\ predicts the highest abundances at $\log_{10}(M_{\rm \star}/\rm{M_\odot})>9$, and \lgalaxies\ tends to predict fewer galaxies, especially at the highest masses ($\log_{10}(M_{\rm \star}/\rm{M_\odot})>10$). At $z>10$, the models tend to under-predict the SMF relative to observations, however we caution that observations are more uncertain at such early times.

\item Looking at the SHMR, we find that central galaxies in low-mass ($\log_{10}(M_{\rm halo}/\rm{M_\odot})<10$) halos are most massive in \lgalaxies, but grow more slowly compared to the other three SAMs. On the other hand, \galform\ has the steepest SHMR, predicting the smallest galaxies in $M_{\rm halo}<10^{10}\ {\rm M_\odot}$ halos and catching up to the other models at higher masses. 

\item The sSFRs predicted by the models are generally comparable. Galaxies in \galform\ tend to be the most highly star-forming despite the model not predicting the highest SMF -- this could be due to the top heavy IMF adopted. In \galform, \scsam\ and \shark, the sSFR of galaxies remains roughly constant at $\log_{10}(M_{\rm \star}/\rm{M_\odot})>8$, while in \lgalaxies, the sSFR appears to increase with stellar mass. Looking at the SFRF, we find that \lgalaxies\ predicts the fewest highly star-forming galaxies (${\rm SFR}>10\ {\rm M_\odot yr^{-1}}$), while \galform\ and \scsam\ tend to predict the most.

\item There are significant differences in the black hole populations predicted by the different models, with \galform\ and \shark\ predicting the highest number densities, and \lgalaxies\ and \scsam\ predicting number densities $\gtrsim2\ {\rm dex}$ lower. The \eagle\ BHMF is somewhere in the middle, closest to observational estimates. However, we note that the observations likely underestimate the total black hole abundances. In agreement with our findings on the BHMF, we note that \galform\ and \shark\ predict higher black hole masses for a galaxy of a given stellar mass than \lgalaxies\ and \scsam.

\item In terms of passive number densities, the model predictions are at the low end of observational constraints. We find a wide range of predictions, with \scsam\ and \shark\ generally predicting the highest passive abundances, but for different reasons (see below). 

\item The majority of the passive galaxies at $\log_{10}(M_{\rm *}/\rm{M_\odot})>9$ in \eagle\ are centrals. Previous analysis by \cite{Lovell_2023} of the fiducial \flares\ suite identifies AGN feedback as the main driver of passivity, be it in centrals or satellites. We find that the fraction of satellites in the passive population is still slightly higher than the fraction of satellites in the general galaxy population. This might be explained by the contribution of additional environmental effects such as satellites being starved of gas \citep{Visser_2025}, however, due to low number counts we are tentative with this suggestion.

\item We show that passive satellites in the SAMs are primarily quenched by environmental processes. Passive centrals in \galform\ and \shark\ are likely caused by AGN feedback, while in \lgalaxies\ and \scsam, we find that AGN feedback is not yet an effective quenching mechanism. Passive galaxies in \scsam\ are especially biased towards being satellites, with a satellite fraction of $>90\%$.

\end{itemize}

In this study, we find model predictions of the SMF and SFRF to be generally within range of high-redshift observations. More discrepancy is observed in the BHMF and passive galaxy abundances. Our findings highlight how differences in the modelling of satellite galaxies and black holes affect the predicted properties of passive galaxies. An in-depth exploration of the physics modelling was not within the scope of this paper, but would certainly be an interesting and feasible topic for future studies, given the resource-friendly nature of SAMs and the \flares\ zoom strategy.


%% file: sections/appendix.tex





\section{Physical Properties}\label{sec:prop_app}

\begin{figure*}
    \includegraphics[width=2\columnwidth]{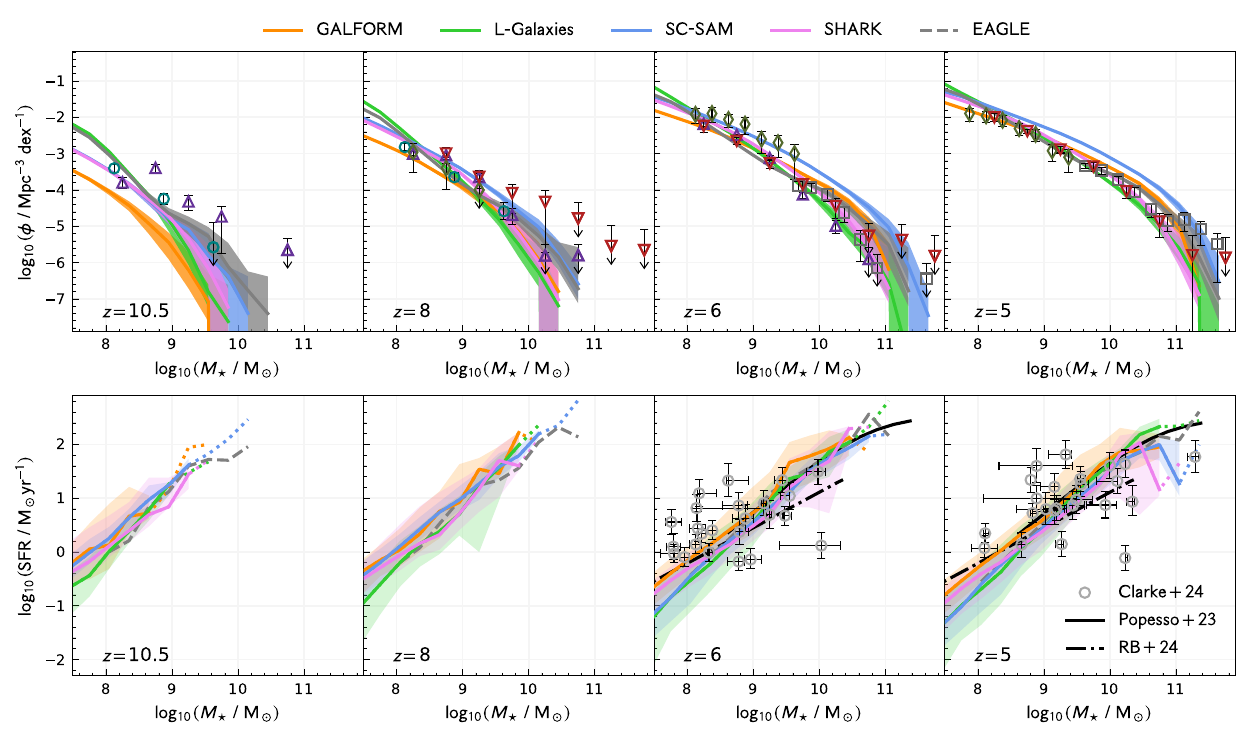}
	\caption{Top row: SMF predicted by the models across a range of redshifts, obtained after convolving stellar masses with errors drawn from a Gaussian distribution of width of 0.3 dex. Shaded regions denote the 1$\sigma$ range for 500 iterations of errors. Observations are shown as scatter points (please see Figure \ref{fig:smf} for the legend). Bottom row: Star formation rate as a function of stellar mass, predicted by the models across a range of redshifts. Note that a stellar mass cut of $M_\star>10^8\ \rm{M_\odot}$ is applied to the \eagle\ results, and that the \eagle\ predictions at $z=10.5$ are in fact at $z=10$, as outputs of the fiducial \flares\ suite are saved at integer redshift. For the SAMs, dotted lines denote bins containing fewer than 10 galaxies. Observational measurements by \protect\cite{Popesso_2023} and \protect\cite{RB_2024} are plotted as solid and dot-dashed black lines respectively, and measurements by \protect\cite{Clarke_2024} are plotted as scatter points.}
    \label{fig:prop_app}
\end{figure*}

The top row of Figure \ref{fig:prop_app} shows the SMF obtained after convolving stellar masses with errors drawn from a Gaussian distribution of width of 0.3 dex. By invoking a relatively small error, we find that the predictions by \lgalaxies\ are within range of observations.

The bottom row of Figure \ref{fig:prop_app} shows the SFS predicted by the models. The slope of the SFS is fairly similar across models. At $z=5$, \lgalaxies\ and \scsam\ exhibit a slightly steeper slope, with a median $\sim0.5\ {\rm dex}$ lower than that of \galform\ and \shark\ at the lowest stellar masses ($M_\star=10^{7.5}\ {\rm M_\odot}$). At higher stellar masses ($M_\star>10^{9}\ {\rm M_\odot}$), the models converge. We also show the best-fitting relation provided by \cite{Popesso_2023}, who convert measurements of the SFS from 50 studies, published between 2007 and 2022, to a \cite{Kennicutt_2012} calibration. We also show a linear fit provided by \cite{RB_2024} to the data of 330 spectroscopically-confirmed galaxies between $5<z<7$. SFRs are estimated from UV luminosity by applying a \cite{Kennicutt_1998} conversion. The models generally show good agreement with the SFS measured by \cite{Popesso_2023}. The gradient of the SFS measured by \cite{RB_2024} is gentler, and favours the higher SFRs at low stellar masses ($M_\star\sim10^{8}\ {\rm M_\odot}$) that are observed in \galform\ and \shark. At $M_\star>10^{9}\ {\rm M_\odot}$, the models predict that galaxies are more highly star-forming than estimated by \cite{RB_2024}. Finally, we also plot measurements by \cite{Clarke_2024} for individual galaxies, using SFRs derived from the H$\alpha$ luminosity, which is sensitive to star formation on short timescales ($\sim5$ Myr).

\section{Convergence tests}\label{sec:conv}

In \S \ref{sec:part_res}, we described the convergence tests performed to determine an appropriate DMO resolution for our analysis. 
The main findings for each SAM are summarised as follows:
\begin{itemize}

\item 
\galform: Overall, we find that the model makes very similar predictions when run on simulations with particle masses in the range $M_{\rm DM}=8.36\times10^7-3.22\times10^8 \rm{M_\odot}$. At the highest redshifts ($z\geq10.5$), the highest resolution ($M_{\rm DM}=1.15\times10^7\ \rm{M_\odot}$) run produces a lower SMF and SFRF than the benchmark run, by $\sim0.5\ {\rm dex}$. Additionally, across all redshifts, the highest resolution run predicts a $\sim0.5-2\ {\rm dex}$ excess of black holes at $M_{\rm BH}<10^{6.5}\ \rm{M_\odot}$, and a lower stellar-to-black hole mass ratio in the same black hole mass range, compared to the benchmark resolution run.

\item 
\lgalaxies: We generally find good convergence across the tested resolution range for the quantities shown, with results in agreement with predictions from the \cite{Yates_2024_LGalaxies} model when run on the complete set of \mill-II trees, and predictions from \cite{Henriques_2020_LGalaxies} when run on the complete set of \mill-I trees. An exception to this is the BHMF, where the result of the highest resolution ($M_{\rm DM}=1.15\times10^7\ \rm{M_\odot}$) run is $\sim1\ {\rm dex}$ lower than that of the lowest resolution run ($M_{\rm DM}=3.22\times10^8\ \rm{M_\odot}$). The BHMF appears to increase with increasing particle mass. It is unclear as to whether the result converges at particle masses beyond the tested range. 
Consequently, the stellar-to-black hole mass relation is also resolution-dependent -- at a given stellar mass, we find a $\sim1-1.5\ {\rm dex}$ difference between the SMBH masses predicted by the highest and lowest resolutions.

\item 
\scsam: We find good convergence across all resolutions, and good agreement with predictions by \cite{Yung_2024_SCSAM}, except when it comes to black hole properties. Interestingly, the resolution-dependence of the BHMF is the reverse of that observed in \lgalaxies\ -- at $M_{\rm BH}<10^{6}\ \rm{M_\odot}$, we find that the higher the resolution, the higher the BHMF. The BHMFs appear to converge at $M_{\rm BH}>10^6\ \rm{M_\odot}$, especially at lower redshifts ($z=5-6$). The stellar-to-black hole mass relation also changes with particle mass. At $M_{\rm BH}<10^{6}\ \rm{M_\odot}$, we find that the lower the particle mass, the smaller the stellar-to-black hole mass ratio. At $M_{\rm BH}>10^{6}\ \rm{M_\odot}$, there is better convergence between the different resolutions.

\item 
\shark: We find that the model predictions generally converge for particle masses in the range $M_{\rm DM}=8.36\times10^7-3.22\times10^8\ \rm{M_\odot}$. The highest resolution ($M_{\rm DM}=1.15\times10^7\ {\rm M_\odot}$) run predicts a SMF and SFRF $\sim0.3-0.5\ {\rm dex}$ lower than the benchmark run. This difference shows up in the SHMR as well, where for a given halo mass, this run tends to produce galaxies with a lower stellar mass, particularly around the characteristic turnover at $M_{\rm halo}\sim10^{11.5}\ {\rm M_\odot}$. Additionally, the $M_{\rm DM}=1.15\times10^7\ {\rm M_\odot}$ run predicts a lower BHMF than the benchmark run. As the stellar-to-black hole mass relation is converged across the range of particle masses tested this difference in the BHMF can be attributed to the aforementioned differences in the SMF.

\end{itemize}

\section{Passive satellites in SAMs}\label{sec:app_passive_sats}

\begin{figure*}
    \begin{minipage}{.9\columnwidth} 
        \includegraphics[width=\columnwidth]{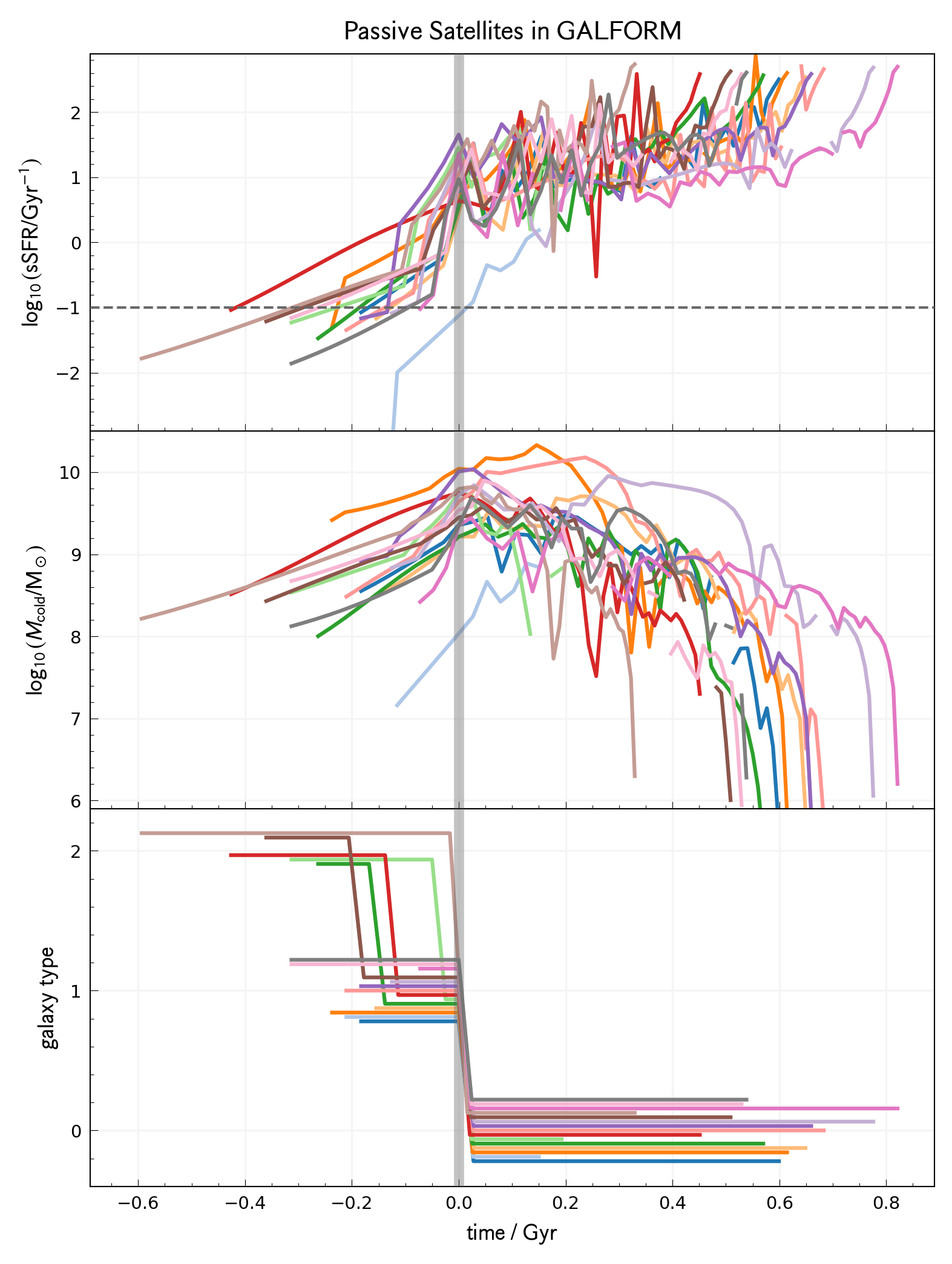}
    \end{minipage}
    \begin{minipage}{.9\columnwidth} 
        \includegraphics[width=\columnwidth]{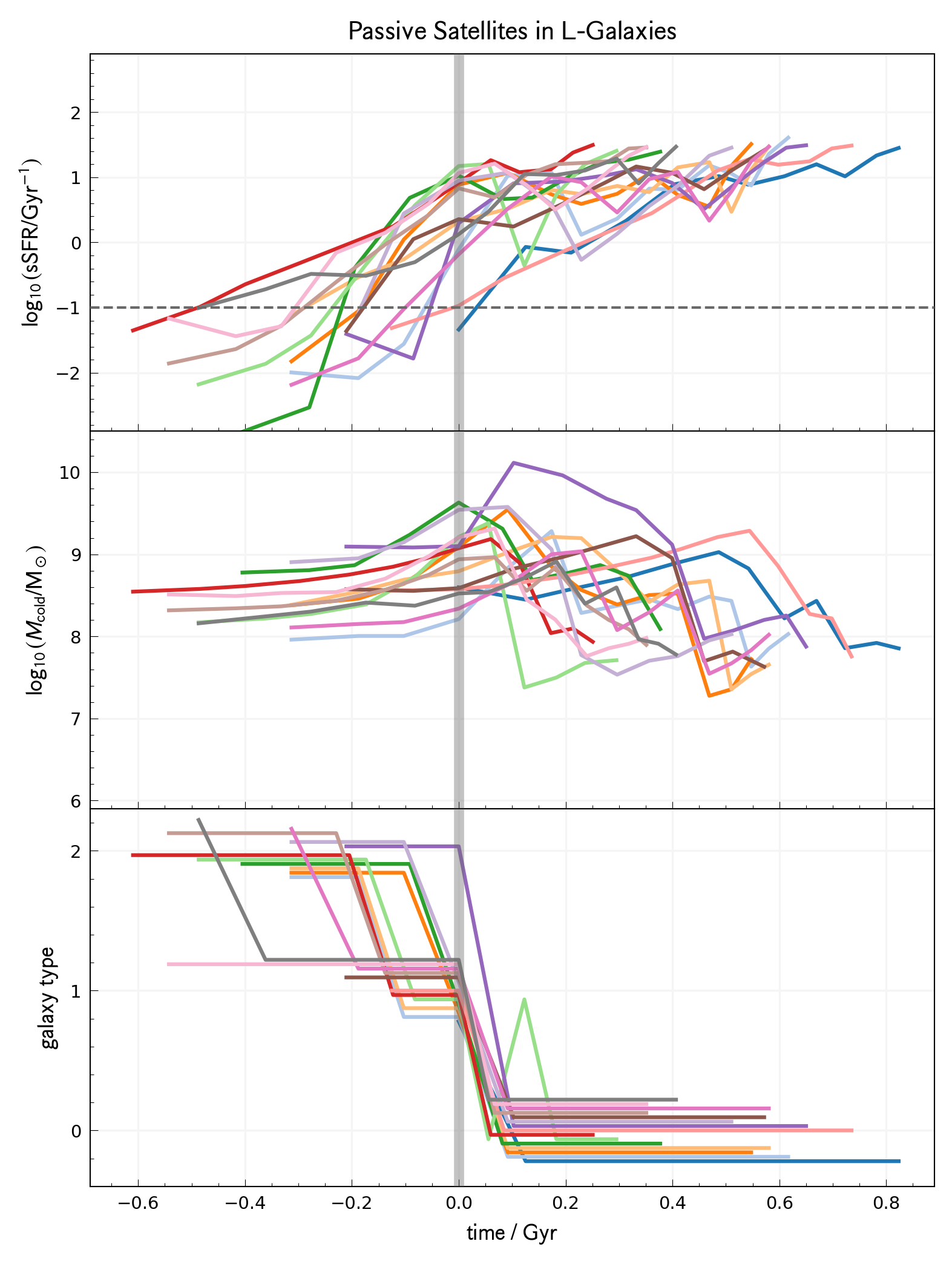}
    \end{minipage}
    \begin{minipage}{.9\columnwidth} 
        \includegraphics[width=\columnwidth]{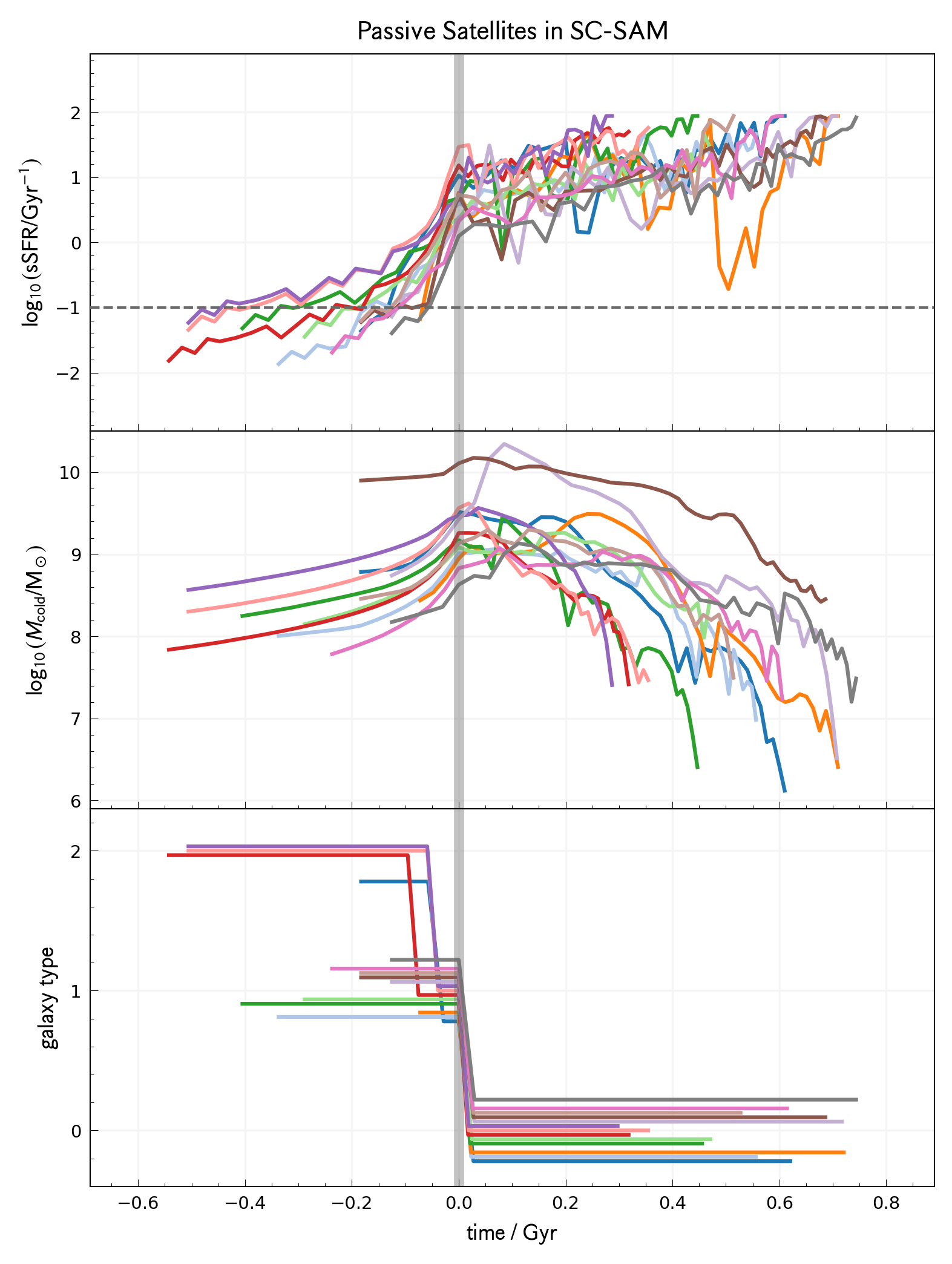}
    \end{minipage}
    \begin{minipage}{.9\columnwidth} 
        \includegraphics[width=\columnwidth]{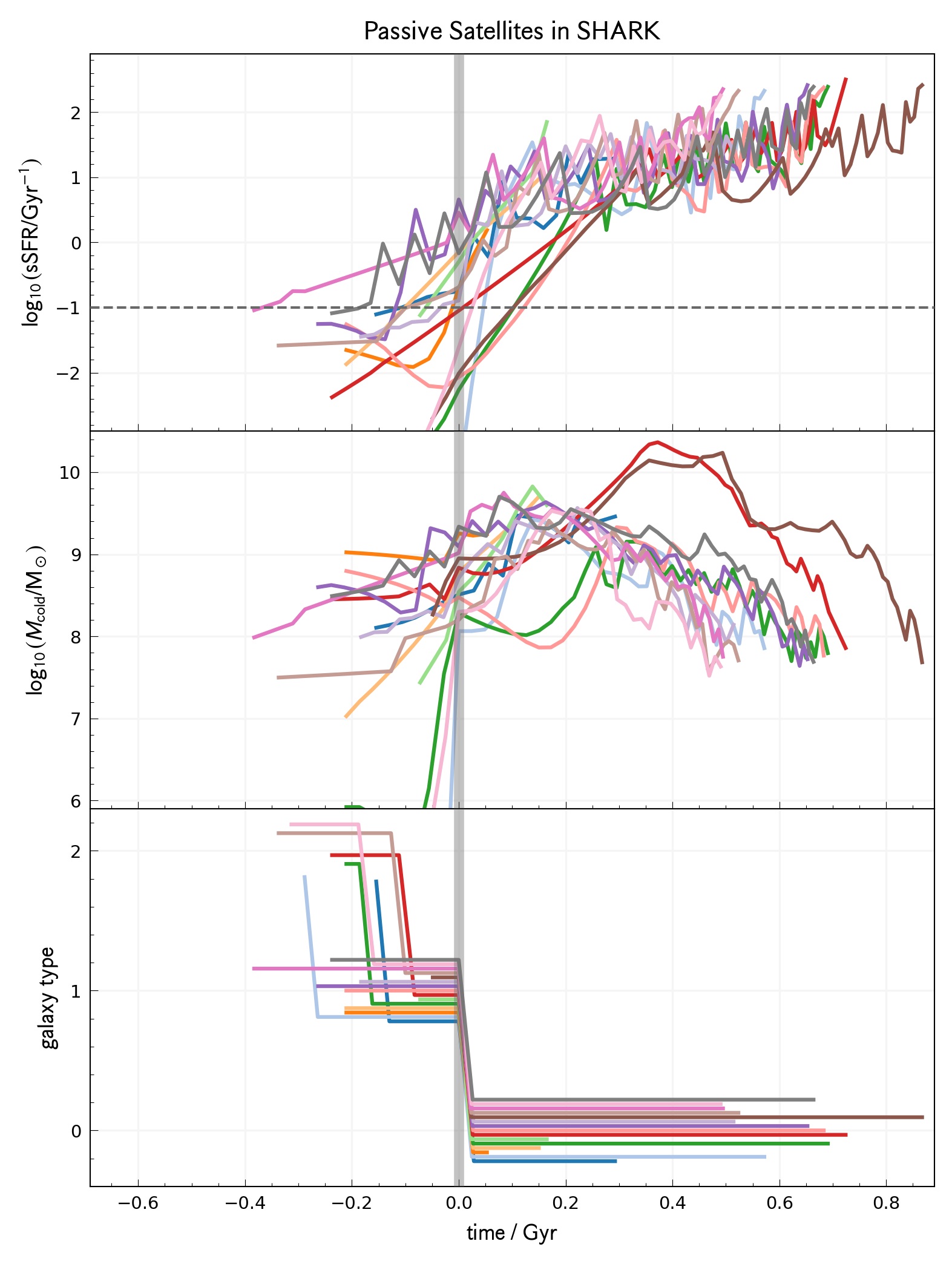}
    \end{minipage}
	\caption{Each plot shows, for a particular SAM, the evolutionary histories of 15 randomly chosen passive satellites with $\log_{10}(M_\star/\rm{M_\odot})>9$. The panels within each plot show, from top to bottom, the sSFR, cold gas mass and galaxy type of each galaxy as a function of time (see text for explanation of types). The dashed horizontal line in the top panel represents the sSFR threshold below which a galaxy is considered passive. Note that the galaxy type of each galaxy is slightly offset for clarity. The galaxies are chosen at $z=5$, but aligned such that the transition from central to satellite occurs at $t=0$, marked by the grey vertical line. The $x$-axis thus shows the time to the transition occurring, with time moving from right to left.}\label{fig:sat_evo}
\end{figure*}

Figure \ref{fig:sat_evo} shows, for each SAM, how the sSFR, cold gas mass and galaxy type evolves with time, for a handful of passive satellite galaxies. Centrals are labelled as type 0. In \scsam, type 1 galaxies are satellites of centrals, while type 2 galaxies are satellites of satellites. In the remaining SAMs, type 1 galaxies are satellites associated with a resolved subhalo, while type 2 satellites are those whose associated subhalo have become sub-resolution. In Figure \ref{fig:sat_evo}, the time axis is adjusted such that the galaxies transition from a central to satellite at $t=0$. Generally, the sSFR of a galaxy tends to trace its cold gas mass. \galform\ and \scsam\ show a steady decline in the cold gas mass and and hence sSFR after a galaxy becomes a satellite. In \shark, the cold gas mass evolves more erratically, dropping sharply in some galaxies while declining more steadily in others. 
As for \lgalaxies, the cold
gas mass is observed to decline after a galaxy becomes a satellite, however, the resultant change in the sSFR is not as obvious, perhaps due to the lower snapshot cadence adopted.

%% file: cite/algorithms.bib
@ARTICLE{Behroozi_2013_CTrees,
       author = {{Behroozi}, Peter S. and {Wechsler}, Risa H. and {Wu}, Hao-Yi and {Busha}, Michael T. and {Klypin}, Anatoly A. and {Primack}, Joel R.},
        title = "{Gravitationally Consistent Halo Catalogs and Merger Trees for Precision Cosmology}",
      journal = {\apj},
         year = 2013,
        month = jan,
       volume = {763},
       number = {1},
          eid = {18},
        pages = {18},
          doi = {10.1088/0004-637X/763/1/18},
archivePrefix = {arXiv},
       eprint = {1110.4370},
 primaryClass = {astro-ph.CO},
       adsurl = {https://ui.adsabs.harvard.edu/abs/2013ApJ...763...18B}
}

@ARTICLE{Behroozi_2013_Rockstar,
       author = {{Behroozi}, Peter S. and {Wechsler}, Risa H. and {Wu}, Hao-Yi},
        title = "{The ROCKSTAR Phase-space Temporal Halo Finder and the Velocity Offsets of Cluster Cores}",
      journal = {\apj},
         year = 2013,
        month = jan,
       volume = {762},
       number = {2},
          eid = {109},
        pages = {109},
          doi = {10.1088/0004-637X/762/2/109},
archivePrefix = {arXiv},
       eprint = {1110.4372},
 primaryClass = {astro-ph.CO},
       adsurl = {https://ui.adsabs.harvard.edu/abs/2013ApJ...762..109B}
}

@ARTICLE{Elahi_2019_TreeFrog,
       author = {{Elahi}, Pascal J. and {Poulton}, Rhys J.~J. and {Tobar}, Rodrigo J. and {Ca{\~n}as}, Rodrigo and {Lagos}, Claudia del P. and {Power}, Chris and {Robotham}, Aaron S.~G.},
        title = "{Climbing halo merger trees with TreeFrog}",
      journal = {\pasa},
         year = 2019,
        month = aug,
       volume = {36},
          eid = {e028},
        pages = {e028},
          doi = {10.1017/pasa.2019.18},
archivePrefix = {arXiv},
       eprint = {1902.01527},
 primaryClass = {astro-ph.IM},
       adsurl = {https://ui.adsabs.harvard.edu/abs/2019PASA...36...28E}
}

@ARTICLE{Jiang_2014_Dhalo,
       author = {{Jiang}, Lilian and {Helly}, John C. and {Cole}, Shaun and {Frenk}, Carlos S.},
        title = "{N-body dark matter haloes with simple hierarchical histories}",
      journal = {\mnras},
         year = 2014,
        month = may,
       volume = {440},
       number = {3},
        pages = {2115-2135},
          doi = {10.1093/mnras/stu390},
archivePrefix = {arXiv},
       eprint = {1311.6649},
 primaryClass = {astro-ph.CO},
       adsurl = {https://ui.adsabs.harvard.edu/abs/2014MNRAS.440.2115J}
}

@ARTICLE{RG_2015_SubLink,
       author = {{Rodriguez-Gomez}, Vicente and {Genel}, Shy and {Vogelsberger}, Mark and {Sijacki}, Debora and {Pillepich}, Annalisa and {Sales}, Laura V. and {Torrey}, Paul and {Snyder}, Greg and {Nelson}, Dylan and {Springel}, Volker and {Ma}, Chung-Pei and {Hernquist}, Lars},
        title = "{The merger rate of galaxies in the Illustris simulation: a comparison with observations and semi-empirical models}",
      journal = {\mnras},
         year = 2015,
        month = may,
       volume = {449},
       number = {1},
        pages = {49-64},
          doi = {10.1093/mnras/stv264},
archivePrefix = {arXiv},
       eprint = {1502.01339},
 primaryClass = {astro-ph.GA},
       adsurl = {https://ui.adsabs.harvard.edu/abs/2015MNRAS.449...49R}
}

@ARTICLE{Roper_2020_MEGA,
       author = {{Roper}, William J. and {Thomas}, Peter A. and {Srisawat}, Chaichalit},
        title = "{MEGA: Merger graphs of structure formation}",
      journal = {\mnras},
         year = 2020,
        month = may,
       volume = {494},
       number = {3},
        pages = {4509-4524},
          doi = {10.1093/mnras/staa982},
archivePrefix = {arXiv},
       eprint = {2003.01187},
 primaryClass = {astro-ph.GA},
       adsurl = {https://ui.adsabs.harvard.edu/abs/2020MNRAS.494.4509R}
}

@ARTICLE{Springel_2001_Subfind,
       author = {{Springel}, Volker and {White}, Simon D.~M. and {Tormen}, Giuseppe and {Kauffmann}, Guinevere},
        title = "{Populating a cluster of galaxies - I. Results at z=0}",
      journal = {\mnras},
         year = 2001,
        month = dec,
       volume = {328},
       number = {3},
        pages = {726-750},
          doi = {10.1046/j.1365-8711.2001.04912.x},
archivePrefix = {arXiv},
       eprint = {astro-ph/0012055},
 primaryClass = {astro-ph},
       adsurl = {https://ui.adsabs.harvard.edu/abs/2001MNRAS.328..726S}
}

@ARTICLE{Springel_2005_LHaloTree,
       author = {{Springel}, Volker and {White}, Simon D.~M. and {Jenkins}, Adrian and {Frenk}, Carlos S. and {Yoshida}, Naoki and {Gao}, Liang and {Navarro}, Julio and {Thacker}, Robert and {Croton}, Darren and {Helly}, John and {Peacock}, John A. and {Cole}, Shaun and {Thomas}, Peter and {Couchman}, Hugh and {Evrard}, August and {Colberg}, J{\"o}rg and {Pearce}, Frazer},
        title = "{Simulations of the formation, evolution and clustering of galaxies and quasars}",
      journal = {\nat},
         year = 2005,
        month = jun,
       volume = {435},
       number = {7042},
        pages = {629-636},
          doi = {10.1038/nature03597},
archivePrefix = {arXiv},
       eprint = {astro-ph/0504097},
 primaryClass = {astro-ph},
       adsurl = {https://ui.adsabs.harvard.edu/abs/2005Natur.435..629S}
}

@ARTICLE{Springel_2021_Gadget4,
       author = {{Springel}, Volker and {Pakmor}, R{\"u}diger and {Zier}, Oliver and {Reinecke}, Martin},
        title = "{Simulating cosmic structure formation with the GADGET-4 code}",
      journal = {\mnras},
         year = 2021,
        month = sep,
       volume = {506},
       number = {2},
        pages = {2871-2949},
          doi = {10.1093/mnras/stab1855},
archivePrefix = {arXiv},
       eprint = {2010.03567},
 primaryClass = {astro-ph.IM},
       adsurl = {https://ui.adsabs.harvard.edu/abs/2021MNRAS.506.2871S}
}

@ARTICLE{Somerville_1999,
       author = {{Somerville}, Rachel S. and {Kolatt}, Tsafrir S.},
        title = "{How to plant a merger tree}",
      journal = {\mnras},
         year = 1999,
        month = may,
       volume = {305},
       number = {1},
        pages = {1-14},
          doi = {10.1046/j.1365-8711.1999.02154.x},
archivePrefix = {arXiv},
       eprint = {astro-ph/9711080},
 primaryClass = {astro-ph},
       adsurl = {https://ui.adsabs.harvard.edu/abs/1999MNRAS.305....1S}
}

@ARTICLE{Chandro_2025,
       author = {{Chandro-G{\'o}mez}, {\'A}ngel and {Lagos}, Claudia del P. and {Power}, Chris and {Forouhar Moreno}, Victor J. and {Helly}, John C. and {Lacey}, Cedric G. and {McGibbon}, Robert J. and {Schaller}, Matthieu and {Schaye}, Joop},
        title = "{On the accuracy of dark matter halo merger trees and the consequences for semi-analytic models of galaxy formation}",
      journal = {\mnras},
         year = 2025,
        month = may,
       volume = {539},
       number = {2},
        pages = {776-807},
          doi = {10.1093/mnras/staf519},
archivePrefix = {arXiv},
       eprint = {2501.07677},
 primaryClass = {astro-ph.GA},
       adsurl = {https://ui.adsabs.harvard.edu/abs/2025MNRAS.539..776C}
}

@ARTICLE{Gomez_2022,
       author = {{G{\'o}mez}, Jonathan S. and {Padilla}, N.~D. and {Helly}, J.~C. and {Lacey}, C.~G. and {Baugh}, C.~M. and {Lagos}, C.~D.~P.},
        title = "{Halo merger tree comparison: impact on galaxy formation models}",
      journal = {\mnras},
         year = 2022,
        month = mar,
       volume = {510},
       number = {4},
        pages = {5500-5519},
          doi = {10.1093/mnras/stab3661},
archivePrefix = {arXiv},
       eprint = {2106.12664},
 primaryClass = {astro-ph.GA},
       adsurl = {https://ui.adsabs.harvard.edu/abs/2022MNRAS.510.5500G}
}

@ARTICLE{Lee_2014,
       author = {{Lee}, Jaehyun and {Yi}, Sukyoung K. and {Elahi}, Pascal J. and {Thomas}, Peter A. and {Pearce}, Frazer R. and {Behroozi}, Peter and {Han}, Jiaxin and {Helly}, John and {Jung}, Intae and {Knebe}, Alexander and {Mao}, Yao-Yuan and {Onions}, Julian and {Rodriguez-Gomez}, Vicente and {Schneider}, Aurel and {Srisawat}, Chaichalit and {Tweed}, Dylan},
        title = "{Sussing merger trees: the impact of halo merger trees on galaxy properties in a semi-analytic model}",
      journal = {\mnras},
         year = 2014,
        month = dec,
       volume = {445},
       number = {4},
        pages = {4197-4210},
          doi = {10.1093/mnras/stu2039},
archivePrefix = {arXiv},
       eprint = {1410.1241},
 primaryClass = {astro-ph.GA},
       adsurl = {https://ui.adsabs.harvard.edu/abs/2014MNRAS.445.4197L}
}


%% file: cite/flares.bib
@ARTICLE{Lovell_2021,
       author = {{Lovell}, Christopher C. and {Vijayan}, Aswin P. and {Thomas}, Peter A. and {Wilkins}, Stephen M. and {Barnes}, David J. and {Irodotou}, Dimitrios and {Roper}, Will},
        title = "{First Light And Reionization Epoch Simulations (FLARES) - I. Environmental dependence of high-redshift galaxy evolution}",
      journal = {\mnras},
         year = 2021,
        month = jan,
       volume = {500},
       number = {2},
        pages = {2127-2145},
          doi = {10.1093/mnras/staa3360},
archivePrefix = {arXiv},
       eprint = {2004.07283},
 primaryClass = {astro-ph.GA},
       adsurl = {https://ui.adsabs.harvard.edu/abs/2021MNRAS.500.2127L}
}

@ARTICLE{Vijayan_2021,
       author = {{Vijayan}, Aswin P. and {Lovell}, Christopher C. and {Wilkins}, Stephen M. and {Thomas}, Peter A. and {Barnes}, David J. and {Irodotou}, Dimitrios and {Kuusisto}, Jussi and {Roper}, William J.},
        title = "{First Light And Reionization Epoch Simulations (FLARES) - II: The photometric properties of high-redshift galaxies}",
      journal = {\mnras},
         year = 2021,
        month = mar,
       volume = {501},
       number = {3},
        pages = {3289-3308},
          doi = {10.1093/mnras/staa3715},
archivePrefix = {arXiv},
       eprint = {2008.06057},
 primaryClass = {astro-ph.GA},
       adsurl = {https://ui.adsabs.harvard.edu/abs/2021MNRAS.501.3289V}
}

@ARTICLE{Lovell_2023,
       author = {{Lovell}, Christopher C. and {Roper}, Will and {Vijayan}, Aswin P. and {Seeyave}, Louise and {Irodotou}, Dimitrios and {Wilkins}, Stephen M. and {Conselice}, Christopher J. and {Fortuni}, Flaminia and {Kuusisto}, Jussi K. and {Merlin}, Emiliano and {Santini}, Paola and {Thomas}, Peter},
        title = "{First light and reionisation epoch simulations (FLARES) - VIII. The emergence of passive galaxies at z {\ensuremath{\geq}} 5}",
      journal = {\mnras},
         year = 2023,
        month = nov,
       volume = {525},
       number = {4},
        pages = {5520-5539},
          doi = {10.1093/mnras/stad2550},
archivePrefix = {arXiv},
       eprint = {2211.07540},
 primaryClass = {astro-ph.GA},
       adsurl = {https://ui.adsabs.harvard.edu/abs/2023MNRAS.525.5520L}
}

@ARTICLE{Thomas_2023,
       author = {{Thomas}, Peter A. and {Lovell}, Christopher C. and {Maltz}, Maxwell G.~A. and {Vijayan}, Aswin P. and {Wilkins}, Stephen M. and {Irodotou}, Dimitrios and {Roper}, William J. and {Seeyave}, Louise},
        title = "{First light and reionization epoch simulations (FLARES) X: environmental galaxy bias and survey variance at high redshift}",
      journal = {\mnras},
         year = 2023,
        month = sep,
       volume = {524},
       number = {1},
        pages = {43-59},
          doi = {10.1093/mnras/stad1819},
archivePrefix = {arXiv},
       eprint = {2301.09510},
 primaryClass = {astro-ph.GA},
       adsurl = {https://ui.adsabs.harvard.edu/abs/2023MNRAS.524...43T}
}

@ARTICLE{Barnes_2017_MACSIS,
       author = {{Barnes}, David J. and {Kay}, Scott T. and {Henson}, Monique A. and {McCarthy}, Ian G. and {Schaye}, Joop and {Jenkins}, Adrian},
        title = "{The redshift evolution of massive galaxy clusters in the MACSIS simulations}",
      journal = {\mnras},
         year = 2017,
        month = feb,
       volume = {465},
       number = {1},
        pages = {213-233},
          doi = {10.1093/mnras/stw2722},
archivePrefix = {arXiv},
       eprint = {1607.04569},
 primaryClass = {astro-ph.CO},
       adsurl = {https://ui.adsabs.harvard.edu/abs/2017MNRAS.465..213B}
}

@ARTICLE{Barnes_2017_CEagle,
       author = {{Barnes}, David J. and {Kay}, Scott T. and {Bah{\'e}}, Yannick M. and {Dalla Vecchia}, Claudio and {McCarthy}, Ian G. and {Schaye}, Joop and {Bower}, Richard G. and {Jenkins}, Adrian and {Thomas}, Peter A. and {Schaller}, Matthieu and {Crain}, Robert A. and {Theuns}, Tom and {White}, Simon D.~M.},
        title = "{The Cluster-EAGLE project: global properties of simulated clusters with resolved galaxies}",
      journal = {\mnras},
         year = 2017,
        month = oct,
       volume = {471},
       number = {1},
        pages = {1088-1106},
          doi = {10.1093/mnras/stx1647},
archivePrefix = {arXiv},
       eprint = {1703.10907},
 primaryClass = {astro-ph.GA},
       adsurl = {https://ui.adsabs.harvard.edu/abs/2017MNRAS.471.1088B}
}

@ARTICLE{Chiang_2013,
       author = {{Chiang}, Yi-Kuan and {Overzier}, Roderik and {Gebhardt}, Karl},
        title = "{Ancient Light from Young Cosmic Cities: Physical and Observational Signatures of Galaxy Proto-clusters}",
      journal = {\apj},
         year = 2013,
        month = dec,
       volume = {779},
       number = {2},
          eid = {127},
        pages = {127},
          doi = {10.1088/0004-637X/779/2/127},
archivePrefix = {arXiv},
       eprint = {1310.2938},
 primaryClass = {astro-ph.CO},
       adsurl = {https://ui.adsabs.harvard.edu/abs/2013ApJ...779..127C}
}

@ARTICLE{Crain_2015,
       author = {{Crain}, Robert A. and {Schaye}, Joop and {Bower}, Richard G. and {Furlong}, Michelle and {Schaller}, Matthieu and {Theuns}, Tom and {Dalla Vecchia}, Claudio and {Frenk}, Carlos S. and {McCarthy}, Ian G. and {Helly}, John C. and {Jenkins}, Adrian and {Rosas-Guevara}, Yetli M. and {White}, Simon D.~M. and {Trayford}, James W.},
        title = "{The EAGLE simulations of galaxy formation: calibration of subgrid physics and model variations}",
      journal = {\mnras},
         year = 2015,
        month = jun,
       volume = {450},
       number = {2},
        pages = {1937-1961},
          doi = {10.1093/mnras/stv725},
archivePrefix = {arXiv},
       eprint = {1501.01311},
 primaryClass = {astro-ph.GA},
       adsurl = {https://ui.adsabs.harvard.edu/abs/2015MNRAS.450.1937C}
}

@ARTICLE{Lovell_2018,
       author = {{Lovell}, Christopher C. and {Thomas}, Peter A. and {Wilkins}, Stephen M.},
        title = "{Characterising and identifying galaxy protoclusters}",
      journal = {\mnras},
         year = 2018,
        month = mar,
       volume = {474},
       number = {4},
        pages = {4612-4628},
          doi = {10.1093/mnras/stx3090},
archivePrefix = {arXiv},
       eprint = {1710.02148},
 primaryClass = {astro-ph.GA},
       adsurl = {https://ui.adsabs.harvard.edu/abs/2018MNRAS.474.4612L}
}

@ARTICLE{Planck_2014,
       author = {{Planck Collaboration} and {Ade}, P.~A.~R. and {Aghanim}, N. and {Alves}, M.~I.~R. and {Armitage-Caplan}, C. and {Arnaud}, M. and {Ashdown}, M. and {Atrio-Barandela}, F. and {Aumont}, J. and {Aussel}, H. and {Baccigalupi}, C. and {Banday}, A.~J. and {Barreiro}, R.~B. and {Barrena}, R. and {Bartelmann}, M. and {Bartlett}, J.~G. and {Bartolo}, N. and {Basak}, S. and {Battaner}, E. and {Battye}, R. and {Benabed}, K. and {Beno{\^\i}t}, A. and {Benoit-L{\'e}vy}, A. and {Bernard}, J. -P. and {Bersanelli}, M. and {Bertincourt}, B. and {Bethermin}, M. and {Bielewicz}, P. and {Bikmaev}, I. and {Blanchard}, A. and {Bobin}, J. and {Bock}, J.~J. and {B{\"o}hringer}, H. and {Bonaldi}, A. and {Bonavera}, L. and {Bond}, J.~R. and {Borrill}, J. and {Bouchet}, F.~R. and {Boulanger}, F. and {Bourdin}, H. and {Bowyer}, J.~W. and {Bridges}, M. and {Brown}, M.~L. and {Bucher}, M. and {Burenin}, R. and {Burigana}, C. and {Butler}, R.~C. and {Calabrese}, E. and {Cappellini}, B. and {Cardoso}, J. -F. and {Carr}, R. and {Carvalho}, P. and {Casale}, M. and {Castex}, G. and {Catalano}, A. and {Challinor}, A. and {Chamballu}, A. and {Chary}, R. -R. and {Chen}, X. and {Chiang}, H.~C. and {Chiang}, L. -Y. and {Chon}, G. and {Christensen}, P.~R. and {Churazov}, E. and {Church}, S. and {Clemens}, M. and {Clements}, D.~L. and {Colombi}, S. and {Colombo}, L.~P.~L. and {Combet}, C. and {Comis}, B. and {Couchot}, F. and {Coulais}, A. and {Crill}, B.~P. and {Cruz}, M. and {Curto}, A. and {Cuttaia}, F. and {Da Silva}, A. and {Dahle}, H. and {Danese}, L. and {Davies}, R.~D. and {Davis}, R.~J. and {de Bernardis}, P. and {de Rosa}, A. and {de Zotti}, G. and {D{\'e}chelette}, T. and {Delabrouille}, J. and {Delouis}, J. -M. and {D{\'e}mocl{\`e}s}, J. and {D{\'e}sert}, F. -X. and {Dick}, J. and {Dickinson}, C. and {Diego}, J.~M. and {Dolag}, K. and {Dole}, H. and {Donzelli}, S. and {Dor{\'e}}, O. and {Douspis}, M. and {Ducout}, A. and {Dunkley}, J. and {Dupac}, X. and {Efstathiou}, G. and {Elsner}, F. and {En{\ss}lin}, T.~A. and {Eriksen}, H.~K. and {Fabre}, O. and {Falgarone}, E. and {Falvella}, M.~C. and {Fantaye}, Y. and {Fergusson}, J. and {Filliard}, C. and {Finelli}, F. and {Flores-Cacho}, I. and {Foley}, S. and {Forni}, O. and {Fosalba}, P. and {Frailis}, M. and {Fraisse}, A.~A. and {Franceschi}, E. and {Freschi}, M. and {Fromenteau}, S. and {Frommert}, M. and {Gaier}, T.~C. and {Galeotta}, S. and {Gallegos}, J. and {Galli}, S. and {Gandolfo}, B. and {Ganga}, K. and {Gauthier}, C. and {G{\'e}nova-Santos}, R.~T. and {Ghosh}, T. and {Giard}, M. and {Giardino}, G. and {Gilfanov}, M. and {Girard}, D. and {Giraud-H{\'e}raud}, Y. and {Gjerl{\o}w}, E. and {Gonz{\'a}lez-Nuevo}, J. and {G{\'o}rski}, K.~M. and {Gratton}, S. and {Gregorio}, A. and {Gruppuso}, A. and {Gudmundsson}, J.~E. and {Haissinski}, J. and {Hamann}, J. and {Hansen}, F.~K. and {Hansen}, M. and {Hanson}, D. and {Harrison}, D.~L. and {Heavens}, A. and {Helou}, G. and {Hempel}, A. and {Henrot-Versill{\'e}}, S. and {Hern{\'a}ndez-Monteagudo}, C. and {Herranz}, D. and {Hildebrandt}, S.~R. and {Hivon}, E. and {Ho}, S. and {Hobson}, M. and {Holmes}, W.~A. and {Hornstrup}, A. and {Hou}, Z. and {Hovest}, W. and {Huey}, G. and {Huffenberger}, K.~M. and {Hurier}, G. and {Ili{\'c}}, S. and {Jaffe}, A.~H. and {Jaffe}, T.~R. and {Jasche}, J. and {Jewell}, J. and {Jones}, W.~C. and {Juvela}, M. and {Kalberla}, P. and {Kangaslahti}, P. and {Keih{\"a}nen}, E. and {Kerp}, J. and {Keskitalo}, R. and {Khamitov}, I. and {Kiiveri}, K. and {Kim}, J. and {Kisner}, T.~S. and {Kneissl}, R. and {Knoche}, J. and {Knox}, L. and {Kunz}, M. and {Kurki-Suonio}, H. and {Lacasa}, F. and {Lagache}, G. and {L{\"a}hteenm{\"a}ki}, A. and {Lamarre}, J. -M. and {Langer}, M. and {Lasenby}, A. and {Lattanzi}, M. and {Laureijs}, R.~J. and {Lavabre}, A. and {Lawrence}, C.~R. and {Le Jeune}, M. and {Leach}, S. and {Leahy}, J.~P.},
        title = "{Planck 2013 results. I. Overview of products and scientific results}",
      journal = {\aap},
         year = 2014,
        month = nov,
       volume = {571},
          eid = {A1},
        pages = {A1},
          doi = {10.1051/0004-6361/201321529},
archivePrefix = {arXiv},
       eprint = {1303.5062},
 primaryClass = {astro-ph.CO},
       adsurl = {https://ui.adsabs.harvard.edu/abs/2014A&A...571A...1P}
}

@ARTICLE{Schaye_2015,
       author = {{Schaye}, Joop and {Crain}, Robert A. and {Bower}, Richard G. and {Furlong}, Michelle and {Schaller}, Matthieu and {Theuns}, Tom and {Dalla Vecchia}, Claudio and {Frenk}, Carlos S. and {McCarthy}, I.~G. and {Helly}, John C. and {Jenkins}, Adrian and {Rosas-Guevara}, Y.~M. and {White}, Simon D.~M. and {Baes}, Maarten and {Booth}, C.~M. and {Camps}, Peter and {Navarro}, Julio F. and {Qu}, Yan and {Rahmati}, Alireza and {Sawala}, Till and {Thomas}, Peter A. and {Trayford}, James},
        title = "{The EAGLE project: simulating the evolution and assembly of galaxies and their environments}",
      journal = {\mnras},
         year = 2015,
        month = jan,
       volume = {446},
       number = {1},
        pages = {521-554},
          doi = {10.1093/mnras/stu2058},
archivePrefix = {arXiv},
       eprint = {1407.7040},
 primaryClass = {astro-ph.GA},
       adsurl = {https://ui.adsabs.harvard.edu/abs/2015MNRAS.446..521S}
}

@Article{Hunter:2007,
  Author    = {Hunter, J. D.},
  Title     = {Matplotlib: A 2D graphics environment},
  Journal   = {Computing in Science \& Engineering},
  Volume    = {9},
  Number    = {3},
  Pages     = {90--95},
  publisher = {IEEE COMPUTER SOC},
  doi       = {10.1109/MCSE.2007.55},
  year      = 2007
}

@Article{harris2020array,
 title         = {Array programming with {NumPy}},
 author        = {Charles R. Harris and K. Jarrod Millman and St{\'{e}}fan J.
                 van der Walt and Ralf Gommers and Pauli Virtanen and David
                 Cournapeau and Eric Wieser and Julian Taylor and Sebastian
                 Berg and Nathaniel J. Smith and Robert Kern and Matti Picus
                 and Stephan Hoyer and Marten H. van Kerkwijk and Matthew
                 Brett and Allan Haldane and Jaime Fern{\'{a}}ndez del
                 R{\'{i}}o and Mark Wiebe and Pearu Peterson and Pierre
                 G{\'{e}}rard-Marchant and Kevin Sheppard and Tyler Reddy and
                 Warren Weckesser and Hameer Abbasi and Christoph Gohlke and
                 Travis E. Oliphant},
 year          = {2020},
 month         = sep,
 journal       = {Nature},
 volume        = {585},
 number        = {7825},
 pages         = {357--362},
 doi           = {10.1038/s41586-020-2649-2},
 publisher     = {Springer Science and Business Media {LLC}},
 url           = {https://doi.org/10.1038/s41586-020-2649-2}
}


%% file: cite/general.bib
@ARTICLE{Adams_2024,
       author = {{Adams}, Nathan J. and {Conselice}, Christopher J. and {Austin}, Duncan and {Harvey}, Thomas and {Ferreira}, Leonardo and {Trussler}, James and {Juod{\v{z}}balis}, Ignas and {Li}, Qiong and {Windhorst}, Rogier and {Cohen}, Seth H. and {Jansen}, Rolf A. and {Summers}, Jake and {Tompkins}, Scott and {Driver}, Simon P. and {Robotham}, Aaron and {D'Silva}, Jordan C.~J. and {Yan}, Haojing and {Coe}, Dan and {Frye}, Brenda and {Grogin}, Norman A. and {Koekemoer}, Anton M. and {Marshall}, Madeline A. and {Pirzkal}, Nor and {Ryan}, Russell E. and {Maksym}, W. Peter and {Rutkowski}, Michael J. and {Willmer}, Christopher N.~A. and {Hammel}, Heidi B. and {Nonino}, Mario and {Bhatawdekar}, Rachana and {Wilkins}, Stephen M. and {Bradley}, Larry D. and {Broadhurst}, Tom and {Cheng}, Cheng and {Dole}, Herv{\'e} and {Hathi}, Nimish P. and {Zitrin}, Adi},
        title = "{EPOCHS. II. The Ultraviolet Luminosity Function from 7.5 < z < 13.5 Using 180 arcmin$^{2}$ of Deep, Blank Fields from the PEARLS Survey and Public JWST Data}",
      journal = {\apj},
         year = 2024,
        month = apr,
       volume = {965},
       number = {2},
          eid = {169},
        pages = {169},
          doi = {10.3847/1538-4357/ad2a7b},
archivePrefix = {arXiv},
       eprint = {2304.13721},
 primaryClass = {astro-ph.GA},
       adsurl = {https://ui.adsabs.harvard.edu/abs/2024ApJ...965..169A}
}

@ARTICLE{Alberts_2024,
       author = {{Alberts}, Stacey and {Williams}, Christina C. and {Helton}, Jakob M. and {Suess}, Katherine A. and {Ji}, Zhiyuan and {Shivaei}, Irene and {Lyu}, Jianwei and {Rieke}, George and {Baker}, William M. and {Bonaventura}, Nina and {Bunker}, Andrew J. and {Carniani}, Stefano and {Charlot}, Stephane and {Curtis-Lake}, Emma and {D'Eugenio}, Francesco and {Eisenstein}, Daniel J. and {de Graaff}, Anna and {Hainline}, Kevin N. and {Hausen}, Ryan and {Johnson}, Benjamin D. and {Maiolino}, Roberto and {Parlanti}, Eleonora and {Rieke}, Marcia J. and {Robertson}, Brant E. and {Sun}, Yang and {Tacchella}, Sandro and {Willmer}, Christopher N.~A. and {Willott}, Chris J.},
        title = "{To High Redshift and Low Mass: Exploring the Emergence of Quenched Galaxies and Their Environments at 3 < z < 6 in the Ultra-deep JADES MIRI F770W Parallel}",
      journal = {\apj},
         year = 2024,
        month = nov,
       volume = {975},
       number = {1},
          eid = {85},
        pages = {85},
          doi = {10.3847/1538-4357/ad66cc},
archivePrefix = {arXiv},
       eprint = {2312.12207},
 primaryClass = {astro-ph.GA},
       adsurl = {https://ui.adsabs.harvard.edu/abs/2024ApJ...975...85A}
}

@ARTICLE{Antwidanso_2023,
       author = {{Antwi-Danso}, Jacqueline and {Papovich}, Casey and {Leja}, Joel and {Marchesini}, Danilo and {Marsan}, Z. Cemile and {Martis}, Nicholas S. and {Labb{\'e}}, Ivo and {Muzzin}, Adam and {Glazebrook}, Karl and {Straatman}, Caroline M.~S. and {Tran}, Kim-Vy H.},
        title = "{Beyond UVJ: Color Selection of Galaxies in the JWST Era}",
      journal = {\apj},
         year = 2023,
        month = feb,
       volume = {943},
       number = {2},
          eid = {166},
        pages = {166},
          doi = {10.3847/1538-4357/aca294},
archivePrefix = {arXiv},
       eprint = {2207.07170},
 primaryClass = {astro-ph.GA},
       adsurl = {https://ui.adsabs.harvard.edu/abs/2023ApJ...943..166A}
}

@ARTICLE{Asquith_2018,
       author = {{Asquith}, Rachel and {Pearce}, Frazer R. and {Almaini}, Omar and {Knebe}, Alexander and {Gonzalez-Perez}, Violeta and {Benson}, Andrew and {Blaizot}, Jeremy and {Carretero}, Jorge and {Castander}, Francisco J. and {Cattaneo}, Andrea and {Cora}, Sof{\'\i}a A. and {Croton}, Darren J. and {Devriendt}, Julien E. and {Fontanot}, Fabio and {Gargiulo}, Ignacio D. and {Hartley}, Will and {Henriques}, Bruno and {Lee}, Jaehyun and {Mamon}, Gary A. and {Onions}, Julian and {Padilla}, Nelson D. and {Power}, Chris and {Srisawat}, Chaichalit and {Stevens}, Adam R.~H. and {Thomas}, Peter A. and {Vega-Mart{\'\i}nez}, Cristian A. and {Yi}, Sukyoung K.},
        title = "{Cosmic CARNage II: the evolution of the galaxy stellar mass function in observations and galaxy formation models}",
      journal = {\mnras},
         year = 2018,
        month = oct,
       volume = {480},
       number = {1},
        pages = {1197-1210},
          doi = {10.1093/mnras/sty1870},
archivePrefix = {arXiv},
       eprint = {1807.03796},
 primaryClass = {astro-ph.GA},
       adsurl = {https://ui.adsabs.harvard.edu/abs/2018MNRAS.480.1197A}
}

@ARTICLE{Ayromlou_2021,
       author = {{Ayromlou}, Mohammadreza and {Nelson}, Dylan and {Yates}, Robert M. and {Kauffmann}, Guinevere and {Renneby}, Malin and {White}, Simon D.~M.},
        title = "{Comparing galaxy formation in the L-GALAXIES semi-analytical model and the IllustrisTNG simulations}",
      journal = {\mnras},
         year = 2021,
        month = mar,
       volume = {502},
       number = {1},
        pages = {1051-1069},
          doi = {10.1093/mnras/staa4011},
archivePrefix = {arXiv},
       eprint = {2004.14390},
 primaryClass = {astro-ph.GA},
       adsurl = {https://ui.adsabs.harvard.edu/abs/2021MNRAS.502.1051A}
}

@ARTICLE{Baker_2025,
       author = {{Baker}, William M. and {Valentino}, Francesco and {Lagos}, Claudia del P. and {Ito}, Kei and {Jespersen}, Christian Kragh and {Gottumukkala}, Rashmi and {Hjorth}, Jens and {Langeroodi}, Danial and {Sedgewick}, Aidan},
        title = "{Exploring over 700 massive quiescent galaxies at z = 2-7: Demographics and stellar mass functions}",
      journal = {arXiv e-prints},
         year = 2025,
        month = jun,
          eid = {arXiv:2506.04119},
        pages = {arXiv:2506.04119},
          doi = {10.48550/arXiv.2506.04119},
archivePrefix = {arXiv},
       eprint = {2506.04119},
 primaryClass = {astro-ph.GA},
       adsurl = {https://ui.adsabs.harvard.edu/abs/2025arXiv250604119B}
}

@ARTICLE{Benson_2010,
       author = {{Benson}, Andrew J.},
        title = "{Galaxy formation theory}",
      journal = {\physrep},
         year = 2010,
        month = oct,
       volume = {495},
       number = {2-3},
        pages = {33-86},
          doi = {10.1016/j.physrep.2010.06.001},
archivePrefix = {arXiv},
       eprint = {1006.5394},
 primaryClass = {astro-ph.CO},
       adsurl = {https://ui.adsabs.harvard.edu/abs/2010PhR...495...33B}
}

@BOOK{Binney_1987,
       author = {{Binney}, James and {Tremaine}, Scott},
        title = "{Galactic dynamics}",
         year = 1987,
       adsurl = {https://ui.adsabs.harvard.edu/abs/1987gady.book.....B}
}

@ARTICLE{Blitz_2006,
       author = {{Blitz}, Leo and {Rosolowsky}, Erik},
        title = "{The Role of Pressure in GMC Formation II: The H$_{2}$-Pressure Relation}",
      journal = {\apj},
         year = 2006,
        month = oct,
       volume = {650},
       number = {2},
        pages = {933-944},
          doi = {10.1086/505417},
archivePrefix = {arXiv},
       eprint = {astro-ph/0605035},
 primaryClass = {astro-ph},
       adsurl = {https://ui.adsabs.harvard.edu/abs/2006ApJ...650..933B}
}

@ARTICLE{Bondi_1944,
       author = {{Bondi}, H. and {Hoyle}, F.},
        title = "{On the mechanism of accretion by stars}",
      journal = {\mnras},
         year = 1944,
        month = jan,
       volume = {104},
        pages = {273},
          doi = {10.1093/mnras/104.5.273},
       adsurl = {https://ui.adsabs.harvard.edu/abs/1944MNRAS.104..273B}
}

@ARTICLE{BK_2008,
       author = {{Boylan-Kolchin}, Michael and {Ma}, Chung-Pei and {Quataert}, Eliot},
        title = "{Dynamical friction and galaxy merging time-scales}",
      journal = {\mnras},
         year = 2008,
        month = jan,
       volume = {383},
       number = {1},
        pages = {93-101},
          doi = {10.1111/j.1365-2966.2007.12530.x},
archivePrefix = {arXiv},
       eprint = {0707.2960},
 primaryClass = {astro-ph},
       adsurl = {https://ui.adsabs.harvard.edu/abs/2008MNRAS.383...93B}
}

@ARTICLE{Bryan_Norman_1998,
       author = {{Bryan}, Greg L. and {Norman}, Michael L.},
        title = "{Statistical Properties of X-Ray Clusters: Analytic and Numerical Comparisons}",
      journal = {\apj},
         year = 1998,
        month = mar,
       volume = {495},
       number = {1},
        pages = {80-99},
          doi = {10.1086/305262},
archivePrefix = {arXiv},
       eprint = {astro-ph/9710107},
 primaryClass = {astro-ph},
       adsurl = {https://ui.adsabs.harvard.edu/abs/1998ApJ...495...80B}
}

@ARTICLE{Carnall_2018,
       author = {{Carnall}, A.~C. and {McLure}, R.~J. and {Dunlop}, J.~S. and {Dav{\'e}}, R.},
        title = "{Inferring the star formation histories of massive quiescent galaxies with BAGPIPES: evidence for multiple quenching mechanisms}",
      journal = {\mnras},
         year = 2018,
        month = nov,
       volume = {480},
       number = {4},
        pages = {4379-4401},
          doi = {10.1093/mnras/sty2169},
archivePrefix = {arXiv},
       eprint = {1712.04452},
 primaryClass = {astro-ph.GA},
       adsurl = {https://ui.adsabs.harvard.edu/abs/2018MNRAS.480.4379C}
}

@ARTICLE{Carnall_2023,
       author = {{Carnall}, A.~C. and {McLeod}, D.~J. and {McLure}, R.~J. and {Dunlop}, J.~S. and {Begley}, R. and {Cullen}, F. and {Donnan}, C.~T. and {Hamadouche}, M.~L. and {Jewell}, S.~M. and {Jones}, E.~W. and {Pollock}, C.~L. and {Wild}, V.},
        title = "{A surprising abundance of massive quiescent galaxies at 3 < z < 5 in the first data from JWST CEERS}",
      journal = {\mnras},
         year = 2023,
        month = apr,
       volume = {520},
       number = {3},
        pages = {3974-3985},
          doi = {10.1093/mnras/stad369},
archivePrefix = {arXiv},
       eprint = {2208.00986},
 primaryClass = {astro-ph.GA},
       adsurl = {https://ui.adsabs.harvard.edu/abs/2023MNRAS.520.3974C}
}

@ARTICLE{Castellano_2023,
       author = {{Castellano}, Marco and {Fontana}, Adriano and {Treu}, Tommaso and {Merlin}, Emiliano and {Santini}, Paola and {Bergamini}, Pietro and {Grillo}, Claudio and {Rosati}, Piero and {Acebron}, Ana and {Leethochawalit}, Nicha and {Paris}, Diego and {Bonchi}, Andrea and {Belfiori}, Davide and {Calabr{\`o}}, Antonello and {Correnti}, Matteo and {Nonino}, Mario and {Polenta}, Gianluca and {Trenti}, Michele and {Boyett}, Kristan and {Brammer}, G. and {Broadhurst}, Tom and {Caminha}, Gabriel B. and {Chen}, Wenlei and {Filippenko}, Alexei V. and {Fortuni}, Flaminia and {Glazebrook}, Karl and {Mascia}, Sara and {Mason}, Charlotte A. and {Menci}, Nicola and {Meneghetti}, Massimo and {Mercurio}, Amata and {Metha}, Benjamin and {Morishita}, Takahiro and {Nanayakkara}, Themiya and {Pentericci}, Laura and {Roberts-Borsani}, Guido and {Roy}, Namrata and {Vanzella}, Eros and {Vulcani}, Benedetta and {Yang}, Lilan and {Wang}, Xin},
        title = "{Early Results from GLASS-JWST. XIX. A High Density of Bright Galaxies at z {\ensuremath{\approx}} 10 in the A2744 Region}",
      journal = {\apjl},
         year = 2023,
        month = may,
       volume = {948},
       number = {2},
          eid = {L14},
        pages = {L14},
          doi = {10.3847/2041-8213/accea5},
archivePrefix = {arXiv},
       eprint = {2212.06666},
 primaryClass = {astro-ph.GA},
       adsurl = {https://ui.adsabs.harvard.edu/abs/2023ApJ...948L..14C}
}

@ARTICLE{Chabrier_2003,
       author = {{Chabrier}, Gilles},
        title = "{Galactic Stellar and Substellar Initial Mass Function}",
      journal = {\pasp},
         year = 2003,
        month = jul,
       volume = {115},
       number = {809},
        pages = {763-795},
          doi = {10.1086/376392},
archivePrefix = {arXiv},
       eprint = {astro-ph/0304382},
 primaryClass = {astro-ph},
       adsurl = {https://ui.adsabs.harvard.edu/abs/2003PASP..115..763C}
}

@ARTICLE{Chaikin_2025,
       author = {{Chaikin}, Evgenii and {Schaye}, Joop and {Schaller}, Matthieu and {Ploeckinger}, Sylvia and {Ben{\'\i}tez-Llambay}, Alejandro and {Frenk}, Carlos S. and {Hu{\v{s}}ko}, Filip and {McGibbon}, Robert and {Richings}, Alexander J. and {Trayford}, James W.},
        title = "{The evolution of the galaxy stellar mass function and star formation rates in the COLIBRE simulations from redshift 17 to 0}",
      journal = {arXiv e-prints},
         year = 2025,
        month = sep,
          eid = {arXiv:2509.07960},
        pages = {arXiv:2509.07960},
          doi = {10.48550/arXiv.2509.07960},
archivePrefix = {arXiv},
       eprint = {2509.07960},
 primaryClass = {astro-ph.GA},
       adsurl = {https://ui.adsabs.harvard.edu/abs/2025arXiv250907960C}
}

@ARTICLE{Clarke_2024,
       author = {{Clarke}, Leonardo and {Shapley}, Alice E. and {Sanders}, Ryan L. and {Topping}, Michael W. and {Brammer}, Gabriel B. and {Bento}, Trinity and {Reddy}, Naveen A. and {Kehoe}, Emily},
        title = "{The Star-forming Main Sequence in JADES and CEERS at z > 1.4: Investigating the Burstiness of Star Formation}",
      journal = {\apj},
         year = 2024,
        month = dec,
       volume = {977},
       number = {1},
          eid = {133},
        pages = {133},
          doi = {10.3847/1538-4357/ad8ba4},
archivePrefix = {arXiv},
       eprint = {2406.05178},
 primaryClass = {astro-ph.GA},
       adsurl = {https://ui.adsabs.harvard.edu/abs/2024ApJ...977..133C}
}

@ARTICLE{Correa_2018,
       author = {{Correa}, Camila A. and {Schaye}, Joop and {van de Voort}, Freeke and {Duffy}, Alan R. and {Wyithe}, J. Stuart B.},
        title = "{The impact of feedback and the hot halo on the rates of gas accretion on to galaxies}",
      journal = {\mnras},
         year = 2018,
        month = jul,
       volume = {478},
       number = {1},
        pages = {255-269},
          doi = {10.1093/mnras/sty871},
archivePrefix = {arXiv},
       eprint = {1804.01537},
 primaryClass = {astro-ph.GA},
       adsurl = {https://ui.adsabs.harvard.edu/abs/2018MNRAS.478..255C}
}

@ARTICLE{Covelo_2025,
       author = {{Covelo-Paz}, Alba and {Giovinazzo}, Emma and {Oesch}, Pascal A. and {Meyer}, Romain A. and {Weibel}, Andrea and {Brammer}, Gabriel and {Fudamoto}, Yoshinobu and {Kerutt}, Josephine and {Lin}, Jamie and {Matharu}, Jasleen and {Naidu}, Rohan P. and {Velichko}, Anna and {Bollo}, Victoria and {Bouwens}, Rychard and {Chisholm}, John and {Illingworth}, Garth D. and {Kramarenko}, Ivan and {Magee}, Daniel and {Maseda}, Michael and {Matthee}, Jorryt and {Nelson}, Erica and {Reddy}, Naveen and {Schaerer}, Daniel and {Stefanon}, Mauro and {Xiao}, Mengyuan},
        title = "{An H{\ensuremath{\alpha}} view of galaxy buildup in the first 2 Gyr: Luminosity functions at z {\ensuremath{\sim}} 4{\ensuremath{-}}6.5 from NIRCam/grism spectroscopy}",
      journal = {\aap},
         year = 2025,
        month = feb,
       volume = {694},
          eid = {A178},
        pages = {A178},
          doi = {10.1051/0004-6361/202452363},
archivePrefix = {arXiv},
       eprint = {2409.17241},
 primaryClass = {astro-ph.GA},
       adsurl = {https://ui.adsabs.harvard.edu/abs/2025A&A...694A.178C}
}

@ARTICLE{Croton_2006,
       author = {{Croton}, Darren J. and {Springel}, Volker and {White}, Simon D.~M. and {De Lucia}, G. and {Frenk}, C.~S. and {Gao}, L. and {Jenkins}, A. and {Kauffmann}, G. and {Navarro}, J.~F. and {Yoshida}, N.},
        title = "{The many lives of active galactic nuclei: cooling flows, black holes and the luminosities and colours of galaxies}",
      journal = {\mnras},
         year = 2006,
        month = jan,
       volume = {365},
       number = {1},
        pages = {11-28},
          doi = {10.1111/j.1365-2966.2005.09675.x},
archivePrefix = {arXiv},
       eprint = {astro-ph/0508046},
 primaryClass = {astro-ph},
       adsurl = {https://ui.adsabs.harvard.edu/abs/2006MNRAS.365...11C}
}

@ARTICLE{Cui_2018,
       author = {{Cui}, Weiguang and {Knebe}, Alexander and {Yepes}, Gustavo and {Pearce}, Frazer and {Power}, Chris and {Dave}, Romeel and {Arth}, Alexander and {Borgani}, Stefano and {Dolag}, Klaus and {Elahi}, Pascal and {Mostoghiu}, Robert and {Murante}, Giuseppe and {Rasia}, Elena and {Stoppacher}, Doris and {Vega-Ferrero}, Jesus and {Wang}, Yang and {Yang}, Xiaohu and {Benson}, Andrew and {Cora}, Sof{\'\i}a A. and {Croton}, Darren J. and {Sinha}, Manodeep and {Stevens}, Adam R.~H. and {Vega-Mart{\'\i}nez}, Cristian A. and {Arthur}, Jake and {Baldi}, Anna S. and {Ca{\~n}as}, Rodrigo and {Cialone}, Giammarco and {Cunnama}, Daniel and {De Petris}, Marco and {Durando}, Giacomo and {Ettori}, Stefano and {Gottl{\"o}ber}, Stefan and {Nuza}, Sebasti{\'a}n E. and {Old}, Lyndsay J. and {Pilipenko}, Sergey and {Sorce}, Jenny G. and {Welker}, Charlotte},
        title = "{The Three Hundred project: a large catalogue of theoretically modelled galaxy clusters for cosmological and astrophysical applications}",
      journal = {\mnras},
         year = 2018,
        month = nov,
       volume = {480},
       number = {3},
        pages = {2898-2915},
          doi = {10.1093/mnras/sty2111},
archivePrefix = {arXiv},
       eprint = {1809.04622},
 primaryClass = {astro-ph.GA},
       adsurl = {https://ui.adsabs.harvard.edu/abs/2018MNRAS.480.2898C}
}

@ARTICLE{Davis_1985,
       author = {{Davis}, M. and {Efstathiou}, G. and {Frenk}, C.~S. and {White}, S.~D.~M.},
        title = "{The evolution of large-scale structure in a universe dominated by cold dark matter}",
      journal = {\apj},
         year = 1985,
        month = may,
       volume = {292},
        pages = {371-394},
          doi = {10.1086/163168},
       adsurl = {https://ui.adsabs.harvard.edu/abs/1985ApJ...292..371D}
}

@ARTICLE{DeLucia_2010,
       author = {{De Lucia}, Gabriella and {Boylan-Kolchin}, Michael and {Benson}, Andrew J. and {Fontanot}, Fabio and {Monaco}, Pierluigi},
        title = "{A semi-analytic model comparison - gas cooling and galaxy mergers}",
      journal = {\mnras},
         year = 2010,
        month = aug,
       volume = {406},
       number = {3},
        pages = {1533-1552},
          doi = {10.1111/j.1365-2966.2010.16806.x},
archivePrefix = {arXiv},
       eprint = {1003.3021},
 primaryClass = {astro-ph.CO},
       adsurl = {https://ui.adsabs.harvard.edu/abs/2010MNRAS.406.1533D}
}

@ARTICLE{Desprez_2024,
       author = {{Desprez}, Guillaume and {Martis}, Nicholas S. and {Asada}, Yoshihisa and {Sawicki}, Marcin and {Willott}, Chris J. and {Muzzin}, Adam and {Abraham}, Roberto G. and {Brada{\v{c}}}, Maru{\v{s}}a and {Brammer}, Gabe and {Estrada-Carpenter}, Vicente and {Iyer}, Kartheik G. and {Matharu}, Jasleen and {Mowla}, Lamiya and {Noirot}, Ga{\"e}l and {Sarrouh}, Ghassan T.~E. and {Strait}, Victoria and {Gledhill}, Rachel and {Rihtar{\v{s}}i{\v{c}}}, Gregor},
        title = "{{\ensuremath{\Lambda}}CDM not dead yet: massive high-z Balmer break galaxies are less common than previously reported}",
      journal = {\mnras},
         year = 2024,
        month = may,
       volume = {530},
       number = {3},
        pages = {2935-2952},
          doi = {10.1093/mnras/stae1084},
archivePrefix = {arXiv},
       eprint = {2310.03063},
 primaryClass = {astro-ph.GA},
       adsurl = {https://ui.adsabs.harvard.edu/abs/2024MNRAS.530.2935D}
}

@ARTICLE{DiMatteo_2005,
       author = {{Di Matteo}, Tiziana and {Springel}, Volker and {Hernquist}, Lars},
        title = "{Energy input from quasars regulates the growth and activity of black holes and their host galaxies}",
      journal = {\nat},
         year = 2005,
        month = feb,
       volume = {433},
       number = {7026},
        pages = {604-607},
          doi = {10.1038/nature03335},
archivePrefix = {arXiv},
       eprint = {astro-ph/0502199},
 primaryClass = {astro-ph},
       adsurl = {https://ui.adsabs.harvard.edu/abs/2005Natur.433..604D}
}

@ARTICLE{DSilva_2025,
       author = {{D'Silva}, Jordan C.~J. and {Driver}, Simon P. and {Lagos}, Claudia D.~P. and {Robotham}, Aaron S.~G. and {Adams}, Nathan J. and {Conselice}, Christopher J. and {Frye}, Brenda and {Hathi}, Nimish P. and {Harvey}, Thomas and {Ortiz}, III, Rafael and {Ricotti}, Massimo and {Robertson}, Clayton and {Silver}, Ross M. and {Wilkins}, Stephen M. and {Willmer}, Christopher N.~A. and {Windhorst}, Rogier A. and {Cohen}, Seth H. and {Jansen}, Rolf A. and {Summers}, Jake and {Koekemoer}, Anton M. and {Coe}, Dan and {Grogin}, Norman A. and {Marshall}, Madeline A. and {Nonino}, Mario and {Pirzkal}, Nor and {Ryan}, Jr., Russell E. and {Yan}, Haojing},
        title = "{Self-Consistent JWST Census of Star Formation and AGN activity at z=5.5-13.5}",
      journal = {arXiv e-prints},
         year = 2025,
        month = mar,
          eid = {arXiv:2503.03431},
        pages = {arXiv:2503.03431},
          doi = {10.48550/arXiv.2503.03431},
archivePrefix = {arXiv},
       eprint = {2503.03431},
 primaryClass = {astro-ph.GA},
       adsurl = {https://ui.adsabs.harvard.edu/abs/2025arXiv250303431D}
}

@ARTICLE{Elmegreen_1993,
       author = {{Elmegreen}, B.~G.},
        title = "{The H to H 2 Transition in Galaxies: Totally Molecular Galaxies}",
      journal = {\apj},
         year = 1993,
        month = jul,
       volume = {411},
        pages = {170},
          doi = {10.1086/172816},
       adsurl = {https://ui.adsabs.harvard.edu/abs/1993ApJ...411..170E}
}

@ARTICLE{Errani_2015,
       author = {{Errani}, R. and {Penarrubia}, J. and {Tormen}, G.},
        title = "{Constraining the distribution of dark matter in dwarf spheroidal galaxies with stellar tidal streams.}",
      journal = {\mnras},
         year = 2015,
        month = apr,
       volume = {449},
        pages = {L46-L50},
          doi = {10.1093/mnrasl/slv012},
archivePrefix = {arXiv},
       eprint = {1501.04968},
 primaryClass = {astro-ph.GA},
       adsurl = {https://ui.adsabs.harvard.edu/abs/2015MNRAS.449L..46E}
}

@ARTICLE{Feldmann_2015,
       author = {{Feldmann}, Robert and {Mayer}, Lucio},
        title = "{The Argo simulation - I. Quenching of massive galaxies at high redshift as a result of cosmological starvation}",
      journal = {\mnras},
         year = 2015,
        month = jan,
       volume = {446},
       number = {2},
        pages = {1939-1956},
          doi = {10.1093/mnras/stu2207},
archivePrefix = {arXiv},
       eprint = {1404.3212},
 primaryClass = {astro-ph.GA},
       adsurl = {https://ui.adsabs.harvard.edu/abs/2015MNRAS.446.1939F}
}

@ARTICLE{Ferland_1998,
       author = {{Ferland}, G.~J. and {Korista}, K.~T. and {Verner}, D.~A. and {Ferguson}, J.~W. and {Kingdon}, J.~B. and {Verner}, E.~M.},
        title = "{CLOUDY 90: Numerical Simulation of Plasmas and Their Spectra}",
      journal = {\pasp},
         year = 1998,
        month = jul,
       volume = {110},
       number = {749},
        pages = {761-778},
          doi = {10.1086/316190},
       adsurl = {https://ui.adsabs.harvard.edu/abs/1998PASP..110..761F}
}

@ARTICLE{Finkelstein_2023,
       author = {{Finkelstein}, Steven L. and {Bagley}, Micaela B. and {Ferguson}, Henry C. and {Wilkins}, Stephen M. and {Kartaltepe}, Jeyhan S. and {Papovich}, Casey and {Yung}, L.~Y. Aaron and {Arrabal Haro}, Pablo and {Behroozi}, Peter and {Dickinson}, Mark and {Kocevski}, Dale D. and {Koekemoer}, Anton M. and {Larson}, Rebecca L. and {Le Bail}, Aur{\'e}lien and {Morales}, Alexa M. and {P{\'e}rez-Gonz{\'a}lez}, Pablo G. and {Burgarella}, Denis and {Dav{\'e}}, Romeel and {Hirschmann}, Michaela and {Somerville}, Rachel S. and {Wuyts}, Stijn and {Bromm}, Volker and {Casey}, Caitlin M. and {Fontana}, Adriano and {Fujimoto}, Seiji and {Gardner}, Jonathan P. and {Giavalisco}, Mauro and {Grazian}, Andrea and {Grogin}, Norman A. and {Hathi}, Nimish P. and {Hutchison}, Taylor A. and {Jha}, Saurabh W. and {Jogee}, Shardha and {Kewley}, Lisa J. and {Kirkpatrick}, Allison and {Long}, Arianna S. and {Lotz}, Jennifer M. and {Pentericci}, Laura and {Pierel}, Justin D.~R. and {Pirzkal}, Nor and {Ravindranath}, Swara and {Ryan}, Russell E. and {Trump}, Jonathan R. and {Yang}, Guang and {Bhatawdekar}, Rachana and {Bisigello}, Laura and {Buat}, V{\'e}ronique and {Calabr{\`o}}, Antonello and {Castellano}, Marco and {Cleri}, Nikko J. and {Cooper}, M.~C. and {Croton}, Darren and {Daddi}, Emanuele and {Dekel}, Avishai and {Elbaz}, David and {Franco}, Maximilien and {Gawiser}, Eric and {Holwerda}, Benne W. and {Huertas-Company}, Marc and {Jaskot}, Anne E. and {Leung}, Gene C.~K. and {Lucas}, Ray A. and {Mobasher}, Bahram and {Pandya}, Viraj and {Tacchella}, Sandro and {Weiner}, Benjamin J. and {Zavala}, Jorge A.},
        title = "{CEERS Key Paper. I. An Early Look into the First 500 Myr of Galaxy Formation with JWST}",
      journal = {\apjl},
         year = 2023,
        month = mar,
       volume = {946},
       number = {1},
          eid = {L13},
        pages = {L13},
          doi = {10.3847/2041-8213/acade4},
archivePrefix = {arXiv},
       eprint = {2211.05792},
 primaryClass = {astro-ph.GA},
       adsurl = {https://ui.adsabs.harvard.edu/abs/2023ApJ...946L..13F}
}

@ARTICLE{Finkelstein_2024,
       author = {{Finkelstein}, Steven L. and {Leung}, Gene C.~K. and {Bagley}, Micaela B. and {Dickinson}, Mark and {Ferguson}, Henry C. and {Papovich}, Casey and {Akins}, Hollis B. and {Arrabal Haro}, Pablo and {Dav{\'e}}, Romeel and {Dekel}, Avishai and {Kartaltepe}, Jeyhan S. and {Kocevski}, Dale D. and {Koekemoer}, Anton M. and {Pirzkal}, Nor and {Somerville}, Rachel S. and {Yung}, L.~Y. Aaron and {Amor{\'\i}n}, Ricardo O. and {Backhaus}, Bren E. and {Behroozi}, Peter and {Bisigello}, Laura and {Bromm}, Volker and {Casey}, Caitlin M. and {Ch{\'a}vez Ortiz}, {\'O}scar A. and {Cheng}, Yingjie and {Chworowsky}, Katherine and {Cleri}, Nikko J. and {Cooper}, M.~C. and {Davis}, Kelcey and {de la Vega}, Alexander and {Elbaz}, David and {Franco}, Maximilien and {Fontana}, Adriano and {Fujimoto}, Seiji and {Giavalisco}, Mauro and {Grogin}, Norman A. and {Holwerda}, Benne W. and {Huertas-Company}, Marc and {Hirschmann}, Michaela and {Iyer}, Kartheik G. and {Jogee}, Shardha and {Jung}, Intae and {Larson}, Rebecca L. and {Lucas}, Ray A. and {Mobasher}, Bahram and {Morales}, Alexa M. and {Morley}, Caroline V. and {Mukherjee}, Sagnick and {P{\'e}rez-Gonz{\'a}lez}, Pablo G. and {Ravindranath}, Swara and {Rodighiero}, Giulia and {Rowland}, Melanie J. and {Tacchella}, Sandro and {Taylor}, Anthony J. and {Trump}, Jonathan R. and {Wilkins}, Stephen M.},
        title = "{The Complete CEERS Early Universe Galaxy Sample: A Surprisingly Slow Evolution of the Space Density of Bright Galaxies at z {\ensuremath{\sim}} 8.5{\textendash}14.5}",
      journal = {\apjl},
         year = 2024,
        month = jul,
       volume = {969},
       number = {1},
          eid = {L2},
        pages = {L2},
          doi = {10.3847/2041-8213/ad4495},
archivePrefix = {arXiv},
       eprint = {2311.04279},
 primaryClass = {astro-ph.GA},
       adsurl = {https://ui.adsabs.harvard.edu/abs/2024ApJ...969L...2F}
}

@ARTICLE{Font_2008,
       author = {{Font}, A.~S. and {Bower}, R.~G. and {McCarthy}, I.~G. and {Benson}, A.~J. and {Frenk}, C.~S. and {Helly}, J.~C. and {Lacey}, C.~G. and {Baugh}, C.~M. and {Cole}, S.},
        title = "{The colours of satellite galaxies in groups and clusters}",
      journal = {\mnras},
         year = 2008,
        month = oct,
       volume = {389},
       number = {4},
        pages = {1619-1629},
          doi = {10.1111/j.1365-2966.2008.13698.x},
archivePrefix = {arXiv},
       eprint = {0807.0001},
 primaryClass = {astro-ph},
       adsurl = {https://ui.adsabs.harvard.edu/abs/2008MNRAS.389.1619F}
}

@ARTICLE{Gabrielpillai_2022,
       author = {{Gabrielpillai}, Austen and {Somerville}, Rachel S. and {Genel}, Shy and {Rodriguez-Gomez}, Vicente and {Pandya}, Viraj and {Yung}, L.~Y. Aaron and {Hernquist}, Lars},
        title = "{Galaxy formation in the Santa Cruz semi-analytic model compared with IllustrisTNG - I. Galaxy scaling relations, dispersions, and residuals at z = 0}",
      journal = {\mnras},
         year = 2022,
        month = dec,
       volume = {517},
       number = {4},
        pages = {6091-6111},
          doi = {10.1093/mnras/stac2297},
archivePrefix = {arXiv},
       eprint = {2111.03077},
 primaryClass = {astro-ph.GA},
       adsurl = {https://ui.adsabs.harvard.edu/abs/2022MNRAS.517.6091G}
}

@ARTICLE{Gardner_2006,
       author = {{Gardner}, Jonathan P. and {Mather}, John C. and {Clampin}, Mark and {Doyon}, Rene and {Greenhouse}, Matthew A. and {Hammel}, Heidi B. and {Hutchings}, John B. and {Jakobsen}, Peter and {Lilly}, Simon J. and {Long}, Knox S. and {Lunine}, Jonathan I. and {McCaughrean}, Mark J. and {Mountain}, Matt and {Nella}, John and {Rieke}, George H. and {Rieke}, Marcia J. and {Rix}, Hans-Walter and {Smith}, Eric P. and {Sonneborn}, George and {Stiavelli}, Massimo and {Stockman}, H.~S. and {Windhorst}, Rogier A. and {Wright}, Gillian S.},
        title = "{The James Webb Space Telescope}",
      journal = {\ssr},
         year = 2006,
        month = apr,
       volume = {123},
       number = {4},
        pages = {485-606},
          doi = {10.1007/s11214-006-8315-7},
archivePrefix = {arXiv},
       eprint = {astro-ph/0606175},
 primaryClass = {astro-ph},
       adsurl = {https://ui.adsabs.harvard.edu/abs/2006SSRv..123..485G}
}

@ARTICLE{Gelli_2024,
       author = {{Gelli}, Viola and {Mason}, Charlotte and {Hayward}, Christopher C.},
        title = "{The Impact of Mass-dependent Stochasticity at Cosmic Dawn}",
      journal = {\apj},
         year = 2024,
        month = nov,
       volume = {975},
       number = {2},
          eid = {192},
        pages = {192},
          doi = {10.3847/1538-4357/ad7b36},
archivePrefix = {arXiv},
       eprint = {2405.13108},
 primaryClass = {astro-ph.GA},
       adsurl = {https://ui.adsabs.harvard.edu/abs/2024ApJ...975..192G}
}

@ARTICLE{Gnedin_2011,
       author = {{Gnedin}, Nickolay Y. and {Kravtsov}, Andrey V.},
        title = "{Environmental Dependence of the Kennicutt-Schmidt Relation in Galaxies}",
      journal = {\apj},
         year = 2011,
        month = feb,
       volume = {728},
       number = {2},
          eid = {88},
        pages = {88},
          doi = {10.1088/0004-637X/728/2/88},
archivePrefix = {arXiv},
       eprint = {1004.0003},
 primaryClass = {astro-ph.CO},
       adsurl = {https://ui.adsabs.harvard.edu/abs/2011ApJ...728...88G}
}

@ARTICLE{Gould_2023,
       author = {{Gould}, Katriona M.~L. and {Brammer}, Gabriel and {Valentino}, Francesco and {Whitaker}, Katherine E. and {Weaver}, John. R. and {Lagos}, Claudia del P. and {Rizzo}, Francesca and {Franco}, Maximilien and {Hsieh}, Bau-Ching and {Ilbert}, Olivier and {Jin}, Shuowen and {Magdis}, Georgios and {McCracken}, Henry J. and {Mobasher}, Bahram and {Shuntov}, Marko and {Steinhardt}, Charles L. and {Strait}, Victoria and {Toft}, Sune},
        title = "{COSMOS2020: Exploring the Dawn of Quenching for Massive Galaxies at 3 < z < 5 with a New Color-selection Method}",
      journal = {\aj},
         year = 2023,
        month = jun,
       volume = {165},
       number = {6},
          eid = {248},
        pages = {248},
          doi = {10.3847/1538-3881/accadc},
archivePrefix = {arXiv},
       eprint = {2302.10934},
 primaryClass = {astro-ph.GA},
       adsurl = {https://ui.adsabs.harvard.edu/abs/2023AJ....165..248G}
}

@ARTICLE{Greene_2024,
       author = {{Greene}, Jenny E. and {Labbe}, Ivo and {Goulding}, Andy D. and {Furtak}, Lukas J. and {Chemerynska}, Iryna and {Kokorev}, Vasily and {Dayal}, Pratika and {Volonteri}, Marta and {Williams}, Christina C. and {Wang}, Bingjie and {Setton}, David J. and {Burgasser}, Adam J. and {Bezanson}, Rachel and {Atek}, Hakim and {Brammer}, Gabriel and {Cutler}, Sam E. and {Feldmann}, Robert and {Fujimoto}, Seiji and {Glazebrook}, Karl and {de Graaff}, Anna and {Khullar}, Gourav and {Leja}, Joel and {Marchesini}, Danilo and {Maseda}, Michael V. and {Matthee}, Jorryt and {Miller}, Tim B. and {Naidu}, Rohan P. and {Nanayakkara}, Themiya and {Oesch}, Pascal A. and {Pan}, Richard and {Papovich}, Casey and {Price}, Sedona H. and {van Dokkum}, Pieter and {Weaver}, John R. and {Whitaker}, Katherine E. and {Zitrin}, Adi},
        title = "{UNCOVER Spectroscopy Confirms the Surprising Ubiquity of Active Galactic Nuclei in Red Sources at z > 5}",
      journal = {\apj},
         year = 2024,
        month = mar,
       volume = {964},
       number = {1},
          eid = {39},
        pages = {39},
          doi = {10.3847/1538-4357/ad1e5f},
archivePrefix = {arXiv},
       eprint = {2309.05714},
 primaryClass = {astro-ph.GA},
       adsurl = {https://ui.adsabs.harvard.edu/abs/2024ApJ...964...39G}
}

@ARTICLE{Griffin_2019,
       author = {{Griffin}, Andrew J. and {Lacey}, Cedric G. and {Gonzalez-Perez}, Violeta and {Lagos}, Claudia del P. and {Baugh}, Carlton M. and {Fanidakis}, Nikos},
        title = "{The evolution of SMBH spin and AGN luminosities for z < 6 within a semi-analytic model of galaxy formation}",
      journal = {\mnras},
         year = 2019,
        month = jul,
       volume = {487},
       number = {1},
        pages = {198-227},
          doi = {10.1093/mnras/stz1216},
archivePrefix = {arXiv},
       eprint = {1806.08370},
 primaryClass = {astro-ph.GA},
       adsurl = {https://ui.adsabs.harvard.edu/abs/2019MNRAS.487..198G}
}

@ARTICLE{Griffin_2020,
       author = {{Griffin}, Andrew J. and {Lacey}, Cedric G. and {Gonzalez-Perez}, Violeta and {Lagos}, Claudia del P. and {Baugh}, Carlton M. and {Fanidakis}, Nikos},
        title = "{AGNs at the cosmic dawn: predictions for future surveys from a {\ensuremath{\Lambda}}CDM cosmological model}",
      journal = {\mnras},
         year = 2020,
        month = feb,
       volume = {492},
       number = {2},
        pages = {2535-2552},
          doi = {10.1093/mnras/staa024},
archivePrefix = {arXiv},
       eprint = {1908.02841},
 primaryClass = {astro-ph.GA},
       adsurl = {https://ui.adsabs.harvard.edu/abs/2020MNRAS.492.2535G}
}

@ARTICLE{Guo_2016,
       author = {{Guo}, Quan and {Gonzalez-Perez}, Violeta and {Guo}, Qi and {Schaller}, Matthieu and {Furlong}, Michelle and {Bower}, Richard G. and {Cole}, Shaun and {Crain}, Robert A. and {Frenk}, Carlos S. and {Helly}, John C. and {Lacey}, Cedric G. and {Lagos}, Claudia del P. and {Mitchell}, Peter and {Schaye}, Joop and {Theuns}, Tom},
        title = "{Galaxies in the EAGLE hydrodynamical simulation and in the Durham and Munich semi-analytical models}",
      journal = {\mnras},
         year = 2016,
        month = oct,
       volume = {461},
       number = {4},
        pages = {3457-3482},
          doi = {10.1093/mnras/stw1525},
archivePrefix = {arXiv},
       eprint = {1512.00015},
 primaryClass = {astro-ph.GA},
       adsurl = {https://ui.adsabs.harvard.edu/abs/2016MNRAS.461.3457G}
}

@ARTICLE{Hadzhiyska_2021,
       author = {{Hadzhiyska}, Boryana and {Liu}, Sonya and {Somerville}, Rachel S. and {Gabrielpillai}, Austen and {Bose}, Sownak and {Eisenstein}, Daniel and {Hernquist}, Lars},
        title = "{Galaxy assembly bias and large-scale distribution: a comparison between IllustrisTNG and a semi-analytic model}",
      journal = {\mnras},
         year = 2021,
        month = nov,
       volume = {508},
       number = {1},
        pages = {698-718},
          doi = {10.1093/mnras/stab2564},
archivePrefix = {arXiv},
       eprint = {2108.00006},
 primaryClass = {astro-ph.CO},
       adsurl = {https://ui.adsabs.harvard.edu/abs/2021MNRAS.508..698H}
}

@ARTICLE{Harikane_2023,
       author = {{Harikane}, Yuichi and {Zhang}, Yechi and {Nakajima}, Kimihiko and {Ouchi}, Masami and {Isobe}, Yuki and {Ono}, Yoshiaki and {Hatano}, Shun and {Xu}, Yi and {Umeda}, Hiroya},
        title = "{A JWST/NIRSpec First Census of Broad-line AGNs at z = 4-7: Detection of 10 Faint AGNs with M $_{BH}$ {}10$^{6}$-{}10$^{8}$ M $_{{\ensuremath{\odot}}}$ and Their Host Galaxy Properties}",
      journal = {\apj},
         year = 2023,
        month = dec,
       volume = {959},
       number = {1},
          eid = {39},
        pages = {39},
          doi = {10.3847/1538-4357/ad029e},
archivePrefix = {arXiv},
       eprint = {2303.11946},
 primaryClass = {astro-ph.GA},
       adsurl = {https://ui.adsabs.harvard.edu/abs/2023ApJ...959...39H}
}

@ARTICLE{Harvey_2024,
       author = {{Harvey}, Thomas and {Conselice}, Christopher and {Adams}, Nathan J. and {Austin}, Duncan and {Juodzbalis}, Ignas and {Trussler}, James and {Li}, Qiong and {Ormerod}, Katherine and {Ferreira}, Leonardo and {Duan}, Qiao and {Westcott}, Lewi and {Harris}, Honor and {Bhatawdekar}, Rachana and {Coe}, Dan and {Cohen}, Seth H. and {Caruana}, Joseph and {Cheng}, Cheng and {Driver}, 9 Simon P. and {Frye}, Brenda and {Furtak}, Lukas J. and {Grogin}, Norman A. and {Hathi}, Nimish P. and {Holwerda}, Benne W. and {Jansen}, Rolf A. and {Koekemoer}, Anton M. and {Lovell}, Christopher J. and {Marshall}, Madeline A. and {Nonino}, Mario and {Smail}, Ian and {Vijayan}, Aswin P. and {Wilkins}, Stephen M. and {Windhorst}, Rogier and {Willmer}, Christopher N.~A. and {Yan}, Haojing and {Zitrin}, Adi},
        title = "{EPOCHS IV: SED Modelling Assumptions and their impact on the Stellar Mass Function at 6.5 < z < 13.5 using PEARLS and public JWST observations}",
      journal = {arXiv e-prints},
         year = 2024,
        month = mar,
          eid = {arXiv:2403.03908},
        pages = {arXiv:2403.03908},
          doi = {10.48550/arXiv.2403.03908},
archivePrefix = {arXiv},
       eprint = {2403.03908},
 primaryClass = {astro-ph.GA},
       adsurl = {https://ui.adsabs.harvard.edu/abs/2024arXiv240303908H}
}

@ARTICLE{Harrold_2024,
       author = {{Harrold}, Jimi E. and {Almaini}, Omar and {Pearce}, Frazer R. and {Yates}, Robert M.},
        title = "{Correcting for the overabundance of low-mass quiescent galaxies in semi-analytic models}",
      journal = {\mnras},
         year = 2024,
        month = jul,
       volume = {532},
       number = {1},
        pages = {L61-L66},
          doi = {10.1093/mnrasl/slae043},
archivePrefix = {arXiv},
       eprint = {2405.06018},
 primaryClass = {astro-ph.GA},
       adsurl = {https://ui.adsabs.harvard.edu/abs/2024MNRAS.532L..61H}
}

@ARTICLE{Henriques_2017,
       author = {{Henriques}, Bruno M.~B. and {White}, Simon D.~M. and {Thomas}, Peter A. and {Angulo}, Raul E. and {Guo}, Qi and {Lemson}, Gerard and {Wang}, Wenting},
        title = "{Galaxy formation in the Planck cosmology - IV. Mass and environmental quenching, conformity and clustering}",
      journal = {\mnras},
         year = 2017,
        month = aug,
       volume = {469},
       number = {3},
        pages = {2626-2645},
          doi = {10.1093/mnras/stx1010},
archivePrefix = {arXiv},
       eprint = {1611.02286},
 primaryClass = {astro-ph.GA},
       adsurl = {https://ui.adsabs.harvard.edu/abs/2017MNRAS.469.2626H}
}

@ARTICLE{Hirschmann_2012,
       author = {{Hirschmann}, Michaela and {Naab}, Thorsten and {Somerville}, Rachel S. and {Burkert}, Andreas and {Oser}, Ludwig},
        title = "{Galaxy formation in semi-analytic models and cosmological hydrodynamic zoom simulations}",
      journal = {\mnras},
         year = 2012,
        month = feb,
       volume = {419},
       number = {4},
        pages = {3200-3222},
          doi = {10.1111/j.1365-2966.2011.19961.x},
archivePrefix = {arXiv},
       eprint = {1104.1626},
 primaryClass = {astro-ph.CO},
       adsurl = {https://ui.adsabs.harvard.edu/abs/2012MNRAS.419.3200H}
}

@ARTICLE{Hopkins_2006,
       author = {{Hopkins}, Philip F. and {Hernquist}, Lars and {Cox}, Thomas J. and {Robertson}, Brant and {Di Matteo}, Tiziana and {Springel}, Volker},
        title = "{The Evolution in the Faint-End Slope of the Quasar Luminosity Function}",
      journal = {\apj},
         year = 2006,
        month = mar,
       volume = {639},
       number = {2},
        pages = {700-709},
          doi = {10.1086/499351},
archivePrefix = {arXiv},
       eprint = {astro-ph/0508299},
 primaryClass = {astro-ph},
       adsurl = {https://ui.adsabs.harvard.edu/abs/2006ApJ...639..700H}
}

@ARTICLE{Hopkins_2009_mergers,
       author = {{Hopkins}, Philip F. and {Cox}, Thomas J. and {Younger}, Joshua D. and {Hernquist}, Lars},
        title = "{How do Disks Survive Mergers?}",
      journal = {\apj},
         year = 2009,
        month = feb,
       volume = {691},
       number = {2},
        pages = {1168-1201},
          doi = {10.1088/0004-637X/691/2/1168},
archivePrefix = {arXiv},
       eprint = {0806.1739},
 primaryClass = {astro-ph},
       adsurl = {https://ui.adsabs.harvard.edu/abs/2009ApJ...691.1168H}
}

@ARTICLE{Izquierdo_2020,
       author = {{Izquierdo-Villalba}, David and {Bonoli}, Silvia and {Dotti}, Massimo and {Sesana}, Alberto and {Rosas-Guevara}, Yetli and {Spinoso}, Daniele},
        title = "{From galactic nuclei to the halo outskirts: tracing supermassive black holes across cosmic history and environments}",
      journal = {\mnras},
         year = 2020,
        month = jul,
       volume = {495},
       number = {4},
        pages = {4681-4706},
          doi = {10.1093/mnras/staa1399},
archivePrefix = {arXiv},
       eprint = {2001.10548},
 primaryClass = {astro-ph.GA},
       adsurl = {https://ui.adsabs.harvard.edu/abs/2020MNRAS.495.4681I}
}

@ARTICLE{Jespersen_2025,
       author = {{Jespersen}, Christian Kragh and {Steinhardt}, Charles L. and {Somerville}, Rachel S. and {Lovell}, Christopher C.},
        title = "{On the Significance of Rare Objects at High Redshift: The Impact of Cosmic Variance}",
      journal = {\apj},
         year = 2025,
        month = mar,
       volume = {982},
       number = {1},
          eid = {23},
        pages = {23},
          doi = {10.3847/1538-4357/adb422},
archivePrefix = {arXiv},
       eprint = {2403.00050},
 primaryClass = {astro-ph.GA},
       adsurl = {https://ui.adsabs.harvard.edu/abs/2025ApJ...982...23J}
}

@ARTICLE{Kauffmann_2000,
       author = {{Kauffmann}, Guinevere and {Haehnelt}, Martin},
        title = "{A unified model for the evolution of galaxies and quasars}",
      journal = {\mnras},
         year = 2000,
        month = jan,
       volume = {311},
       number = {3},
        pages = {576-588},
          doi = {10.1046/j.1365-8711.2000.03077.x},
archivePrefix = {arXiv},
       eprint = {astro-ph/9906493},
 primaryClass = {astro-ph},
       adsurl = {https://ui.adsabs.harvard.edu/abs/2000MNRAS.311..576K}
}

@ARTICLE{Kennicutt_1983,
       author = {{Kennicutt}, Jr., R.~C.},
        title = "{The rate of star formation in normal disk galaxies.}",
      journal = {\apj},
         year = 1983,
        month = sep,
       volume = {272},
        pages = {54-67},
          doi = {10.1086/161261},
       adsurl = {https://ui.adsabs.harvard.edu/abs/1983ApJ...272...54K}
}

@ARTICLE{Kennicutt_1989,
       author = {{Kennicutt}, Jr., Robert C.},
        title = "{The Star Formation Law in Galactic Disks}",
      journal = {\apj},
         year = 1989,
        month = sep,
       volume = {344},
        pages = {685},
          doi = {10.1086/167834},
       adsurl = {https://ui.adsabs.harvard.edu/abs/1989ApJ...344..685K}
}

@ARTICLE{Kennicutt_1998,
       author = {{Kennicutt}, Jr., Robert C.},
        title = "{Star Formation in Galaxies Along the Hubble Sequence}",
      journal = {\araa},
         year = 1998,
        month = jan,
       volume = {36},
        pages = {189-232},
          doi = {10.1146/annurev.astro.36.1.189},
archivePrefix = {arXiv},
       eprint = {astro-ph/9807187},
 primaryClass = {astro-ph},
       adsurl = {https://ui.adsabs.harvard.edu/abs/1998ARA&A..36..189K}
}

@ARTICLE{Kennicutt_2012,
       author = {{Kennicutt}, Robert C. and {Evans}, Neal J.},
        title = "{Star Formation in the Milky Way and Nearby Galaxies}",
      journal = {\araa},
         year = 2012,
        month = sep,
       volume = {50},
        pages = {531-608},
          doi = {10.1146/annurev-astro-081811-125610},
archivePrefix = {arXiv},
       eprint = {1204.3552},
 primaryClass = {astro-ph.GA},
       adsurl = {https://ui.adsabs.harvard.edu/abs/2012ARA&A..50..531K}
}

@ARTICLE{Krumholz_2009,
       author = {{Krumholz}, Mark R. and {McKee}, Christopher F. and {Tumlinson}, Jason},
        title = "{The Atomic-to-Molecular Transition in Galaxies. II: H I and H$_{2}$ Column Densities}",
      journal = {\apj},
         year = 2009,
        month = mar,
       volume = {693},
       number = {1},
        pages = {216-235},
          doi = {10.1088/0004-637X/693/1/216},
archivePrefix = {arXiv},
       eprint = {0811.0004},
 primaryClass = {astro-ph},
       adsurl = {https://ui.adsabs.harvard.edu/abs/2009ApJ...693..216K}
}

@ARTICLE{Lagos_2011,
       author = {{Lagos}, Claudia Del P. and {Lacey}, Cedric G. and {Baugh}, Carlton M. and {Bower}, Richard G. and {Benson}, Andrew J.},
        title = "{On the impact of empirical and theoretical star formation laws on galaxy formation}",
      journal = {\mnras},
         year = 2011,
        month = sep,
       volume = {416},
       number = {2},
        pages = {1566-1584},
          doi = {10.1111/j.1365-2966.2011.19160.x},
archivePrefix = {arXiv},
       eprint = {1011.5506},
 primaryClass = {astro-ph.CO},
       adsurl = {https://ui.adsabs.harvard.edu/abs/2011MNRAS.416.1566L}
}

@ARTICLE{Lagos_2013,
       author = {{Lagos}, Claudia del P. and {Lacey}, Cedric G. and {Baugh}, Carlton M.},
        title = "{A dynamical model of supernova feedback: gas outflows from the interstellar medium}",
      journal = {\mnras},
         year = 2013,
        month = dec,
       volume = {436},
       number = {2},
        pages = {1787-1817},
          doi = {10.1093/mnras/stt1696},
archivePrefix = {arXiv},
       eprint = {1303.6635},
 primaryClass = {astro-ph.CO},
       adsurl = {https://ui.adsabs.harvard.edu/abs/2013MNRAS.436.1787L}
}

@ARTICLE{Lagos_2025,
       author = {{Lagos}, Claudia del P. and {Valentino}, Francesco and {Wright}, Ruby J. and {de Graaff}, Anna and {Glazebrook}, Karl and {De Lucia}, Gabriella and {Robotham}, Aaron S.~G. and {Nanayakkara}, Themiya and {Chandro-Gomez}, Angel and {Bravo}, Mat{\'\i}as and {Baugh}, Carlton M. and {Harborne}, Katherine E. and {Hirschmann}, Michaela and {Fontanot}, Fabio and {Xie}, Lizhi and {Chittenden}, Harry},
        title = "{The diverse star formation histories of early massive, quenched galaxies in modern galaxy formation simulations}",
      journal = {\mnras},
         year = 2025,
        month = jan,
       volume = {536},
       number = {3},
        pages = {2324-2354},
          doi = {10.1093/mnras/stae2626},
archivePrefix = {arXiv},
       eprint = {2409.16916},
 primaryClass = {astro-ph.GA},
       adsurl = {https://ui.adsabs.harvard.edu/abs/2025MNRAS.536.2324L}
}

@ARTICLE{Leung_2023,
       author = {{Leung}, Gene C.~K. and {Bagley}, Micaela B. and {Finkelstein}, Steven L. and {Ferguson}, Henry C. and {Koekemoer}, Anton M. and {P{\'e}rez-Gonz{\'a}lez}, Pablo G. and {Morales}, Alexa and {Kocevski}, Dale D. and {Yang}, Guang and {Somerville}, Rachel S. and {Wilkins}, Stephen M. and {Yung}, L.~Y. Aaron and {Fujimoto}, Seiji and {Larson}, Rebecca L. and {Papovich}, Casey and {Pirzkal}, Nor and {Berg}, Danielle A. and {Lotz}, Jennifer M. and {Castellano}, Marco and {Ch{\'a}vez Ortiz}, {\'O}scar A. and {Cheng}, Yingjie and {Dickinson}, Mark and {Giavalisco}, Mauro and {Hathi}, Nimish P. and {Hutchison}, Taylor A. and {Jung}, Intae and {Kartaltepe}, Jeyhan S. and {Natarajan}, Priyamvada and {Rothberg}, Barry},
        title = "{NGDEEP Epoch 1: The Faint End of the Luminosity Function at z   9-12 from Ultradeep JWST Imaging}",
      journal = {\apjl},
         year = 2023,
        month = sep,
       volume = {954},
       number = {2},
          eid = {L46},
        pages = {L46},
          doi = {10.3847/2041-8213/acf365},
archivePrefix = {arXiv},
       eprint = {2306.06244},
 primaryClass = {astro-ph.GA},
       adsurl = {https://ui.adsabs.harvard.edu/abs/2023ApJ...954L..46L}
}

@ARTICLE{Lu_2014,
       author = {{Lu}, Yu and {Wechsler}, Risa H. and {Somerville}, Rachel S. and {Croton}, Darren and {Porter}, Lauren and {Primack}, Joel and {Behroozi}, Peter S. and {Ferguson}, Henry C. and {Koo}, David C. and {Guo}, Yicheng and {Safarzadeh}, Mohammadtaher and {Finlator}, Kristian and {Castellano}, Marco and {White}, Catherine E. and {Sommariva}, Veronica and {Moody}, Chris},
        title = "{Semi-analytic Models for the CANDELS Survey: Comparison of Predictions for Intrinsic Galaxy Properties}",
      journal = {\apj},
         year = 2014,
        month = nov,
       volume = {795},
       number = {2},
          eid = {123},
        pages = {123},
          doi = {10.1088/0004-637X/795/2/123},
archivePrefix = {arXiv},
       eprint = {1312.3233},
 primaryClass = {astro-ph.CO},
       adsurl = {https://ui.adsabs.harvard.edu/abs/2014ApJ...795..123L}
}

@ARTICLE{Malbon_2007,
       author = {{Malbon}, Rowena K. and {Baugh}, C.~M. and {Frenk}, C.~S. and {Lacey}, C.~G.},
        title = "{Black hole growth in hierarchical galaxy formation}",
      journal = {\mnras},
         year = 2007,
        month = dec,
       volume = {382},
       number = {4},
        pages = {1394-1414},
          doi = {10.1111/j.1365-2966.2007.12317.x},
archivePrefix = {arXiv},
       eprint = {astro-ph/0607424},
 primaryClass = {astro-ph},
       adsurl = {https://ui.adsabs.harvard.edu/abs/2007MNRAS.382.1394M}
}

@ARTICLE{Maraston_2005,
       author = {{Maraston}, Claudia},
        title = "{Evolutionary population synthesis: models, analysis of the ingredients and application to high-z galaxies}",
      journal = {\mnras},
         year = 2005,
        month = sep,
       volume = {362},
       number = {3},
        pages = {799-825},
          doi = {10.1111/j.1365-2966.2005.09270.x},
archivePrefix = {arXiv},
       eprint = {astro-ph/0410207},
 primaryClass = {astro-ph},
       adsurl = {https://ui.adsabs.harvard.edu/abs/2005MNRAS.362..799M}
}

@ARTICLE{Matthee_2024,
       author = {{Matthee}, Jorryt and {Naidu}, Rohan P. and {Brammer}, Gabriel and {Chisholm}, John and {Eilers}, Anna-Christina and {Goulding}, Andy and {Greene}, Jenny and {Kashino}, Daichi and {Labbe}, Ivo and {Lilly}, Simon J. and {Mackenzie}, Ruari and {Oesch}, Pascal A. and {Weibel}, Andrea and {Wuyts}, Stijn and {Xiao}, Mengyuan and {Bordoloi}, Rongmon and {Bouwens}, Rychard and {van Dokkum}, Pieter and {Illingworth}, Garth and {Kramarenko}, Ivan and {Maseda}, Michael V. and {Mason}, Charlotte and {Meyer}, Romain A. and {Nelson}, Erica J. and {Reddy}, Naveen A. and {Shivaei}, Irene and {Simcoe}, Robert A. and {Yue}, Minghao},
        title = "{Little Red Dots: An Abundant Population of Faint Active Galactic Nuclei at z {\ensuremath{\sim}} 5 Revealed by the EIGER and FRESCO JWST Surveys}",
      journal = {\apj},
         year = 2024,
        month = mar,
       volume = {963},
       number = {2},
          eid = {129},
        pages = {129},
          doi = {10.3847/1538-4357/ad2345},
archivePrefix = {arXiv},
       eprint = {2306.05448},
 primaryClass = {astro-ph.GA},
       adsurl = {https://ui.adsabs.harvard.edu/abs/2024ApJ...963..129M}
}

@ARTICLE{McCarthy_2008,
       author = {{McCarthy}, I.~G. and {Frenk}, C.~S. and {Font}, A.~S. and {Lacey}, C.~G. and {Bower}, R.~G. and {Mitchell}, N.~L. and {Balogh}, M.~L. and {Theuns}, T.},
        title = "{Ram pressure stripping the hot gaseous haloes of galaxies in groups and clusters}",
      journal = {\mnras},
         year = 2008,
        month = jan,
       volume = {383},
       number = {2},
        pages = {593-605},
          doi = {10.1111/j.1365-2966.2007.12577.x},
archivePrefix = {arXiv},
       eprint = {0710.0964},
 primaryClass = {astro-ph},
       adsurl = {https://ui.adsabs.harvard.edu/abs/2008MNRAS.383..593M}
}

@ARTICLE{Mckee_2010,
       author = {{McKee}, Christopher F. and {Krumholz}, Mark R.},
        title = "{The Atomic-to-Molecular Transition in Galaxies. III. A New Method for Determining the Molecular Content of Primordial and Dusty Clouds}",
      journal = {\apj},
         year = 2010,
        month = jan,
       volume = {709},
       number = {1},
        pages = {308-320},
          doi = {10.1088/0004-637X/709/1/308},
archivePrefix = {arXiv},
       eprint = {0908.0330},
 primaryClass = {astro-ph.GA},
       adsurl = {https://ui.adsabs.harvard.edu/abs/2010ApJ...709..308M}
}

@ARTICLE{Meier_2002,
       author = {{Meier}, David L.},
        title = "{Grand unification of AGN and the accretion and spin paradigms}",
      journal = {New Astronomy Reviews},
         year = 2002,
        month = may,
       volume = {46},
       number = {2-7},
        pages = {247-255},
          doi = {10.1016/S1387-6473(01)00189-0},
archivePrefix = {arXiv},
       eprint = {astro-ph/9908283},
 primaryClass = {astro-ph},
       adsurl = {https://ui.adsabs.harvard.edu/abs/2002NewAR..46..247M}
}

@ARTICLE{Merlin_2012,
       author = {{Merlin}, E. and {Chiosi}, C. and {Piovan}, L. and {Grassi}, T. and {Buonomo}, U. and {La Barbera}, F.},
        title = "{Formation and evolution of early-type galaxies - III. Dependence of the star formation history on the total mass and initial overdensity}",
      journal = {\mnras},
         year = 2012,
        month = dec,
       volume = {427},
       number = {2},
        pages = {1530-1554},
          doi = {10.1111/j.1365-2966.2012.21965.x},
archivePrefix = {arXiv},
       eprint = {1204.5118},
 primaryClass = {astro-ph.CO},
       adsurl = {https://ui.adsabs.harvard.edu/abs/2012MNRAS.427.1530M}
}

@ARTICLE{Mitchell_2018,
       author = {{Mitchell}, Peter D. and {Lacey}, Cedric G. and {Lagos}, Claudia D.~P. and {Frenk}, Carlos S. and {Bower}, Richard G. and {Cole}, Shaun and {Helly}, John C. and {Schaller}, Matthieu and {Gonzalez-Perez}, Violeta and {Theuns}, Tom},
        title = "{Comparing galaxy formation in semi-analytic models and hydrodynamical simulations}",
      journal = {\mnras},
         year = 2018,
        month = feb,
       volume = {474},
       number = {1},
        pages = {492-521},
          doi = {10.1093/mnras/stx2770},
archivePrefix = {arXiv},
       eprint = {1709.08647},
 primaryClass = {astro-ph.GA},
       adsurl = {https://ui.adsabs.harvard.edu/abs/2018MNRAS.474..492M}
}

@ARTICLE{Mitchell_2020,
       author = {{Mitchell}, Peter D. and {Schaye}, Joop and {Bower}, Richard G. and {Crain}, Robert A.},
        title = "{Galactic outflow rates in the EAGLE simulations}",
      journal = {\mnras},
         year = 2020,
        month = may,
       volume = {494},
       number = {3},
        pages = {3971-3997},
          doi = {10.1093/mnras/staa938},
archivePrefix = {arXiv},
       eprint = {1910.09566},
 primaryClass = {astro-ph.GA},
       adsurl = {https://ui.adsabs.harvard.edu/abs/2020MNRAS.494.3971M}
}

@ARTICLE{Monaco_2014,
       author = {{Monaco}, Pierluigi and {Benson}, Andrew J. and {De Lucia}, Gabriella and {Fontanot}, Fabio and {Borgani}, Stefano and {Boylan-Kolchin}, Michael},
        title = "{A semi-analytic model comparison: testing cooling models against hydrodynamical simulations}",
      journal = {\mnras},
         year = 2014,
        month = jul,
       volume = {441},
       number = {3},
        pages = {2058-2077},
          doi = {10.1093/mnras/stu655},
archivePrefix = {arXiv},
       eprint = {1404.0811},
 primaryClass = {astro-ph.CO},
       adsurl = {https://ui.adsabs.harvard.edu/abs/2014MNRAS.441.2058M}
}

@ARTICLE{Narayanan_2024,
       author = {{Narayanan}, Desika and {Lower}, Sidney and {Torrey}, Paul and {Brammer}, Gabriel and {Cui}, Weiguang and {Dav{\'e}}, Romeel and {Iyer}, Kartheik G. and {Li}, Qi and {Lovell}, Christopher C. and {Sales}, Laura V. and {Stark}, Daniel P. and {Marinacci}, Federico and {Vogelsberger}, Mark},
        title = "{Outshining by Recent Star Formation Prevents the Accurate Measurement of High-z Galaxy Stellar Masses}",
      journal = {\apj},
         year = 2024,
        month = jan,
       volume = {961},
       number = {1},
          eid = {73},
        pages = {73},
          doi = {10.3847/1538-4357/ad0966},
archivePrefix = {arXiv},
       eprint = {2306.10118},
 primaryClass = {astro-ph.GA},
       adsurl = {https://ui.adsabs.harvard.edu/abs/2024ApJ...961...73N}
}

@ARTICLE{Navarro_1996,
       author = {{Navarro}, Julio F. and {Frenk}, Carlos S. and {White}, Simon D.~M.},
        title = "{The Structure of Cold Dark Matter Halos}",
      journal = {\apj},
         year = 1996,
        month = may,
       volume = {462},
        pages = {563},
          doi = {10.1086/177173},
archivePrefix = {arXiv},
       eprint = {astro-ph/9508025},
 primaryClass = {astro-ph},
       adsurl = {https://ui.adsabs.harvard.edu/abs/1996ApJ...462..563N}
}

@ARTICLE{Navarro_2024,
       author = {{Navarro-Carrera}, Rafael and {Rinaldi}, Pierluigi and {Caputi}, Karina I. and {Iani}, Edoardo and {Kokorev}, Vasily and {van Mierlo}, Sophie E.},
        title = "{Constraints on the Faint End of the Galaxy Stellar Mass Function at z ≃ 4{\textendash}8 from Deep JWST Data}",
      journal = {\apj},
         year = 2024,
        month = feb,
       volume = {961},
       number = {2},
          eid = {207},
        pages = {207},
          doi = {10.3847/1538-4357/ad0df6},
archivePrefix = {arXiv},
       eprint = {2305.16141},
 primaryClass = {astro-ph.GA},
       adsurl = {https://ui.adsabs.harvard.edu/abs/2024ApJ...961..207N}
}

@ARTICLE{Nulsen_2000,
       author = {{Nulsen}, P.~E.~J. and {Fabian}, A.~C.},
        title = "{Fuelling quasars with hot gas}",
      journal = {\mnras},
         year = 2000,
        month = jan,
       volume = {311},
       number = {2},
        pages = {346-356},
          doi = {10.1046/j.1365-8711.2000.03038.x},
archivePrefix = {arXiv},
       eprint = {astro-ph/9908282},
 primaryClass = {astro-ph},
       adsurl = {https://ui.adsabs.harvard.edu/abs/2000MNRAS.311..346N}
}

@ARTICLE{Overzier_2016,
       author = {{Overzier}, Roderik A.},
        title = "{The realm of the galaxy protoclusters. A review}",
      journal = {\aapr},
         year = 2016,
        month = nov,
       volume = {24},
       number = {1},
          eid = {14},
        pages = {14},
          doi = {10.1007/s00159-016-0100-3},
archivePrefix = {arXiv},
       eprint = {1610.05201},
 primaryClass = {astro-ph.GA},
       adsurl = {https://ui.adsabs.harvard.edu/abs/2016A&ARv..24...14O}
}

@ARTICLE{Pacifici_2016,
       author = {{Pacifici}, Camilla and {Kassin}, Susan A. and {Weiner}, Benjamin J. and {Holden}, Bradford and {Gardner}, Jonathan P. and {Faber}, Sandra M. and {Ferguson}, Henry C. and {Koo}, David C. and {Primack}, Joel R. and {Bell}, Eric F. and {Dekel}, Avishai and {Gawiser}, Eric and {Giavalisco}, Mauro and {Rafelski}, Marc and {Simons}, Raymond C. and {Barro}, Guillermo and {Croton}, Darren J. and {Dav{\'e}}, Romeel and {Fontana}, Adriano and {Grogin}, Norman A. and {Koekemoer}, Anton M. and {Lee}, Seong-Kook and {Salmon}, Brett and {Somerville}, Rachel and {Behroozi}, Peter},
        title = "{The Evolution of Star Formation Histories of Quiescent Galaxies}",
      journal = {\apj},
         year = 2016,
        month = nov,
       volume = {832},
       number = {1},
          eid = {79},
        pages = {79},
          doi = {10.3847/0004-637X/832/1/79},
archivePrefix = {arXiv},
       eprint = {1609.03572},
 primaryClass = {astro-ph.GA},
       adsurl = {https://ui.adsabs.harvard.edu/abs/2016ApJ...832...79P}
}

@ARTICLE{Pandya_2020,
       author = {{Pandya}, Viraj and {Somerville}, Rachel S. and {Angl{\'e}s-Alc{\'a}zar}, Daniel and {Hayward}, Christopher C. and {Bryan}, Greg L. and {Fielding}, Drummond B. and {Forbes}, John C. and {Burkhart}, Blakesley and {Genel}, Shy and {Hernquist}, Lars and {Kim}, Chang-Goo and {Tonnesen}, Stephanie and {Starkenburg}, Tjitske},
        title = "{First Results from SMAUG: The Need for Preventative Stellar Feedback and Improved Baryon Cycling in Semianalytic Models of Galaxy Formation}",
      journal = {\apj},
         year = 2020,
        month = dec,
       volume = {905},
       number = {1},
          eid = {4},
        pages = {4},
          doi = {10.3847/1538-4357/abc3c1},
archivePrefix = {arXiv},
       eprint = {2006.16317},
 primaryClass = {astro-ph.GA},
       adsurl = {https://ui.adsabs.harvard.edu/abs/2020ApJ...905....4P}
}

@ARTICLE{Popesso_2023,
       author = {{Popesso}, P. and {Concas}, A. and {Cresci}, G. and {Belli}, S. and {Rodighiero}, G. and {Inami}, H. and {Dickinson}, M. and {Ilbert}, O. and {Pannella}, M. and {Elbaz}, D.},
        title = "{The main sequence of star-forming galaxies across cosmic times}",
      journal = {\mnras},
         year = 2023,
        month = feb,
       volume = {519},
       number = {1},
        pages = {1526-1544},
          doi = {10.1093/mnras/stac3214},
archivePrefix = {arXiv},
       eprint = {2203.10487},
 primaryClass = {astro-ph.GA},
       adsurl = {https://ui.adsabs.harvard.edu/abs/2023MNRAS.519.1526P}
}

@ARTICLE{Porter_2014,
       author = {{Porter}, L.~A. and {Somerville}, R.~S. and {Primack}, J.~R. and {Johansson}, P.~H.},
        title = "{Understanding the structural scaling relations of early-type galaxies}",
      journal = {\mnras},
         year = 2014,
        month = oct,
       volume = {444},
       number = {1},
        pages = {942-960},
          doi = {10.1093/mnras/stu1434},
archivePrefix = {arXiv},
       eprint = {1407.0594},
 primaryClass = {astro-ph.GA},
       adsurl = {https://ui.adsabs.harvard.edu/abs/2014MNRAS.444..942P}
}

@ARTICLE{Poulton_2021,
       author = {{Poulton}, R.~J.~J. and {Power}, C. and {Robotham}, A.~S.~G. and {Elahi}, P.~J. and {Lagos}, C.~D.~P.},
        title = "{Extracting galaxy merger time-scales II: a new fitting formula}",
      journal = {\mnras},
         year = 2021,
        month = feb,
       volume = {501},
       number = {2},
        pages = {2810-2820},
          doi = {10.1093/mnras/staa3247},
archivePrefix = {arXiv},
       eprint = {2010.08786},
 primaryClass = {astro-ph.GA},
       adsurl = {https://ui.adsabs.harvard.edu/abs/2021MNRAS.501.2810P}
}

@ARTICLE{RB_2024,
       author = {{Roberts-Borsani}, Guido and {Treu}, Tommaso and {Shapley}, Alice and {Fontana}, Adriano and {Pentericci}, Laura and {Castellano}, Marco and {Morishita}, Takahiro and {Bergamini}, Pietro and {Rosati}, Piero},
        title = "{Between the Extremes: A JWST Spectroscopic Benchmark for High-redshift Galaxies Using {\ensuremath{\sim}}500 Confirmed Sources at z {\ensuremath{\geq}} 5}",
      journal = {\apj},
         year = 2024,
        month = dec,
       volume = {976},
       number = {2},
          eid = {193},
        pages = {193},
          doi = {10.3847/1538-4357/ad85d3},
archivePrefix = {arXiv},
       eprint = {2403.07103},
 primaryClass = {astro-ph.GA},
       adsurl = {https://ui.adsabs.harvard.edu/abs/2024ApJ...976..193R}
}

@ARTICLE{Robertson_2006,
       author = {{Robertson}, Brant and {Cox}, Thomas J. and {Hernquist}, Lars and {Franx}, Marijn and {Hopkins}, Philip F. and {Martini}, Paul and {Springel}, Volker},
        title = "{The Fundamental Scaling Relations of Elliptical Galaxies}",
      journal = {\apj},
         year = 2006,
        month = apr,
       volume = {641},
       number = {1},
        pages = {21-40},
          doi = {10.1086/500360},
archivePrefix = {arXiv},
       eprint = {astro-ph/0511053},
 primaryClass = {astro-ph},
       adsurl = {https://ui.adsabs.harvard.edu/abs/2006ApJ...641...21R}
}

@ARTICLE{Robertson_2024,
       author = {{Robertson}, Brant and {Johnson}, Benjamin D. and {Tacchella}, Sandro and {Eisenstein}, Daniel J. and {Hainline}, Kevin and {Arribas}, Santiago and {Baker}, William M. and {Bunker}, Andrew J. and {Carniani}, Stefano and {Cargile}, Phillip A. and {Carreira}, Courtney and {Charlot}, Stephane and {Chevallard}, Jacopo and {Curti}, Mirko and {Curtis-Lake}, Emma and {D'Eugenio}, Francesco and {Egami}, Eiichi and {Hausen}, Ryan and {Helton}, Jakob M. and {Jakobsen}, Peter and {Ji}, Zhiyuan and {Jones}, Gareth C. and {Maiolino}, Roberto and {Maseda}, Michael V. and {Nelson}, Erica and {P{\'e}rez-Gonz{\'a}lez}, Pablo G. and {Pusk{\'a}s}, D{\'a}vid and {Rieke}, Marcia and {Smit}, Renske and {Sun}, Fengwu and {{\"U}bler}, Hannah and {Whitler}, Lily and {Williams}, Christina C. and {Willmer}, Christopher N.~A. and {Willott}, Chris and {Witstok}, Joris},
        title = "{Earliest Galaxies in the JADES Origins Field: Luminosity Function and Cosmic Star Formation Rate Density 300 Myr after the Big Bang}",
      journal = {\apj},
         year = 2024,
        month = jul,
       volume = {970},
       number = {1},
          eid = {31},
        pages = {31},
          doi = {10.3847/1538-4357/ad463d},
archivePrefix = {arXiv},
       eprint = {2312.10033},
 primaryClass = {astro-ph.GA},
       adsurl = {https://ui.adsabs.harvard.edu/abs/2024ApJ...970...31R}
}

@ARTICLE{Russell_2024,
       author = {{Russell}, Tobias A. and {Dobric}, Neva and {Adams}, Nathan J. and {Conselice}, Christopher J. and {Austin}, Duncan and {Harvey}, Thomas and {Trussler}, James and {Ferreira}, Leonardo and {Westcott}, Lewi and {Harris}, Honor and {Windhorst}, Rogier A. and {Coe}, Dan and {Cohen}, Seth H. and {Driver}, Simon P. and {Frye}, Brenda and {Grogin}, Norman A. and {Hathi}, Nimish P. and {Jansen}, Rolf A. and {Koekemoer}, Anton M. and {Marshall}, Madeline A. and {Ortiz}, III, Rafael and {Pirzkal}, Nor and {Robotham}, Aaron and {Ryan}, Jr, Russell E. and {Summers}, Jake and {D'Silva}, Jordan C.~J. and {Willmer}, Christopher N.~A. and {Yan}, Haojing},
        title = "{Cosmic Stillness: High Quiescent Galaxy Fractions Across Upper Mass Scales in the Early Universe to z = 7 with JWST}",
      journal = {arXiv e-prints},
         year = 2024,
        month = dec,
          eid = {arXiv:2412.11861},
        pages = {arXiv:2412.11861},
          doi = {10.48550/arXiv.2412.11861},
archivePrefix = {arXiv},
       eprint = {2412.11861},
 primaryClass = {astro-ph.GA},
       adsurl = {https://ui.adsabs.harvard.edu/abs/2024arXiv241211861R}
}

@ARTICLE{Saghiha_2017,
       author = {{Saghiha}, Hananeh and {Simon}, Patrick and {Schneider}, Peter and {Hilbert}, Stefan},
        title = "{Confronting semi-analytic galaxy models with galaxy-matter correlations observed by CFHTLenS}",
      journal = {\aap},
         year = 2017,
        month = may,
       volume = {601},
          eid = {A98},
        pages = {A98},
          doi = {10.1051/0004-6361/201629608},
archivePrefix = {arXiv},
       eprint = {1608.08629},
 primaryClass = {astro-ph.CO},
       adsurl = {https://ui.adsabs.harvard.edu/abs/2017A&A...601A..98S}
}

@ARTICLE{Salpeter_1955,
       author = {{Salpeter}, Edwin E.},
        title = "{The Luminosity Function and Stellar Evolution.}",
      journal = {\apj},
         year = 1955,
        month = jan,
       volume = {121},
        pages = {161},
          doi = {10.1086/145971},
       adsurl = {https://ui.adsabs.harvard.edu/abs/1955ApJ...121..161S}
}

@ARTICLE{Schaye_2004,
       author = {{Schaye}, Joop},
        title = "{Star Formation Thresholds and Galaxy Edges: Why and Where}",
      journal = {\apj},
         year = 2004,
        month = jul,
       volume = {609},
       number = {2},
        pages = {667-682},
          doi = {10.1086/421232},
archivePrefix = {arXiv},
       eprint = {astro-ph/0205125},
 primaryClass = {astro-ph},
       adsurl = {https://ui.adsabs.harvard.edu/abs/2004ApJ...609..667S}
}

@ARTICLE{Schaye_2008,
       author = {{Schaye}, Joop and {Dalla Vecchia}, Claudio},
        title = "{On the relation between the Schmidt and Kennicutt-Schmidt star formation laws and its implications for numerical simulations}",
      journal = {\mnras},
         year = 2008,
        month = jan,
       volume = {383},
       number = {3},
        pages = {1210-1222},
          doi = {10.1111/j.1365-2966.2007.12639.x},
archivePrefix = {arXiv},
       eprint = {0709.0292},
 primaryClass = {astro-ph},
       adsurl = {https://ui.adsabs.harvard.edu/abs/2008MNRAS.383.1210S}
}

@ARTICLE{Schaye_2025,
       author = {{Schaye}, Joop and {Chaikin}, Evgenii and {Schaller}, Matthieu and {Ploeckinger}, Sylvia and {Hu{\v{s}}ko}, Filip and {McGibbon}, Rob and {Trayford}, James W. and {Ben{\'\i}tez-Llambay}, Alejandro and {Correa}, Camila and {Frenk}, Carlos S. and {Richings}, Alexander J. and {Forouhar Moreno}, Victor J. and {Bah{\'e}}, Yannick M. and {Borrow}, Josh and {Durrant}, Anna and {Gebek}, Andrea and {Helly}, John C. and {Jenkins}, Adrian and {Lacey}, Cedric G. and {Ludlow}, Aaron and {Nobels}, Folkert S.~J.},
        title = "{The COLIBRE project: cosmological hydrodynamical simulations of galaxy formation and evolution}",
      journal = {arXiv e-prints},
         year = 2025,
        month = aug,
          eid = {arXiv:2508.21126},
        pages = {arXiv:2508.21126},
          doi = {10.48550/arXiv.2508.21126},
archivePrefix = {arXiv},
       eprint = {2508.21126},
 primaryClass = {astro-ph.GA},
       adsurl = {https://ui.adsabs.harvard.edu/abs/2025arXiv250821126S}
}

@ARTICLE{Shakura_1973,
       author = {{Shakura}, N.~I. and {Sunyaev}, R.~A.},
        title = "{Black holes in binary systems. Observational appearance.}",
      journal = {\aap},
         year = 1973,
        month = jan,
       volume = {24},
        pages = {337-355},
       adsurl = {https://ui.adsabs.harvard.edu/abs/1973A&A....24..337S}
}

@ARTICLE{Simha_2017,
       author = {{Simha}, Vimal and {Cole}, Shaun},
        title = "{Modelling galaxy merger time-scales and tidal destruction}",
      journal = {\mnras},
         year = 2017,
        month = dec,
       volume = {472},
       number = {2},
        pages = {1392-1400},
          doi = {10.1093/mnras/stx1942},
archivePrefix = {arXiv},
       eprint = {1609.09520},
 primaryClass = {astro-ph.GA},
       adsurl = {https://ui.adsabs.harvard.edu/abs/2017MNRAS.472.1392S}
}

@ARTICLE{Silva_1998,
       author = {{Silva}, Laura and {Granato}, Gian Luigi and {Bressan}, Alessandro and {Danese}, Luigi},
        title = "{Modeling the Effects of Dust on Galactic Spectral Energy Distributions from the Ultraviolet to the Millimeter Band}",
      journal = {\apj},
         year = 1998,
        month = dec,
       volume = {509},
       number = {1},
        pages = {103-117},
          doi = {10.1086/306476},
       adsurl = {https://ui.adsabs.harvard.edu/abs/1998ApJ...509..103S}
}

@ARTICLE{Simpson_2018,
       author = {{Simpson}, Christine M. and {Grand}, Robert J.~J. and {G{\'o}mez}, Facundo A. and {Marinacci}, Federico and {Pakmor}, R{\"u}diger and {Springel}, Volker and {Campbell}, David J.~R. and {Frenk}, Carlos S.},
        title = "{Quenching and ram pressure stripping of simulated Milky Way satellite galaxies}",
      journal = {\mnras},
         year = 2018,
        month = jul,
       volume = {478},
       number = {1},
        pages = {548-567},
          doi = {10.1093/mnras/sty774},
archivePrefix = {arXiv},
       eprint = {1705.03018},
 primaryClass = {astro-ph.GA},
       adsurl = {https://ui.adsabs.harvard.edu/abs/2018MNRAS.478..548S}
}

@ARTICLE{Somerville_2015,
       author = {{Somerville}, Rachel S. and {Dav{\'e}}, Romeel},
        title = "{Physical Models of Galaxy Formation in a Cosmological Framework}",
      journal = {\araa},
         year = 2015,
        month = aug,
       volume = {53},
        pages = {51-113},
          doi = {10.1146/annurev-astro-082812-140951},
archivePrefix = {arXiv},
       eprint = {1412.2712},
 primaryClass = {astro-ph.GA},
       adsurl = {https://ui.adsabs.harvard.edu/abs/2015ARA&A..53...51S}
}

@ARTICLE{Somerville_2025,
       author = {{Somerville}, Rachel S. and {Yung}, L.~Y. Aaron and {Lancaster}, Lachlan and {Menon}, Shyam and {Sommovigo}, Laura and {Finkelstein}, Steven L.},
        title = "{Density modulated star formation efficiency: implications for the observed abundance of ultra-violet luminous galaxies at z>10}",
      journal = {arXiv e-prints},
         year = 2025,
        month = may,
          eid = {arXiv:2505.05442},
        pages = {arXiv:2505.05442},
          doi = {10.48550/arXiv.2505.05442},
archivePrefix = {arXiv},
       eprint = {2505.05442},
 primaryClass = {astro-ph.GA},
       adsurl = {https://ui.adsabs.harvard.edu/abs/2025arXiv250505442S}
}

@ARTICLE{Steinhardt_2023,
       author = {{Steinhardt}, Charles L. and {Kokorev}, Vasily and {Rusakov}, Vadim and {Garcia}, Ethan and {Sneppen}, Albert},
        title = "{Templates for Fitting Photometry of Ultra-high-redshift Galaxies}",
      journal = {\apjl},
         year = 2023,
        month = jul,
       volume = {951},
       number = {2},
          eid = {L40},
        pages = {L40},
          doi = {10.3847/2041-8213/acdef6},
archivePrefix = {arXiv},
       eprint = {2208.07879},
 primaryClass = {astro-ph.GA},
       adsurl = {https://ui.adsabs.harvard.edu/abs/2023ApJ...951L..40S}
}

@ARTICLE{Suess_2022,
       author = {{Suess}, Katherine A. and {Leja}, Joel and {Johnson}, Benjamin D. and {Bezanson}, Rachel and {Greene}, Jenny E. and {Kriek}, Mariska and {Lower}, Sidney and {Narayanan}, Desika and {Setton}, David J. and {Spilker}, Justin S.},
        title = "{Recovering the Star Formation Histories of Recently Quenched Galaxies: The Impact of Model and Prior Choices}",
      journal = {\apj},
         year = 2022,
        month = aug,
       volume = {935},
       number = {2},
          eid = {146},
        pages = {146},
          doi = {10.3847/1538-4357/ac82b0},
archivePrefix = {arXiv},
       eprint = {2207.02883},
 primaryClass = {astro-ph.GA},
       adsurl = {https://ui.adsabs.harvard.edu/abs/2022ApJ...935..146S}
}

@ARTICLE{Sun_2023,
       author = {{Sun}, Guochao and {Faucher-Gigu{\`e}re}, Claude-Andr{\'e} and {Hayward}, Christopher C. and {Shen}, Xuejian and {Wetzel}, Andrew and {Cochrane}, Rachel K.},
        title = "{Bursty Star Formation Naturally Explains the Abundance of Bright Galaxies at Cosmic Dawn}",
      journal = {\apjl},
         year = 2023,
        month = oct,
       volume = {955},
       number = {2},
          eid = {L35},
        pages = {L35},
          doi = {10.3847/2041-8213/acf85a},
archivePrefix = {arXiv},
       eprint = {2307.15305},
 primaryClass = {astro-ph.GA},
       adsurl = {https://ui.adsabs.harvard.edu/abs/2023ApJ...955L..35S}
}

@ARTICLE{Sutherland_1993,
       author = {{Sutherland}, Ralph S. and {Dopita}, M.~A.},
        title = "{Cooling Functions for Low-Density Astrophysical Plasmas}",
      journal = {\apjs},
         year = 1993,
        month = sep,
       volume = {88},
        pages = {253},
          doi = {10.1086/191823},
       adsurl = {https://ui.adsabs.harvard.edu/abs/1993ApJS...88..253S}
}

@ARTICLE{Taylor_2024,
       author = {{Taylor}, Anthony J. and {Finkelstein}, Steven L. and {Kocevski}, Dale D. and {Jeon}, Junehyoung and {Bromm}, Volker and {Amorin}, Ricardo O. and {Arrabal Haro}, Pablo and {Backhaus}, Bren E. and {Bagley}, Micaela B. and {Ba{\~n}ados}, Eduardo and {Bhatawdekar}, Rachana and {Brooks}, Madisyn and {Calabro}, Antonello and {Chavez Ortiz}, Oscar A. and {Cheng}, Yingjie and {Cleri}, Nikko J. and {Cole}, Justin W. and {Davis}, Kelcey and {Dickinson}, Mark and {Donnan}, Callum and {Dunlop}, James S. and {Ellis}, Richard S. and {Fernandez}, Vital and {Fontana}, Adriano and {Fujimoto}, Seiji and {Giavalisco}, Mauro and {Grazian}, Andrea and {Guo}, Jingsong and {Hathi}, Nimish P. and {Holwerda}, Benne W. and {Hirschmann}, Michaela and {Inayoshi}, Kohei and {Kartaltepe}, Jeyhan S. and {Khusanova}, Yana and {Koekemoer}, Anton M. and {Kokorev}, Vasily and {Larson}, Rebecca L. and {Leung}, Gene C.~K. and {Lucas}, Ray A. and {McLeod}, Derek J. and {Napolitano}, Lorenzo and {Onoue}, Masafusa and {Pacucci}, Fabio and {Papovich}, Casey and {P{\'e}rez-Gonz{\'a}lez}, Pablo G. and {Pirzkal}, Nor and {Somerville}, Rachel S. and {Trump}, Jonathan R. and {Wilkins}, Stephen M. and {Yung}, L.~Y. Aaron and {Zhang}, Haowen},
        title = "{Broad-Line AGN at $3.5<z<6$: The Black Hole Mass Function and a Connection with Little Red Dots}",
      journal = {arXiv e-prints},
         year = 2024,
        month = sep,
          eid = {arXiv:2409.06772},
        pages = {arXiv:2409.06772},
          doi = {10.48550/arXiv.2409.06772},
archivePrefix = {arXiv},
       eprint = {2409.06772},
 primaryClass = {astro-ph.GA},
       adsurl = {https://ui.adsabs.harvard.edu/abs/2024arXiv240906772T}
}

@ARTICLE{Trussler_2020,
       author = {{Trussler}, James and {Maiolino}, Roberto and {Maraston}, Claudia and {Peng}, Yingjie and {Thomas}, Daniel and {Goddard}, Daniel and {Lian}, Jianhui},
        title = "{Both starvation and outflows drive galaxy quenching}",
      journal = {\mnras},
         year = 2020,
        month = feb,
       volume = {491},
       number = {4},
        pages = {5406-5434},
          doi = {10.1093/mnras/stz3286},
archivePrefix = {arXiv},
       eprint = {1811.09283},
 primaryClass = {astro-ph.GA},
       adsurl = {https://ui.adsabs.harvard.edu/abs/2020MNRAS.491.5406T}
}

@ARTICLE{Trussler_2024,
       author = {{Trussler}, James A.~A. and {Conselice}, Christopher J. and {Adams}, Nathan and {Austin}, Duncan and {Ferreira}, Leonardo and {Harvey}, Tom and {Li}, Qiong and {Vijayan}, Aswin P. and {Wilkins}, Stephen M. and {Windhorst}, Rogier A. and {Bhatawdekar}, Rachana and {Cheng}, Cheng and {Coe}, Dan and {Cohen}, Seth H. and {Driver}, Simon P. and {Frye}, Brenda and {Grogin}, Norman A. and {Hathi}, Nimish and {Jansen}, Rolf A. and {Koekemoer}, Anton and {Marshall}, Madeline A. and {Nonino}, Mario and {Ortiz}, Rafael and {Pirzkal}, Nor and {Robotham}, Aaron and {Ryan}, Russell E. and {D'Silva}, Jordan C.~J. and {Summers}, Jake and {Tompkins}, Scott and {Willmer}, Christopher N.~A. and {Yan}, Haojing},
        title = "{EPOCHS IX. When cosmic dawn breaks: evidence for evolved stellar populations in 7 < z < 12 galaxies from PEARLS GTO and public NIRCam imaging}",
      journal = {\mnras},
         year = 2024,
        month = feb,
       volume = {527},
       number = {4},
        pages = {11627-11650},
          doi = {10.1093/mnras/stad3877},
archivePrefix = {arXiv},
       eprint = {2308.09665},
 primaryClass = {astro-ph.GA},
       adsurl = {https://ui.adsabs.harvard.edu/abs/2024MNRAS.52711627T}
}

@ARTICLE{Turner_2025,
       author = {{Turner}, Jack C. and {Roper}, Will J. and {Vijayan}, Aswin P. and {Newman}, Sophie L. and {Wilkins}, Stephen M. and {Lovell}, Christopher C. and {Liao}, Shihong and {Seeyave}, Louise T.~C.},
        title = "{The Nature of High-Redshift Massive Quiescent Galaxies -- Searching for RUBIES-UDS-QG-z7 in FLARES}",
      journal = {arXiv e-prints},
         year = 2025,
        month = sep,
          eid = {arXiv:2509.16111},
        pages = {arXiv:2509.16111},
archivePrefix = {arXiv},
       eprint = {2509.16111},
 primaryClass = {astro-ph.GA},
       adsurl = {https://ui.adsabs.harvard.edu/abs/2025arXiv250916111T}
}

@ARTICLE{Valentino_2023,
       author = {{Valentino}, Francesco and {Brammer}, Gabriel and {Gould}, Katriona M.~L. and {Kokorev}, Vasily and {Fujimoto}, Seiji and {Jespersen}, Christian Kragh and {Vijayan}, Aswin P. and {Weaver}, John R. and {Ito}, Kei and {Tanaka}, Masayuki and {Ilbert}, Olivier and {Magdis}, Georgios E. and {Whitaker}, Katherine E. and {Faisst}, Andreas L. and {Gallazzi}, Anna and {Gillman}, Steven and {Gim{\'e}nez-Arteaga}, Clara and {G{\'o}mez-Guijarro}, Carlos and {Kubo}, Mariko and {Heintz}, Kasper E. and {Hirschmann}, Michaela and {Oesch}, Pascal and {Onodera}, Masato and {Rizzo}, Francesca and {Lee}, Minju and {Strait}, Victoria and {Toft}, Sune},
        title = "{An Atlas of Color-selected Quiescent Galaxies at z > 3 in Public JWST Fields}",
      journal = {\apj},
         year = 2023,
        month = apr,
       volume = {947},
       number = {1},
          eid = {20},
        pages = {20},
          doi = {10.3847/1538-4357/acbefa},
archivePrefix = {arXiv},
       eprint = {2302.10936},
 primaryClass = {astro-ph.GA},
       adsurl = {https://ui.adsabs.harvard.edu/abs/2023ApJ...947...20V}
}

@ARTICLE{Vandevoort_2017,
       author = {{van de Voort}, Freeke and {Bah{\'e}}, Yannick M. and {Bower}, Richard G. and {Correa}, Camila A. and {Crain}, Robert A. and {Schaye}, Joop and {Theuns}, Tom},
        title = "{The environmental dependence of gas accretion on to galaxies: quenching satellites through starvation}",
      journal = {\mnras},
         year = 2017,
        month = apr,
       volume = {466},
       number = {3},
        pages = {3460-3471},
          doi = {10.1093/mnras/stw3356},
archivePrefix = {arXiv},
       eprint = {1611.03870},
 primaryClass = {astro-ph.GA},
       adsurl = {https://ui.adsabs.harvard.edu/abs/2017MNRAS.466.3460V}
}

@ARTICLE{Vani_2025,
       author = {{Vani}, Akash and {Ayromlou}, Mohammadreza and {Kauffmann}, Guinevere and {Springel}, Volker},
        title = "{Probing galaxy evolution from z = 0 to z ≃ 10 through galaxy scaling relations in three L-GALAXIES flavours}",
      journal = {\mnras},
         year = 2025,
        month = jan,
       volume = {536},
       number = {1},
        pages = {777-806},
          doi = {10.1093/mnras/stae2625},
archivePrefix = {arXiv},
       eprint = {2408.00824},
 primaryClass = {astro-ph.GA},
       adsurl = {https://ui.adsabs.harvard.edu/abs/2025MNRAS.536..777V}
}

@ARTICLE{Visser_2025,
       author = {{Visser-Zadvornyi}, Anatolii I. and {Carstairs}, Mary E. and {Oman}, Kyle A. and {Verheijen}, Marc A.~W.},
        title = "{Star formation and stellar \& AGN feedback in the absence of accretion, not gas stripping, set the quenching timescale in satellite galaxies}",
      journal = {arXiv e-prints},
         year = 2025,
        month = mar,
          eid = {arXiv:2503.15183},
        pages = {arXiv:2503.15183},
          doi = {10.48550/arXiv.2503.15183},
archivePrefix = {arXiv},
       eprint = {2503.15183},
 primaryClass = {astro-ph.GA},
       adsurl = {https://ui.adsabs.harvard.edu/abs/2025arXiv250315183V}
}

@ARTICLE{White_1991,
       author = {{White}, Simon D.~M. and {Frenk}, Carlos S.},
        title = "{Galaxy Formation through Hierarchical Clustering}",
      journal = {\apj},
         year = 1991,
        month = sep,
       volume = {379},
        pages = {52},
          doi = {10.1086/170483},
       adsurl = {https://ui.adsabs.harvard.edu/abs/1991ApJ...379...52W}
}
